\documentclass[reprint,onecolumn,superscript address, nofootinbib]{revtex4-2}
\usepackage{amsmath}
\usepackage{mathtools}
\usepackage{graphicx}
\usepackage{amsfonts}
\usepackage{color}
\usepackage{microtype} 
\usepackage{subcaption} 
\usepackage{array}
\usepackage{tabularx}
\usepackage{booktabs} 
\usepackage{cryptocode}
\usepackage{xspace}
\usepackage{enumitem}  
\usepackage{verbatim} 

\usepackage{xparse} 
\usepackage{xspace}
\usepackage{braket}
\usepackage{physics}
\usepackage{bm}

\usepackage{float}
\usepackage{mdframed}

\makeindex
\usepackage{amsthm}
\usepackage{thmtools}
\usepackage{thm-restate}
\usepackage{tikz}
\usetikzlibrary{positioning, calc, shapes.geometric, backgrounds,fit}

\usepackage{titlesec}

\makeatletter
\renewcommand{\p@subsection}{}
\renewcommand{\p@subsubsection}{}
\makeatother
\titleformat*{\paragraph}{\bfseries}

\usepackage{hyperref} 
\usepackage[capitalize]{cleveref} 

\crefname{enumi}{}{} 
\hypersetup{
	colorlinks = true,
	linkcolor =blue,
	citecolor=blue, 
	urlcolor=blue 
}

\newlength\colsep   
\newlength\offsetX  
\newlength\offsetY  

\newtheorem{theorem}{Theorem}[section]
\newtheorem{lemma}{Lemma}[section]
\newtheorem{corollary}{Corollary}[section]

\newtheorem{remark}{Remark}[section]

\theoremstyle{definition} 
\newtheorem{definition}{Definition}[section]
\newtheorem{prot}{Protocol}

\newcommand{\Renyi}{R\'{e}nyi} 
\newcommand{\mbf}[1]{\mathbf{#1}} 
\newcommand{\bsym}[1]{\boldsymbol{#1}} 

\newcommand{\sourcesymbol}{\sigma} 
\newcommand{\testgenflag}{z}

\newcommand{\term}[1]{\textbf{#1}} 
\newcommand{\defvar}{\coloneqq} 

\newcommand{\todo}[1]{{\color{red}#1}}

\newcommand{\Gammacc}{M_{\mathrm{o}}} 
\newcommand{\finGammacc}{F_{\mathrm{o}}} 
\newcommand{\pibar}{{\overline{\Pi}}} 
\newcommand{\pibarFlag}{\pibar^{\mathrm{Sq}}} 

\newcommand{\id}{\mathbb{I}} 
\newcommand{\Pos}{\operatorname{Pos}} 
\newcommand{\dop}[1]{\operatorname{S}_{#1}} 

\newcommand{\supp}{\operatorname{supp}} 

\newcommand{\floor}[1]{\left\lfloor #1 \right\rfloor} 
\newcommand{\ceil}[1]{\left\lceil #1 \right\rceil} 
\newcommand{\realistic}{practical}

\newcommand{\eps}{\varepsilon} 
\newcommand{\epssecure}{\eps^\mathrm{secure}}
\newcommand{\epscorr}{\eps^\mathrm{correct}}
\newcommand{\epssecret}{\eps^\mathrm{secret}}
\newcommand{\epsEV}{\eps_\mathrm{EV}}
\newcommand{\epsPA}{\eps_\mathrm{PA}}

\newcommand{\CAs}{{C_{A \rightarrow E}}}
\newcommand{\CBs}{{C_{B \rightarrow E}}}
\newcommand{\CAr}{{C_{E \rightarrow A}}}
\newcommand{\CBr}{{C_{E \rightarrow A}}}

\newcommand{\complement}{\mathrm{C}} 

\newcommand{\Hinf}{\mathcal{H}_{\infty}} 
 
\newcommand{\CEC}{C_\mathrm{EC}}
\newcommand{\CEV}{C_\mathrm{EV}}
\newcommand{\Cauth}{\bm{C_\mathrm{auth}}}
\newcommand{\HEV}{H_\mathrm{EV}}
\newcommand{\HPA}{H_\mathrm{PA}}

\newcommand{\leak}{\lambda_\mathrm{EC}}
\newcommand{\leakfixed}{\lambda_\mathrm{EC}^\mathrm{fixed}} 
\newcommand{\lkey}{\ell} 
\newcommand{\lfixed}{\ell_{\mathrm{fixed}}} 
\newcommand{\PAstring}{\mbf{S}} 

\newcommand{\block}{M_\mathrm{block}}
\newcommand{\freq}{\operatorname{freq}}

\newcommand{\Xbasis}{\mathsf{X}}
\newcommand{\Zbasis}{\mathsf{Z}}

\newcommand{\alphCP}{\widehat{\mathcal{C}}}
\newcommand{\cP}{\hat{c}}

\newcommand{\randregtilde}[1]{\tilde{R}^\mathrm{#1}}
\newcommand{\gen}{\monofont{gen}}

\newcommand{\accept}{\monofont{accept}\xspace}
\newcommand{\abort}{\monofont{abort}\xspace}
\newcommand{\authabort}{\monofont{auth-abort}\xspace}

\newcommand{\flagkey}{F_\mathrm{K}}
\newcommand{\flagEV}{F_\mathrm{EV}}
\newcommand{\flagEC}{F_\mathrm{EC}}

\newcommand{\Sigmab}{\overline{\Sigma}}

\newcommand{\Sacc}{S_{\mathrm{acc}}} 
\newcommand{\convSacc}{S^{\mathrm{conv}}_{\mathrm{acc}}} 

\newcommand{\beamsplit}{t} 
\newcommand{\Sclick}{S_\mathrm{click}} 
\newcommand{\lambdamin}{{\lambda_{\min}}}

\newcommand{\Ameas}{\bar{A}}
\newcommand{\Ashield}{\hat{A}}

\newcommand{\tracedist}[1]{\frac{1}{2} \norm{#1}_1}

\newcommand{\ndecoy}{{N_{\mathrm{ph}}}} 

\newcommand*\widebar[1]{%
  \hbox{%
    \vbox{%
      \hrule height 0.5pt 
      \kern0.5ex
      \hbox{%
        \kern-0.1em
        \ensuremath{#1}%
        \kern-0.1em
      }%
    }%
  }%
}

\newcommand{\cobs}{\cP_1^n}

\newcommand{\monofont}[1]{\texttt{\textbf{#1}}}

\renewcommand{\selectlanguage}[1]{} 
\UseRawInputEncoding 

\usepackage{tikz}

\usepackage{tikz-cd}

\usetikzlibrary{positioning,arrows.meta,shapes.geometric}

\tikzset{
  pics/detector/.style args={#1,#2,#3}{
    code={
      \draw[line width=0.5mm] (0,-1) -- (0,1);
      \draw[line width=0.5mm] (0,1) -- ++(0.1,0);
      \draw[line width=0.5mm] (0,-1) -- ++(0.1,0);
      \draw[line width=0.5mm] (0.1,1) arc (90:-90:1); 
      \node[#3, align=center] (#1) at (0.5,0) {#2};
    }
  }
}

\tikzset{
  pics/beamsplitter/.style args={#1,#2,#3}{
    code={
    \coordinate (A) at (0.3,0.3);
    \coordinate (B) at (-0.3,-0.3);
    \coordinate (#1) at (0,0);
    
    \draw[line width=0.3mm] (A) -- (B); 
    \node[align=center,font=\fontsize{7}{7}\selectfont] at #3 {#2};
    }
  }
}

\tikzset{
  pics/pulse/.style args = {#1}{
    code ={
    \draw[#1] plot[smooth,tension=1] coordinates {(0,0) (0.3,0.2) (0.5,1) (0.7,0.2) (1,0)};
    }
  }
}

\tikzstyle{process} = [rectangle, line width=0.3mm, minimum width=1cm, minimum height=1cm, text centered, text width=1cm, draw=black]

\newcommand{\doublewidehat}[1]{%
  \begingroup
    \setbox0=\hbox{$#1$}%
    \rlap{\raise.47ex\hbox{$\smash{\widehat{\phantom{\copy0}}}$}}
    \rlap{\hbox{$\smash{\widehat{\copy0}}$}}
    \copy0
  \endgroup
}

\begin{document}
	\title{A rigorous and complete security proof of decoy-state BB84 quantum key distribution}
 \author{Devashish Tupkary}
\email{djtupkar@uwaterloo.ca}
	 \affiliation{Institute for Quantum Computing and Department of Physics and Astronomy, University of Waterloo, Waterloo, Ontario, Canada, N2L 3G1}	
 \author{Shlok Nahar}
	 \email{sanahar@uwaterloo.ca}
	 \affiliation{Institute for Quantum Computing and Department of Physics and Astronomy, University of Waterloo, Waterloo, Ontario, Canada, N2L 3G1}	
 \author{Amir Arqand}
	 \email{aarqand@uwaterloo.ca}
	 \affiliation{Institute for Quantum Computing and Department of Physics and Astronomy, University of Waterloo, Waterloo, Ontario, Canada, N2L 3G1}	
  \author{Ernest Y.-Z. Tan}
	 \email{yzetan@uwaterloo.ca}
	 \affiliation{Institute for Quantum Computing and Department of Physics and Astronomy, University of Waterloo, Waterloo, Ontario, Canada, N2L 3G1}	
      \affiliation{Department of Physics, National University of Singapore, Singapore, Singapore, Singapore, 117542}
	 \author{Norbert L\"utkenhaus}
	 \email{lutkenhaus.office@uwaterloo.ca}
	 \affiliation{Institute for Quantum Computing and Department of Physics and Astronomy, University of Waterloo, Waterloo, Ontario, Canada, N2L 3G1}
 
\begin{abstract}
We present a rigorous and complete security proof of the decoy-state BB84 quantum key distribution (QKD) protocol. Our analysis aims to achieve a high standard of mathematical rigour and completeness, thereby providing the necessary foundation for certification and standardization efforts. Beyond establishing the security of a specific protocol, this work develops a general and modular framework that can be readily adapted to a broad class of QKD protocols, including both prepare-and-measure and entanglement-based variants. Our framework unifies all major ingredients required for the analysis of realistic QKD protocols, including the analysis of classical authentication and classical processing, source-replacement schemes, finite-size analysis, source maps, squashing maps, and decoy-state techniques. In doing so, this work consolidates a diverse range of techniques scattered across the QKD literature into a unified formalism, representing a general and rigorous treatment of QKD security. Finally, it outlines a clear path towards incorporating practical imperfections within the same framework, thereby laying the groundwork for addressing implementation security in future analysis. 
\end{abstract}
\maketitle

\tableofcontents

\section{Introduction}

Over the past few decades, quantum key distribution (QKD) has evolved from a theoretical concept \cite{Bennett_2014} into a mature technology poised for deployment and commercial use. A critical milestone that must be achieved before QKD can be widely adopted is the certification of QKD devices by a relevant certification authority. A key ingredient of this step is the existence of a rigorous and complete security proof for the underlying QKD protocol, which can be scrutinized and vetted by the wider community \cite{bundPositionPaper}. Such a proof must satisfy several key criteria, as outlined in Refs.~\cite{bundPositionPaper,tupkary2025qkdsecurityproofsdecoystate}:
\begin{itemize}
    \item It must specify a complete protocol relevant to practical implementations.
    \item It must clearly state all assumptions about the devices used. 
    \item It must precisely state the security criterion to be satisfied.
    \item It must provide a rigorous mathematical proof that, under the stated assumptions, the protocol achieves the desired level of security with the specified parameters.
\end{itemize}
As discussed in Ref.~\cite{tupkary2025qkdsecurityproofsdecoystate}, no existing security analysis fully meets all of the above requirements simultaneously. A recent attempt to address this gap is Ref.~
\cite{mizutani2025protocolleveldescriptionselfcontainedsecurity}, which is currently under active scrutiny by the community. Ref.~\cite{tomamichel_largely_2017} presents another exemplary and fully rigorous security proof; however, it applies to an idealized qubit-based BB84 protocol and does not address the practical decoy-state setting.

In this work, we present a rigorous and complete security proof for a fully specified instance of the decoy-state BB84 protocol \cite{Lo_decoystate_2005,Hwang_qkdiwthhighloss_2003,Ma_practicaldecoy_2005,Bennett_2014}. To support this goal, our approach extends far beyond this specific case: we consolidate several key elements required for a modern QKD security analysis into a unified and modular framework. Our treatment encompasses all the major components essential for the analysis of realistic QKD protocols, including classical authentication and postprocessing \cite{portmann_key_2014,fung_practical_2010}, finite-size effects \cite{renner_security_2005,scarani_security_2008,hayashi_upper_2007,cai_finitekey_2009,curty_finitekey_2014,hayashi_concise_2012,hayashi_security_2014,lim_concise_2014,tomamichel_largely_2017,george_numerical_2021,metger_security_2023,dupuis_entropy_2020}, source-replacement schemes \cite{bennett_quantum_1992,curty_entanglement_2004}, source maps \cite{gottesman_security_2004}, squashing maps \cite{tsurumaru_security_2008,beaudry_squashing_2008,tsurumaru_squash_2010,gittsovich_squashing_2014,zhang_security_2021} and decoy-state methods. Consequently, the framework we develop can be readily adapted to obtain security proofs for other QKD protocols, as we illustrate throughout this work. In particular, \cref{subsubsec:recipe} provides a concrete recipe that can be followed to derive security proofs for a broad class of QKD protocols. In this sense, this work offers a general, modular, and rigorous analysis of practical QKD protocols. We emphasize that this analysis relies heavily on prior work in which many of the individual ingredients required for QKD security proofs were developed (and often combined). The primary contribution of this work is the clean and fully explicit integration of these ingredients into a single, coherent, rigorous and complete security analysis.

Modern QKD theory offers several distinct proof techniques \cite{koashi_simple_2005,koashi_simple_2009,nahar_postselection_2024,christandl_postselection_2009,tomamichel_uncertainty_2011,dupuis_entropy_2020,metger_generalised_2022,arqand_marginal_2025,inprep_vanhimbeeck_tight_2024}, each with its own advantages and limitations (see \cite[Section VII]{tupkary2025qkdsecurityproofsdecoystate}). In this work, we employ the recently developed marginal-constrained entropy accumulation theorem (MEAT) \cite{arqand_marginal_2025} for our analysis. This approach is motivated by its ability to yield tight key rates \cite{kamin_renyi_2025}, accommodate a broad range of protocol variations, and maintain a modular structure (see \cref{subsec:discussionabstractproof}). 
Moreover, it enables the framework developed here to be readily extended to incorporate device imperfections \cite{kamin_renyi_2025} , providing a systematic foundation for future implementation-level security analyses.
The main drawbacks of this approach are its relatively recent development (limited external scrutiny) and its reliance on numerical optimization to extract key rates. We believe, however, that the former limitation will naturally diminish with time and further community validation.   Furthermore, the required numerical optimizations are performed only once during the security analysis phase, and not during the execution of the QKD protocol itself.  Overall, the use of the MEAT enables us to incorporate 
a number of important features (explained in more detail later in this work) such as 
on-the-fly announcements, fully adaptive key-rates, ability to handle channel variability, and robustness to source and detector imperfections, while maintaining compatibility with a wide range of state preparation, measurement, and classical post-processing choices (which are also allowed to vary across rounds).

We note that the MEAT was first applied to the security analysis of decoy-state BB84 in Ref.~\cite{kamin_renyi_2025}, which also developed numerical techniques for the evaluation of finite-size key rates (see also Ref.~\cite{navarro_finite_2025}), and which we leverage in this work. The key rates obtained in the present work match those reported in Ref.~\cite{kamin_renyi_2025}. Ref.~\cite{kamin_renyi_2025} treats both variable-length and fixed-length protocols, and explicitly analyzes phase and intensity imperfections. In comparison, the analysis presented here does not explicitly model device imperfections (it only provides a sketch). However, it is able to handle on-the-fly announcements (which arises from the way the MEAT is invoked in the security analysis) and is more general in its treatment of several structural aspects of the security proof, such as squashing maps and source maps. Moreover, it addresses certain technical issues related to classical authentication and the treatment of infinite-dimensional systems. 
The emphasis of the present work is therefore complementary: we aim to provide a fully specified protocol together with a rigorous, self-contained security analysis in which all assumptions, protocol steps, and proof components are made explicit and assembled into a single coherent framework, which is not the goal of Ref.~\cite{kamin_renyi_2025}.

\subsubsection*{Organization of this work}
This work is organized as follows. In \cref{sec:notation}, we begin with several tables of notation designed to serve as a convenient reference throughout this work. In \cref{sec:securitydefinition}, we state the security definition of QKD used in this work and provide the mathematical background and notation required to understand it.
In \cref{sec:protocoldesc}, we describe the QKD protocol in two stages. First, in \cref{subsec:abstractprotdesc}, we present a general formulation of a QKD protocol without fixing the specific states, measurements, or classical post-processing steps --- leaving these components abstract. Then, in \cref{subsec:detailprotdesc}, we instantiate this framework with the decoy-state BB84 protocol by specifying more details about the states sent, measurements performed, and classical processing steps. This two-stage presentation allows us to perform the security analysis in a correspondingly modular way, enhancing both clarity and generality.
In \cref{sec:reductionstatement}, we address the problem of authentication aborts and use the results of Ref.~\cite{inprep_authentication} to show that, without loss of generality, the QKD security analysis can be carried out under the assumption of honest authentication: that is, authentication that never aborts and transmits messages faithfully, and does not tamper with message timings.
In \cref{sec:backgroundinfotheory}, we summarize the relevant tools and definitions from quantum information theory that are used in our security analysis of the QKD protocol. In \cref{sec:sketch}, we provide a brief sketch of the proof for fixed-length protocols to introduce the main ideas and intuition gradually. The full proof appears in \cref{sec:proof}, where we rigorously establish the security of the generic QKD protocol (as defined in \cref{subsec:abstractprotdesc}) under the assumption that Alice sends finite-dimensional states and Bob performs finite-dimensional measurements. This culminates in \cref{theorem:abstractsecuritystatement}, which states the central security result.
In \cref{sec:optics}, we address the infinite-dimensional nature of practical implementations and introduce the notions of source maps and squashing maps to reduce the analysis to the finite-dimensional case. This is then followed by the final security proof for the decoy-state BB84 protocol (described in \cref{subsec:detailprotdesc}), which is presented in \cref{theorem:decoystatebb84securitystatement}.
To extract concrete key rates, our results require solving a finite-dimensional convex 
optimization problem. In practice, any suitable numerical method may be employed, provided 
it satisfies certain requirements - namely, that it guarantees a reliable lower bound for the 
relevant minimization problem. In \cref{sec:numerics}, we outline these necessary requirements 
and caveats, and we also describe the main ideas behind one concrete approach based on the method of 
Ref.~\cite{kamin_renyi_2025,winick_reliable_2018}.  In \cref{sec:imperfections}, we discuss how the framework developed here can be extended to handle device imperfections; we provide various possible paths and leave a detailed analysis to future work. Finally, in \cref{sec:conclusion}, we present our concluding remarks. Additional details are provided in the Appendices.

\clearpage 
\section{Notation} \label{sec:notation}

\newcommand{\singleRoundBot}{\bot}

\newcommand{\CPTP}{\operatorname{CPTP}} 

\NewDocumentCommand{\Halpha}{o}{%
  \ensuremath{\widetilde{H}^{\uparrow\IfValueT{#1}{\,,#1}}_\alpha}%
  }

  \NewDocumentCommand{\HalphaDown}{o}{%
\ensuremath{\widetilde{H}^{\downarrow\IfValueT{#1}{\,,#1}}_\alpha}%
  }

\newcommand{\HalphaP}{\bar{H}_\alpha}

\newcommand{\hashfamily}[2]{\mathcal{F}_{\mathrm{hash}}\left(#1,#2\right)}
\newcommand{\idealhashfamily}[2]{\mathcal{F}^\mathrm{ideal}_{\mathrm{hash}}\left(#1,#2\right)}

\newcommand{\pf}{\operatorname{\mathtt{Pur}}}
\newcommand{\pfc}{\operatorname{\mathtt{Pur}_{\mathtt{C}}}}
\newcommand{\isoV}{\mathcal{V}}
\newcommand{\keyspace}{\mathcal{K}}
\newcommand{\idealmap}{\mathcal{R}_\mathrm{ideal}}

\newcommand{\OlAlB}{\Omega_{l_A,l_B}}
\newcommand{\measChannel}[1]{\mathcal{M}^{\mathrm{meas}}_{#1}}

\begin{table}[h]
    \centering
    \renewcommand{\arraystretch}{1.3} 
    \begin{tabularx}{\textwidth}{>{\raggedright\arraybackslash}p{4cm} X}
        \toprule
        \textbf{Symbol} & \textbf{Meaning} \\
        \midrule
        $A_j^k$ & registers $A_j\cdots A_k$.\\
            $\CPTP(A,B)$ & Set of CPTP maps from registers $A$ to $B$.\\
              $\Pos(A)$ & Set of positive operators on $A$. \\  
            $\dop{=}(A)$ & Set of normalized states (density operators) on register $A$. \\   
            $\dop{\leq}(A)$ & Set of sub-normalized states on register $A$. \\
            $\norm{X}_p$ & Schatten $p$-norm of $X$.\\
            $\operatorname{T}(\rho,\sigma)$ & Trace distance between $\rho$ and $\sigma$.\\
            $\rho_{A|\Omega}$ & States conditioned on the event $\Omega$.\\
            $K_A, K_B$  & Alice and Bob's key registers at the end of a QKD protocol. \\
            $\idealmap$ & Map acting on the real output states to give ideal output states. \\
            $\keyspace$ & Set of possible output key length combinations. \\ 
            $\tau^{l_A, l_B}_{K_A K_B}$ & Ideal output state on the key registers, storing keys of length $l_A,l_B$. \\
            $\OlAlB$ & Event that protocol ends with Alice and Bob producing keys of length $l_A$ and $l_B$ respectively. \\
            $\Omega_\mathrm{acc}$  & Event that a fixed-length protocol accepts and produces a key. \\
            $\lfixed$ & The output key length of a fixed-length QKD protocol (if protocol accepts). \\
            $\epssecure$ & Security parameter of the QKD protocol. \\
            $\measChannel{\{\Gamma_k\}} \in \CPTP(A,X)$ & A measurement channel that measures a register $A$ using POVM $\{\Gamma_k\}$, and stores the outcome in the register X. \\
        \bottomrule
    \end{tabularx}
    \caption{Notation introduced in \cref{sec:securitydefinition}.}
\end{table}

\newcommand{\aliceauthmap}  {\mathcal{E}^\mathrm{A}_\mathrm{auth-replace}}
\newcommand{\bobauthmap}{\mathcal{E}^\mathrm{B}_\mathrm{auth-replace}} 

\newcommand{\authmap}
{\mathcal{E}^\mathrm{repl}_\mathrm{auth}}

\newcommand{\identityMap}{\mathord{\rm id}}

\newcommand{\attack}[1]{\mathcal{A}_{#1}} 

\newcommand{\EfinalQKD}{\bm{E_\mathrm{fin}}}
 \newcommand{\CfinalQKD}{\bm{C_\mathrm{fin}}}  
 \newcommand{\Efinal}{\bm{E^\prime_{\mathrm{fin}}}} 
\newcommand{\Cfinal}{\bm{C^\prime_{\mathrm{fin}}}} 
\newcommand{\Ccorr}{\widetilde{C}}

\newcommand{\Esecdef}{\bm{E}}

\newcommand{\OmegaEV}{\Omega_\mathrm{EV}}
\newcommand{\OmegaAT}{\Omega_\mathrm{AT}}
\newcommand{\Omegadiff}{\Omega_\mathrm{diff}} 


\newcommand{\OlAlBprime}{\Omega_{l^\prime_A,l^\prime_B}}
\newcommand{\Onice}{\Omega_{\mathrm{auth\text{-}hon}}}
\newcommand{\Ocp}{\Omega_{\hat{c}_1^n}}
\newcommand{\CP}{\widehat{C}} 

\newcommand{\CPhat}{\widetilde{C}}
\newcommand{\QKDprotocol}{\mathcal{P}_\mathrm{QKD}}
\newcommand{\coreQKDprotocol}{\widetilde{\mathcal{P}}_\mathrm{QKD}}
\newcommand{\APPprotocol}{\mathcal{P}_\mathrm{APP}}
\newcommand{\worldreal}{\mathcal{W}^\mathrm{real}_\mathrm{auth}}
\newcommand{\worldhonest}{\mathcal{W}^\mathrm{hon}_\mathrm{auth}}

\newcommand{\announcementfunction}{f_\mathrm{ann}}
\newcommand{\keymapfunction}{f_\mathrm{kmap}}

\newcommand{\lprePA}{l_\mathrm{pre\text{-}PA}}

\begin{table}[h]
    \centering
    \renewcommand{\arraystretch}{1.3} 
    \begin{tabularx}{\textwidth}{>{\raggedright\arraybackslash}p{4cm} X}
        \toprule
        \textbf{Symbol} & \textbf{Meaning} \\
        \midrule
        \( n \in \mathbb{N}_0\) 			    &    Total number of rounds. \\
	\(\{\sigma^{(j)}_k\}_{k=1 \dots d_A}, p^{(j)}_k \)	& 	States sent by Alice in the $j$th round, along with their probabilities. Alice sends one of $d_A$ possible states states. \\
$A'_j$, $B_j$ & Registers in which Alice sends her $j$th round state, and Bob measures the $j$ round. \\
$\{M^{(B_j)}_k\}_{k = 1 \dots d_B}$ & POVM used by Bob for measurement in the $j$th round. \\
$X_j,Y_j$ & Registers in which Alice stores the labels of states sent, and Bob stores his measurement outcome. With alphabets $\mathcal{X},\mathcal{Y}$. \\
$\CP_j$ & Register storing Alice and Bob's public announcements for round $j$. With alphabet $\mathcal{\CP}$. \\
$E_n$ & Eve's quantum sideinformation register after all $n$ states are sent and received. (May contain copy of public announcements). \\
$\cobs$ & The value stored in the public announcement register $\CP_1^n$. \\
$\announcementfunction^{(j)}(\cdot,\cdot) : \mathcal{X} \times \mathcal{Y} \rightarrow \mathcal{\CP} $ & Function mapping Alice's and Bob's local data, to public announcements, for round $j$. \\
  \( \block \) & Blocksize of public announcements. \\
$t^\mathrm{A}_j$ & The (global) time when Alice sends the $j$th-round state.  \\
$t^\mathrm{B}_j$ & The (global) time when Bob measures the $j$th round state. \\
$t^\mathrm{ann}_j$ & The (global) time when Alice and Bobs start public announcements for the $j$th round. \\
$\keymapfunction^{(j)}(\cdot,\cdot) : \mathcal{X} \times \mathcal{\CP} \rightarrow \mathcal{S} $ & Function mapping Alice's local data stored in $X_j$ to $S_j$, based on public announcements stored in $\CP_j$, for round $j$.\\
$S_j$ & The pre-amplification string register, after key map but before sifting and discarding. Has alphabet \(\mathcal{S}\).				\\
$\PAstring$ & The pre-amplification string register after keymap, sifting and discarding. \\
	$\lkey(\cobs)$, $\flagkey$ & Function that determines length of output key based on public announcements, Register that stores this value. \\
    $\leak(\cobs)$, $\flagEC$ & Function that determines number of possible transcripts of the error-correction protocol based on public announcements, Register that stores this value.\\
    $\Ocp$ & Event that $\cobs$ is observed in public announcements. \\
    $\PAstring_B$& Bob's error-corrected pre-amplification string. \\
      $\lprePA$ & Length of Alice and Bob's pre-amplification strings. \\
      $\HEV$ & Register storing publicly announced hash choice for error-verification. \\
          $\OmegaEV$ & Event that error-verification accepts. \\
      $\CEV$ & Register storing publicly announced hash value during error-verification. \\
      $\flagEV$ & Register storing publicly announced result of error-verification. \\
      $\HPA$ & Register storing publicly announced hash choice for privacy amplification. \\
      $\Cauth$ & Classical register storing public communication during authentication postprocessing step of QKD protocol. \\
      \abort, \accept & Special messages sent by  a party when they wish to abort, accept a QKD protocol. \\
      \authabort & Special message received when authentication fails. Only received by the receiving party. \\
           $\hashfamily{l_\mathrm{in}}{l_\mathrm{out}}$ & universal$_2$ hash family from $l_\mathrm{in}$ bits to $l_\mathrm{out}$ bits. \\
           $\idealhashfamily{l_\mathrm{in}}{l_\mathrm{out}}$  & ideal universal$_2$ hash family from $l_\mathrm{in}$ bits to $l_\mathrm{out}$ bits. \\
               \bottomrule
    \end{tabularx}
    \caption{Notation introduced in \cref{subsec:abstractprotdesc}.} \label{table:abstractprodesc}
\end{table}

\begin{table}[h]
    \centering
    \renewcommand{\arraystretch}{1.3} 
    \begin{tabularx}{\textwidth}{>{\raggedright\arraybackslash}p{4cm} X}
        \toprule
        \textbf{Symbol} & \textbf{Meaning} \\
        \midrule
            $a,\mu,z$ & Choice of signal state : polarization, intensity, test vs gen. \\
        $\sigma_{(a,\mu,\testgenflag)}$ & Signal state for the decoy-state BB84 protocol. \\
           $\gamma$ & Probability of a test round. \\
           $p_{a,\mu | \mathrm{test}}$ , ($p_{a,\mu | \mathrm{gen}}$) & Probability of sending state in polarization $a$, and intensity $\mu$, conditioned on round being a $\mathrm{test}$ ($\mathrm{gen}$) round ).\\
           $t$ & beamsplitting ratio of Bob's beamsplitter in the passive detection setup. \\
           $\hat{a}_H, \hat{a}_V$ & annihilation operators for the horizontal and vertically polarized modes entering Bob's detection setup. \\
           $\keymapfunction^\mathrm{Bob}(\cdot,\cdot) : \mathcal{Y} \otimes \mathcal{\CP} \rightarrow \mathcal{S}$ & Function mapping Bob's local data $Y_j$ to $S_j$, based on public announcements stored in $\CP_j$, for round $j$. \\
    $f_\text{synd}^{\lprePA,\leak(\cobs)}(\cdot)$ & Function computing error syndrome for classical error correction. Maps $\{0,1\}^{\lprePA}$ to $\{0,1\}^{\leak(\cobs)}$. \\
$f_\text{corr}^{\lprePA,\leak(\cobs)}(\cdot,\cdot)$ & Function mapping Bob's sifted local data, and received error syndrome to the corrected pre-amplification string. Maps $ \{0,1\}^{\lprePA} \times \{0,1\}^{\leak(\cobs)} $ to $\{0,1\}^{\lprePA}$.  \\
        \bottomrule
    \end{tabularx}
    \caption{Notation introduced in \cref{subsec:detailprotdesc}.}
\end{table}

\begin{table}[h]
    \centering
    \renewcommand{\arraystretch}{1.3} 
    \begin{tabularx}{\textwidth}{>{\raggedright\arraybackslash}p{3cm} X}
        \toprule
        \textbf{Symbol} & \textbf{Meaning} \\
        \midrule
        $C^{(i)}_{A \rightarrow E}$ & Register storing $i$th message sent from Alice at (global) time $t^{(i)}_{A \rightarrow E}$. \\
         $C^{(i)}_{B \rightarrow E}$ & Register storing $i$th message sent from Bob, at (global) time  $t^{(i)}_{B \rightarrow E}$. \\  $C^{(i)}_{E \rightarrow A}$ & Register storing $i$th message received by Alice at (global) time $t^{(i)}_{E \rightarrow A}$. \\
         $C^{(i)}_{E \rightarrow B}$ & Register storing $i$th message received by Bob, at (global) time  $t^{(i)}_{E \rightarrow B}$. \\
       $\coreQKDprotocol$ & The core QKD protocol as described in \nameref{prot:abstractqkdprotocol}, excluding the final authentication post-processing step. Involves classical communication in registers $\CAs, \CAr, \CBs, \CBr$. \\       $\rho^\mathrm{real}_{K_A K_B \CfinalQKD \EfinalQKD}$ & The output state at the end of $\coreQKDprotocol$, before authentication postprocessing. \\
        $\CfinalQKD$ & Register storing all public communication during $\coreQKDprotocol$. \\
         $\EfinalQKD$ & Register storing Eve's quantum side-info at the end of $\coreQKDprotocol$, before authentication postprocessing.  \\  
         $\APPprotocol$ & The final step of \nameref{prot:abstractqkdprotocol}, called the authentication postprocessing step. \\
         $\QKDprotocol$ & The entirety of \nameref{prot:abstractqkdprotocol}. Is given by $\APPprotocol \circ \coreQKDprotocol$. \\
     $\Cfinal$ & Register storing all classical communication at the end of $\QKDprotocol$. \\
    $\Efinal$& Register storing Eve's quantum side-info at the end of the QKD protocol at the end of $\QKDprotocol$.\\ 
  $\rho^\mathrm{real,final}_{K_A K_B \Cfinal \Efinal}$ & State at the end of $\QKDprotocol$, i.e, at the end of \nameref{prot:abstractqkdprotocol}. \\
     $\worldreal$ & Set of final output states  when authentication can result in asymmetric aborts, and timings may be tampered. \\
    $ \worldhonest$ & Set of output states after the when authentication is not allowed to abort, and timings cannot be tampered. \\
        \bottomrule
    \end{tabularx}
    \caption{Notation introduced in \cref{sec:reductionstatement}.}
\end{table}

\newcommand{\kappafuncgeneric}[4]{\kappa\left( #1 ,\; #2, \; #3, \; #4 \right)}

\newcommand{\kappaQKDfunc}[3]{\kappa^\mathrm{QKD}\left( #1 , #2,  #3 \right)}

\newcommand{\fhat}{\hat{f}}
\newcommand{\fhatfull}{\fhat_\mathrm{full}}

\newcommand{\fhatfullQKD}{\fhat^\mathrm{QKD}_\mathrm{full}}

\begin{table}[h]
    \centering
    \renewcommand{\arraystretch}{1.3} 
    \begin{tabularx}{\textwidth}{>{\raggedright\arraybackslash}p{3cm} X}
        \toprule
        \textbf{Symbol} & \textbf{Meaning} \\
        \midrule
        $\Halpha,\HalphaDown$ & Sandwiched {\Renyi} entropies. \\
        $\Halpha[f]$,$\HalphaDown[f]$ & $f$-weighted {\Renyi} entropies. \\
            $\pf$ & Purifying function.\\
            $\freq_{z_1^n}$ & Frequency distribution of $z_1^n$. \\
     \bottomrule
    \end{tabularx}
    \caption{Notation introduced in \cref{sec:backgroundinfotheory}.}
\end{table}

\newcommand{\attackSquash}[1]{{\attack{#1}^{\mathrm{Sq}}}}
\newcommand{\nFSS}{N_\mathrm{FSS}}
\newcommand{\preservedSubspace}{\Pi_{m\leq \nFSS}}
\newcommand{\nonpreservedSubspace}{\Pi_{m> \nFSS}}

\newcommand{\QKDGmap}[1]{\mathcal{G}_{#1}} 

\newcommand{\QKDGmapfull}[1]{{\mathcal{G}}^{\mathrm{full}}_{#1}} 

\newcommand{\QKDGmapfullbeforeSR}[1]{\widetilde{{\mathcal{G}}}^{\mathrm{full}}_{#1}}

\newcommand{\QKDpostprocessingmap}{\mathcal{M}_\mathrm{CPP}} 
\newcommand{\sigmaconstraint}[1]{\sigma_{A_{#1}}^{(#1)}}

\newcommand{\attackset}[1]{\bm{\mathcal{A}}^\mathrm{set}_{#1}} 
\newcommand{\Qset}[1]{\bm{\mathcal{Q}}^\mathrm{set}_{#1}} 
\newcommand{\Qattack}[1]
{\mathcal{Q}_{#1}}

\begin{table}[h]
    \centering
    \renewcommand{\arraystretch}{1.3} 
    \begin{tabularx}{\textwidth}{>{\raggedright\arraybackslash}p{5cm} X}
        \toprule
        \textbf{Symbol} & \textbf{Meaning} \\
        \midrule
        $\CPhat_j$ & Register storing a copy of Alice and Bob's announcements for the $j$th round, that is given to Eve. \\
        $E_j$ & Eve's total side-information after $j$th round of the \nameref{prot:abstractqkdprotocol} (with modified timings as described in \cref{subsec:expressingprotocolassequence}). \\
        $E'_j$ & The side-information register of the  output of Eve's attack channel $\attack{j}$. We have $E'_j \CPhat_j = E_j$. \\
             $ \QKDGmapfullbeforeSR{j} \in \CPTP(X_j B_j, S_j X_j Y_j \CP_j \CPhat_j)$ & Map representing Alice and Bob's operations in the $j$th round, before source replacement. \\
             $\attack{l} \in \CPTP(E_{j-1}, B_j E'_j)$ & Map representing Eve's attack before $l$th round of the quantum phase of the QKD protocol. \\ 
               $\QKDpostprocessingmap$ & Map performing classical phase of the QKD protocol \\
                $\Ameas_j$ & Register that Alice measures, after source-replacement.\\
        $\Ashield_j$ & Register for Alice's shield system, which is prepared by Alice but cannot be touched by Eve. \\
        $\sourcesymbol^{(j)}_{\Ameas_j \Ashield_j A'_j}$ & The source-replaced state that is prepared by Alice in the $j$th round, in the \cref{prot:virtualprotocoll}. \\
                  $\QKDGmapfull{j} \in \CPTP(A_j B_j, S_j X_j Y_j \CP_j \CPhat_j)$ & Map representing Alice and Bob's measurement, public announcements and sifting operation, after using source-replacement scheme.\\
        $ \QKDGmap{j} \in \CPTP(A_j B_j, S_j \CP_j \CPhat_j)$ & Map representing Alice and Bob's operations in the $l$th round, given by $\Tr_{X_j Y_j} \circ \QKDGmapfull{j} $. \\
        $d_j \in \mathbb{N}$ & The dimension of $E'_j$, which denotes Eve's side-information register after her $j$th attack channel.  \\
       $\fhatfull(\cobs)$ & The global tradeoff function, which results in the global f-weighted {\Renyi} entropy being positive.\\  
       $\fhatfullQKD$ & The global tradeoff function used for QKD key rate computations, which is such that it lower bounds $\fhatfull$. \\
     $f_{|\cP_1^{j-1}}$ 
& Tradeoff function for round $j$, chosen as a function of the previous announcements $\cP_1^{j-1}$. 
\\
$\sigmaconstraint{j}$ & Alice's marginal state in the $j$th round. $A=\Ameas_j \Ashield_j$ denotes the registers in the source-replaced state that do not leave Alice's lab. \\ 
$\kappafuncgeneric{f_{|\cP_1^{j-1}}}{\sigma^{(j)}_{A_j}}{\QKDGmap{j}}{\attack{j}}$ 
& Quantity defined in \cref{theorem:MEATQKDfirst}, obtained by optimizing over all admissible single-round output states.  
When the final argument is a set of channels $\attackset{j}$, the value includes an additional infimum over all $\attack{j} \in \attackset{j}$. 
\\

$\Sigma_j(\attack{j})$ 
& The set of single-round output states whose infimum appears in  
$\kappafuncgeneric{f_{|\cP_1^{j-1}}}{\sigma^{(j)}_{A_j}}{\QKDGmap{j}}{\attack{j}}$.  If the argument is a set of channels $\attackset{j}$, then the set of states is the obtained for all possible $\attack{j} \in \attackset{j}$.
\\
           $\attackset{j}$ & Set of maps representing Eve's attack before $j$th round of the QKD protocol. \\ 
              $\Qattack{j}$ & Map representing Eve's attack before $j$th round of the QKD protocol. \\ 
           $\Qset{j}$ & Set of maps representing Eve's attack before $j$th round of the QKD protocol. \\ 
 
        \bottomrule
    \end{tabularx}
    \caption{Notation introduced in \cref{sec:proof}.}
\end{table}

\newcommand{\QKDmapfullwithoutBobMeas}[1]{{\bar{\mathcal{G}}}^{\mathrm{full}}_{#1}}

\begin{table}[h]
    \centering
    \renewcommand{\arraystretch}{1.3} 
    \begin{tabularx}{\textwidth}{>{\raggedright\arraybackslash}p{5cm} X}
        \toprule
        \textbf{Symbol} & \textbf{Meaning} \\
        \midrule
   $\xi_{X_1^n (A'')_1^n}$ & Virtual state preparation by Alice, after the usage of source maps. \\
   $\xi_{(a,\mu,\testgenflag)}
$ & Virtual states created by Alice. The tagging source map takes these states to $\sigma_{(a,\mu,\testgenflag)}$. \\
$\Psi_\mathrm{tag}$ & The tagging source maps, which tags higher photon numbers with classical flags. \\
$\QKDmapfullwithoutBobMeas{j} \in \CPTP(X_j Y_j, S_j X_j Y_j \CP_j \CPhat_j) $ & The QKD protocol map for round $j$, representing all operations after Bob's measurement, before source-replacement.\\ 
$\Lambda_j \in \CPTP(B_j, Q_j)$ & The squashing map for $j$th round. \\
$\attackSquash{j} \in \CPTP(E_{j-1}, Q_j E'_j) $ & Map representing Eve's attack on the $j$th round of the QKD protocol which uses squashed POVMs.  \\
$\nFSS$ & Photon number cutoff chosen for the flag state squasher.  \\
$M_{i,m \leq \nFSS},M_{i,m> \nFSS}$     & The two blocks of the Bob's POVM element, corresponding to outcome $i$, corresponding to photon number being $\leq \nFSS$ and $> \nFSS$. \\
$\preservedSubspace,\nonpreservedSubspace$ & Projectors onto the space of photon numbers being $\leq \nFSS$ and $> \nFSS \in \mathbb{N}$. \\
$ \pibarFlag$ & Projector onto the flag space of $Q$. \\
$\Gammacc^{(B)}$ & A specific POVM element, before squashing, used to bound weight of the incoming state in the higher photon number subspace ($\geq \nFSS$). \\
$\finGammacc^{(Q)} $ & The squashed POVM element, corresponding to $\Gammacc$. \\
$\lambdamin \in \mathbb{N}  $ &  Minimum non-zero eigenvalue of $\nonpreservedSubspace \Gammacc^{(B)} \nonpreservedSubspace$, outside the preserved subspace. \\
      \bottomrule
    \end{tabularx}
    \caption{Notation Introduced in \cref{sec:optics}.}
\end{table}

\newcommand{\choiform}{J}
\newcommand{\choiset}{\bm{J}^\mathrm{set}}
\newcommand{\funcJ}[1]{\phi_{\sigma_{AA'}}(#1)}
\newcommand{\ZMap}{\mathcal{Z}}
\newcommand{\QKDnumericsMap}{\mathcal{G}_\mathrm{nmr}}

\newcommand{\testset}{\mathcal{C}_\mathrm{test}}
\newcommand{\genset}{\mathcal{C}_\mathrm{gen}}

\newcommand{\testevent}{\Omega_\mathrm{test}}
\newcommand{\genevent}{\Omega_\mathrm{gen}}

\clearpage

\section{Security Definition of QKD} 
\label{sec:securitydefinition}
In this section, we state the security definition of QKD used in this work. We begin by introducing the necessary notation and mathematical background required to state this definition, and the rest of this work.

\subsection{Mathematical Machinery} \label{subsec:mathematical}
We typically use uppercase letters such as $X, A, B$ to denote registers. For a sequence of registers, we use subscripts to indicate indices: $X_j$ denotes the $j$-th register, while $X_i^j$ denotes the collection of registers $X_i, X_{i+1}, \ldots, X_j$.  We use $S_=(A)$ to denote the set of density operators on the register $A$, and $S_{\leq}(A)$ to denote the set of subnormalized density operators on the register $A$. We use $\Pos(A)$ to denote the set of positive semi-definite operators on the register $A$. Subscripts specify the registers on which the state exists; for example, $\rho_A$ denotes a state on register $A$. For a multipartite state $\rho_{ABC\cdots}$, we write $\rho_A$ to denote its marginal on register $A$. Conversely, for a state $\rho_A$, we use $\rho_{AB}$ to denote some extension of $\rho_A$ to an additional register $B$.  Whenever we write an entropic quantity, such as $\mathbb{H}(A|B)\rho$ (where $\mathbb{H}$ may represent any of the entropy measures defined in this work), we implicitly assume that it is evaluated on the appropriate marginal of the underlying state, for example, $\rho_{AB}$ in this case. We denote by $\CPTP(A,B)$ the set of CPTP maps from registers $A$ to $B$.

\begin{definition}
(Classical registers) Throughout this work, we suppose that every Hilbert space is additionally equipped with a particular basis that we shall call its \term{classical basis}. We say that a state $\rho \in \dop{\leq}(CQ)$ is \term{classical on $C$}, or alternatively that \term{$C$ is classical} in that state, if it has the form 
\begin{align}
\rho_{CQ} = \sum_c \lambda_c \ketbra{c} \otimes \sigma_c,
\label{eq:cq}
\end{align}
for some normalized states $\sigma_c \in \dop{=}(Q) $ and weights $\lambda_c \geq 0$, with $\ket{c}$ being the classical basis on $C$. In most circumstances, when writing states in the above form we will not explicitly declare that $\ket{c}$ (or similar) is the classical basis, leaving it as implicitly understood. Throughout this work, all registers denoted using the letter $C$ (including variants such as $\hat{C}$, $\tilde{C}$, etc.) will be classical registers.

\end{definition}
\begin{definition}\label{def:cond}
(Conditioning on classical events) For a state $\rho \in \dop{\leq}(CQ)$ classical on $C$, written in the form
$\rho_{CQ} = \sum_c \lambda_c \ketbra{c}{c} \otimes \sigma_c$ 
for some $\sigma_c \in \dop{=}(Q)$ and $\lambda_c \geq 0$,
and an event $\Omega$ defined on the register $C$, we will define a corresponding \term{conditional state} as,
\begin{align}
\rho_{|\Omega} \defvar \frac{\Tr[\rho]}{\Tr[\rho_{\land\Omega}]} \rho_{\land\Omega} = \frac{
\sum_{c} \Tr[ \lambda_c \sigma_c]
}{\sum_{c\in\Omega} \Tr[ \lambda_c \sigma_c]} \rho_{\land\Omega}, \qquad\qquad  \text{ where }
\rho_{\land\Omega} \defvar \sum_{c\in\Omega} \lambda_c \ketbra{c}{c} \otimes \sigma_c.
\end{align}

The notation corresponding to events without normalization ($\wedge \Omega)$ is commutative and associative, in the sense that for any events $\Omega,\Omega'$ we have $(\rho_{\land\Omega})_{\land\Omega'} = (\rho_{\land\Omega'})_{\land\Omega} = \rho_{\land(\Omega\land\Omega')}$; hence for brevity we will denote all of these expressions as
\begin{align}
\rho_{\land\Omega\land\Omega'} \defvar (\rho_{\land\Omega})_{\land\Omega'} = (\rho_{\land\Omega'})_{\land\Omega} = \rho_{\land(\Omega\land\Omega')}.
\end{align}
On the other hand, some disambiguating parentheses are needed when combined with taking conditional states (due to the normalization factors). 

\end{definition}
Given these definitions, a normalized state $\rho \in \dop{=}(CQ)$ that is classical on $C$ can be written in the form
\begin{align}\label{eq:cqstateprobs}
\rho_{CQ} = \sum_c \rho(c) \ketbra{c}{c} \otimes \rho_{Q|c},
\end{align}
where $\rho(c)$ denotes the probability of $C=c$ according to $\rho$, and $\rho_{Q|c}$ can indeed be interpreted as the conditional state on $Q$ corresponding to $C=c$, i.e.~$\rho_{Q|c} = \tr_C[{\rho_{QC|\Omega_{C=c}}}]$ where $\Omega_{C=c}$ is the event $C=c$. 
We also denote the distribution on $C$ induced by $\rho$ as the tuple
\begin{align}\label{eq:stateprobvec}
\bsym{\rho}_C \defvar \left(\rho(1),\rho(2),\dots\right).
\end{align}

\begin{definition}(Schatten $p$-norm)\label{def:schatten}
     For  every linear operator $X$, we define its \term{Schatten $p$-norm} as
    \begin{align}
        \label{eq:schatten}
        \norm{X}_p\defvar\left(\Tr \left[\abs{X}^p \right]\right)^{1/p},
    \end{align}
    where $p\in[1,\infty)$ and $\abs{X}=\sqrt{X^\dagger X}$.
\end{definition}
\begin{definition}\label{def:trace_dist}(Trace distance)
    For $\rho,\sigma \in \dop{=}(A)$, the \term{trace distance} is defined as
    \begin{align}
        \operatorname{T}(\rho,\sigma)=\frac{1}{2}\norm{\rho-\sigma}_1.
    \end{align}
\end{definition}

\begin{definition}\label{def:2universal}($\delta$-almost-universal$_2$ and universal$_2$ hash families) 
Given a domain $D$ and a finite codomain $D'$, a \term{$\delta$-almost-universal$_2$} ($\delta$-AU$_2$) hash family consists of a set $\mathcal{H}$ of functions $h:D \to D'$, together with a probability distribution $P_\mathrm{HASH}$ over $\mathcal{H}$. The defining property is that, if $h$ is sampled according to $P_\mathrm{HASH}$, then for all distinct $x,y \in D$, the probability of the hash function resulting in a collision is upper-bounded by $\delta$, that is, 
\begin{align}\label{eq:AUdefn}
\Pr[h(x) = h(y)] \leq \delta,
\end{align}
where the probability is taken over the choice of $h$. A special case is \term{universal$_2$ hashing}, which corresponds to
\begin{align}
\delta = \frac{1}{|D'|}.
\end{align}
A further special case is \term{ideal universal$_2$} hashing, which corresponds to $\delta = \frac{1}{|D'|}$ and equality holding in \cref{eq:AUdefn}, i.e.~for all distinct $x,y \in D$,
\begin{align}
\Pr[h(x) = h(y)] = \frac{1}{|D'|}.
\end{align}
\end{definition}
We note that Toeplitz hashing is known to be ideal universal$_2$ \cite{mansour_computational_1993}. 

\begin{definition}[POVM] \label{def:POVM}
    A \term{POVM} on a register $A$ is a set $\{\Gamma_k | \Gamma_k \in \Pos(A) \}_k$ of positive semidefinite operators that sum to the identity operator. That is, $\Gamma_k \in \Pos(A)$, and $\sum_{k} \Gamma_k = \id_A$.
\end{definition}

\begin{definition}[Measurement channel] \label{def:measurementchannels}
Let $\{\Gamma_k\}^m_{k=1}\subset \Pos(A)$ be a POVM. Let $X$ be a classical register of dimension $m$. We define the \term{measurement channel} $\measChannel{\{\Gamma_k\}}\in\CPTP(A,X)$ corresponding to the POVM $\{\Gamma_k\}_{k=1}^m$ to be the quantum channel which acts on any state as
$$\measChannel{\{\Gamma_k\}} \left[ \rho \right] = \sum_{k=1}^m \Tr[\rho\Gamma_k]\ketbra{k}_X.$$

\end{definition}

Having established the necessary machinery to state the QKD security definition, we will now briefly discuss the role of authentication in QKD, and its impact on the security definition of QKD. 
\subsection{Authentication in QKD}
 QKD protocols require access to an authenticated classical channel between the honest parties. 
The security definition of QKD used in the literature 
\cite{tupkary2025qkdsecurityproofsdecoystate,portmann_security_2022,ben-or_universal_2004}, 
and, more importantly, the associated security proofs, are tailored to an idealized setting in 
which authentication is assumed to behave “honestly”: it never aborts, and all classical messages are delivered faithfully with their original timing preserved\footnote{We call this the ``honest'' authentication setting, since it corresponds to the case where the eavesdropper does not attack the authentication}.  This assumption is unrealistic, since such a channel (or even something close to its functionality) cannot be constructed in practice. 

What can be constructed in practice is an authenticated channel that is close to the following ideal functionality: this ideal channel either transmits the message faithfully \emph{or} delivers a special symbol $\authabort$ to the receiving party (see \cref{sec:reductionstatement}), representing that authentication has failed \cite{portmann_security_2022,portmann_key_2014}. Notice that only the receiving party learns of the authentication failure, which can occur, for instance, when the authentication tag attached to a message does not match the expected value. This could arise either because the message was tampered with, or because an adversary attempted to inject a new message without the correct tag. In either case, the receiver discards the message and registers $\authabort$. Furthermore, Eve is permitted to delay messages, alter their ordering, or perform other timing-related manipulations (some of which may result in an $\authabort$ to the receiving party).

Thus in {\realistic} settings, authentication can fail asymmetrically (only the receiving party learns of an authentication abort). This allows the adversary to force asymmetric aborts on the underlying QKD protocol at will. This can be done for instance, by attacking the last message sent in the QKD protocol, which results in an \authabort. The receiving party then aborts the QKD protocol, while the sending party does not.\footnote{This example assumes that the honest parties abort the QKD protocol whenever they receive an \authabort. }

As a result, the standard QKD security definition used in the literature cannot be satisfied, since it accounts only for symmetric abort scenarios. Moreover, note that the adversary can influence the timing of classical messages. In the absence of the synchronized clocks assumption, Alice and Bob typically rely on exchanged messages to indicate the completion of various protocol steps; if the timing of these messages is perturbed, the relative ordering of their actions is also affected. Thus, in the {\realistic} authentication setting, one can no longer assume a fixed ordering of operations performed by Alice and Bob. This problem has already been noted \cite{portmann_security_2022}, and a resolution proposed in Ref.~\cite{inprep_authentication}, which we utilize in this work. 

 We proceed as follows. In \cref{subsec:securitydefinition}, we introduce the security definition for QKD from Ref.~\cite{ferradini2025definingsecurityquantumkey} that remains valid even when the protocol may experience asymmetric aborts. This definition reduces to the standard one commonly used in the literature \cite{portmann_security_2022,ben-or_universal_2004} under the assumption of 
honest authentication (and appropriate protocol design). By honest authentication, we mean the assumption that all classical messages are received \emph{correctly} by the receiving party, at some time \emph{after} they are sent by the sending party.\footnote{
Technically, we only require the authenticated channel to inform \textit{both} parties of authentication aborts for the QKD security definition used here to reduce to the usual one found in the literature. However, we state the honest authentication assumption here, which is stronger than merely requiring that both parties learn of authentication aborts, since QKD security analyses are typically performed under this assumption. Moreover, in \cref{sec:reductionstatement}, we reduce our security analysis to exactly this setting. }

Furthermore, we make use of Ref.~\cite{inprep_authentication}, which shows that security of  QKD protocol (under some minor conditions) in the {\realistic} authentication setting reduces to the security of essentially the same QKD protocol in the honest authentication setting.  The application of this result to our analysis is explained in \cref{sec:reductionstatement}.

\subsection{Security definition of QKD
} \label{subsec:securitydefinition}
Let us focus on the output state of a generic QKD protocol, defined on the registers $K_A K_B \Esecdef$, in the setting where asymmetric aborts are possible.  Alice and Bob possess classical registers $K_A$ and $K_B$. 
The classical registers $K_A$, $K_B$ encode keys of arbitrary length for Alice and Bob --- this is formalized by having $K_A$ consist of a direct sum $K_A = \bigoplus_{l_A} K_A^{l_A}$, where $K_A^{l_A}$ is a classical register holding keys of length $l_A$, and analogously for $K_B$. 
We treat any party aborting as them storing a key of length $0$ in their registers, which we denote with a special symbol $\bot$. The $\Esecdef$ register denotes all of Eve's information at the end of the QKD protocol, and may include a copy of the classical communications that occurred in the protocol. The output state can be written as
\begin{equation} \label{cref:realoutputstate}
\rho^\text{real}_{K_A K_B \Esecdef} \coloneq \bigoplus_{(l_A, l_B) \in \keyspace} \Pr(\OlAlB) \rho^\mathrm{real}_{K_A^{l_A} K_B^{l_B}  \Esecdef | \OlAlB}
\end{equation}
where $\OlAlB$ denotes the event that Alice and Bob produce a key of length $l_A$ and $l_B$ respectively, and $\keyspace$ denotes the set of possible output key length combinations and is given by
\begin{equation}
    \keyspace =  \{ (l_A,l_B) \; | \; l_A = l_B \;  \lor \; l_A = 0 \; \lor \; l_B = 0\},
\end{equation}
that is, Alice and Bob either share keys of the same length, or at least one of them abort the protocol.

The ideal output state $\rho^\mathrm{ideal}_{K_A K_B  \Esecdef}$ is defined to be the one obtained by acting a map $\idealmap \in \CPTP( K_A K_B, K_A K_B)$ acting on the real output state $\rho^\mathrm{real}_{K_A K_B  \Esecdef}$ as
\begin{equation}
       \rho^\mathrm{ideal}_{K_A K_B  \Esecdef} \defvar \idealmap\left[\rho^\mathrm{real}_{K_A K_B  \Esecdef} \right]. 
\end{equation}
The map $\idealmap$ does the following operations:
\begin{itemize}
    \item It looks at the length of the keys stored in registers $K_A, K_B$ to compute $l_A, l_B$. 
    \item It replaces the $K_A,K_B$ registers with the state 
  \begin{equation}\label{eq:taukAkB}
\begin{aligned}
\tau^{l_A, l_B}_{K_A K_B} &\defvar
\begin{cases}
\displaystyle \frac{1}{2^{l_A}} \sum_{k \in \{0,1\}^{l_A}} \ketbra{kk}_{K_A K_B}, 
& \text{if } l_A = l_B, \\[1.2em]
\tau^{l_A}_{K_A} \otimes \tau^{l_B}_{K_B}
& \text{if } l_A \neq l_B,
\end{cases} \\[1.2em]
\tau^{l_A}_{K_A} &\coloneq \frac{1}{2^{l_A}} \sum_{k \in \{0,1\}^{l_A}} \ketbra{k}_{K_A}, \qquad 
\tau^{l_B}_{K_B} \coloneq \frac{1}{2^{l_B}} \sum_{k \in \{0,1\}^{l_B}} \ketbra{k}_{K_B} .
\end{aligned}
\end{equation}
    That is, if the output lengths are the same, the key registers are replaced with perfectly uniform identical keys of that length, independent of all other registers. If the output key lengths are not the same, the key registers are \emph{individually} replaced with perfectly uniform keys of the corresponding lengths, independent of all other registers.  Note that we have
    \begin{equation}
        \tau^{l_A=0}_{K_A} \coloneq \ketbra{\bot}_{K_A},
    \end{equation}
    with Bob's key register defined analogously. 
\end{itemize}

Thus intuitively, any key obtained from the ideal state is safe to use, since it is ``secret'' against any side-information registers. The ideal state can be written as \begin{equation}
    \begin{aligned}
          \rho^\text{ideal}_{K_A K_B  \Esecdef}  &= \idealmap \left[ \rho^\text{real}_{K_A K_B \Esecdef}   \right] \\
        &= \sum_{l_A = l_B = l}  \Pr(\OlAlB)  \tau^{l,l}_{K_A K_B} \otimes \rho^\text{real}_{  | \OlAlB} + \sum_{l_A \neq l_B} \Pr(\OlAlB) \tau^{l_A}_{K_B} \otimes \tau^{l_B}_{K_B} \otimes \rho^\text{real}_{  \Esecdef | \OlAlB},
    \end{aligned}
\end{equation}
Moreover, $\idealmap$ can be written as 
\begin{equation} \label{eq:idealmapdecomp}
    \idealmap = \bigoplus_{l_A, l_B} \idealmap^{(l_A,l_B)},  \qquad \text{ where $\idealmap^{(l_A,l_B)} \in \CPTP(K_A^{l_A} K_B^{l_B} , K_A^{l_A} K_B^{l_B} )$.}
\end{equation}
We can now state the security definition of QKD. Our definition is formulated in terms of a QKD protocol $\QKDprotocol$, which describes all operations performed by Alice and Bob, and the corresponding set of possible output states, denoted by $\mathcal{W}(\QKDprotocol)$. This set contains all output states that can arise when $\QKDprotocol$ is executed, over all possible attacks by the adversary.

\begin{definition}[QKD Security with asymmetric aborts \cite{ferradini2025definingsecurityquantumkey}] \label{def:qkdsecurityasymmetric}
Let $\QKDprotocol$ be a QKD protocol, and let $\rho^\text{real}_{K_A K_B \Esecdef}$ be the output state of the QKD protocol, and let $\mathcal{W}\left( \QKDprotocol \right)$ denote the set of possible output states of the QKD protocol.  Let $\rho^\text{ideal}_{K_A K_B  \Esecdef}$ be the ideal output state, obtained by acting the map $\idealmap$ on the actual output state. That is
\begin{equation} \label{eq:secdefrealideal}
    \begin{aligned}
        \rho^\text{real}_{K_A K_B  \Esecdef} &\coloneq \bigoplus_{l_A, l_B} \Pr(\OlAlB) \rho^\text{real}_{K_A^{l_A} K_B^{l_B} \Esecdef| \OlAlB} \\
        \rho^\text{ideal}_{K_A K_B \Esecdef}  &\defvar \idealmap \left[ \rho^\text{real}_{K_A K_B  \Esecdef}   \right].
    \end{aligned}
\end{equation}
Then, the QKD protocol is $\epssecure$-secure if, for all output state $\rho^\text{real}_{K_A K_B \Esecdef} \in \mathcal{W}(\QKDprotocol)$, the following inequality is satisfied\footnote{Note that in Ref.~\cite{ferradini2025definingsecurityquantumkey}, the trace norm appearing in the security definition is not divided by $2$. In contrast, the typical security definition \cite{ben-or_universal_2004,portmann_security_2022} includes the explicit factor of $1/2$.  The definition used in Ref.~\cite{ferradini2025definingsecurityquantumkey} is deliberate and well motivated within that work; we stress that the difference amounts only to an overall factor of $2$ in the  security parameter.}:
\begin{equation}
    \tracedist{ \rho^\text{real}_{K_A K_B  \Esecdef} -  \rho^\text{ideal}_{K_A K_B  \Esecdef}} \leq \epssecure.
    \end{equation}
\end{definition}
We take this opportunity to clarify some concepts. 
Let us consider a protocol \(\mathcal{P}\). Notice that the security definition depends on the set of output states that can be obtained, which in turn depends on the various assumptions $\mathcal{W}$ under which we analyze the protocol (made more precise later in \cref{sec:reductionstatement}). For example, an assumption may be that Alice and Bob use unauthenticated classical communication, and Eve is allowed to tamper with classical messages. A different assumption may be that they use authenticated classical communication, and no tampering of the classical communication is allowed. These give rise to different sets of possible output states. In this work, we consider only those assumptions that are relevant to the classical communication.\footnote{However, one could imagine alternative scenarios in which the assumptions concern, for instance, the quality of randomness available to Alice or Bob.} Thus, these assumptions fix the set of possible attacks Eve can perform, and the protocol $\mathcal{P}$ along with Eve's attack together determine a channel mapping the relevant input states to output states. This channel represents the implementation of the protocol along with Eve's attack. In this sense, $\mathcal{W}$ can also be thought of as determining the set of all possible attack strategies Eve is allowed to perform. When we write \(\mathcal{W}(\mathcal{P})\), we refer to the set of all output states that can arise under all possible choices of Eve’s attack. Moreover, when we compose two protocols via \(\mathcal{P}_2 \circ \mathcal{P}_1\), the composed protocol is understood in the sense of channel composition, where the resulting channel is fixed by the protocol descriptions and Eve's attack on both protocols.  This level of formalism is sufficient for our purposes.

Note that \cref{def:qkdsecurityasymmetric} generalizes the standard definition of QKD security \cite{ben-or_universal_2004,portmann_security_2022,ferradini2025definingsecurityquantumkey} (which corresponds to the case where $\Pr(\OlAlB ) = 0$ whenever $l_A \neq l_B$ in \cref{eq:secdefrealideal})
  to the case where the output state may have asymmetric aborts. Furthermore, observe that the definition depends on the set of possible output states $\mathcal{W}(\QKDprotocol)$, which in turn depends on the authentication setting under consideration. Under the assumption of honest authentication (and appropriate protocol design), asymmetric aborts cannot occur. In this case, we obtain the following (standard) definition of QKD security. We state the definition of both fixed-length QKD protocols (where the length of the output key is some fixed value $\lfixed$ or $0$), and variable-length QKD protocols (where the length of the output key can take many values). 
\begin{definition}[QKD Security with symmetric aborts ] \label{def:qkdsecuritysymmetric}
Let $\QKDprotocol$ be a QKD protocol, and let $\rho^\text{real}_{K_A K_B  \Esecdef}$ be the output state of the QKD protocol, and let $\mathcal{W}\left( \QKDprotocol \right)$ denote the set of possible output states of the QKD protocol. Suppose that the output key lengths for all output states in $\mathcal{W}\left( \QKDprotocol \right)$ are always equal.  Let $\rho^\text{ideal}_{K_A K_B  \Esecdef}$ be the ideal output state, obtained by acting the map $\idealmap$ on the actual output state. That is
\begin{equation} \label{eq:secdefrealidealsymmetric}
    \begin{aligned}
        \rho^\text{real}_{K_A K_B  \Esecdef} &\coloneq \bigoplus_{l} \Pr(\Omega_{l,l}) \rho^\text{real}_{K_A^{l} K_B^{l} \Esecdef| \Omega_{l,l}}, \\
        \rho^\text{ideal}_{K_A K_B  \Esecdef}  &\defvar \idealmap \left[ \rho^\text{real}_{K_A K_B  \Esecdef}   \right].
    \end{aligned}
\end{equation}
Then, the (variable-length) QKD protocol is $\epssecure$-secure if, for all output state $\rho^\text{real}_{K_A K_B \Esecdef} \in \mathcal{W}(\QKDprotocol)$, the following inequality is satisfied:
\begin{equation} \label{eq:tracedistsecdefsymmetric}
    \tracedist{ \rho^\text{real}_{K_A K_B \Esecdef} -  \rho^\text{ideal}_{K_A K_B  \Esecdef}} \leq \epssecure.
    \end{equation}
If one specializes to a case where the QKD protocol outputs a key of some \emph{fixed} length $l_\mathrm{fixed}$ conditioned on a particular acceptance event $\Omega_\mathrm{acc}$, and aborts otherwise (i.e.~produces a key of zero bits upon the event $\Omega^\complement_\mathrm{acc}$ which denotes the complement of the event $\Omega_\mathrm{acc}$),  then we refer to it as a fixed-length protocol. In this case, the states in \cref{eq:secdefrealidealsymmetric} simplify further to the special form 
\begin{equation}
    \begin{aligned}
        \rho^\text{real}_{K_A K_B \Esecdef} &\coloneq  \Pr(\Omega_\mathrm{acc}) \rho^\text{real}_{K_A K_B  \Esecdef| \Omega_\mathrm{acc}} + \Pr(\Omega^\complement_\mathrm{acc}) \rho^\text{real}_{K_A K_B \Esecdef| \Omega^\complement_\mathrm{acc}},  \\
        \rho^\text{ideal}_{K_A K_B  \Esecdef}  &\defvar \idealmap \left[ \rho^\text{real}_{K_A K_B \Esecdef}   \right],
    \end{aligned}
\end{equation}
and thus the above $\epssecure$-security condition (for a fixed-length QKD protocol) simplifies to the condition that all output states $\rho^\text{real}_{K_A K_B  \Esecdef} \in \mathcal{W}(\QKDprotocol)$ satisfy
\begin{equation} \label{eq:tracedistsecdefsymmetric}
    \tracedist{ \rho^\text{real}_{K_A K_B  \Esecdef} -  \rho^\text{ideal}_{K_A K_B  \Esecdef}}  = \Pr(\Omega_\mathrm{acc})  \tracedist{ \rho^\text{real}_{K_A K_B  \Esecdef | \Omega_\mathrm{acc}} -  \rho^\text{ideal}_{K_A K_B \Esecdef| \Omega_\mathrm{acc}}} \leq \epssecure.
    \end{equation} 
\end{definition}

\section{Protocol Specification} \label{sec:protocoldesc}
In this section, we first describe a generic QKD protocol in \cref{subsec:abstractprotdesc}, whose security will be proven later in \cref{sec:proof}. When supplemented with detailed specifications provided in \cref{subsec:detailprotdesc}, and after fixing the values of various free parameters, this generic protocol becomes a particular instance of the decoy-state BB84 protocol. We emphasize that our analysis concerns the security of the mathematical protocol described below, not the security of its practical implementation. Accordingly, all mathematical descriptions introduced here are taken as given. We nevertheless make explicit a few assumptions below.

We assume that the protocol begins with Alice and Bob having access to  local randomness stored in classical registers, which they use to implement the steps of the QKD protocol. The adversary is uncorrelated to this randomness, which is only accessible to the honest parties, and is never leaked to the adversary (even after the termination of the QKD protocol).\footnote{While this randomness is sometimes used to generate states or classical messages that are publicly released at some point (when explicitly specified in the protocol), what we mean by this requirement is that the ``raw'' values of the randomness are never made accessible to the adversary unless explicitly specified in the protocol.}  
In this work, we do not explicitly describe the generation and use of these local random numbers in the QKD protocol execution; they are implicitly utilized in choosing the signal states sent, basis choices, and choosing seeds for hashing. Thus, the assumption that Alice and Bob correctly implement the prescribed state preparation and measurements (and later, hash choices) implicitly assumes that they have access to adequate randomness for doing so. We note that the assumption of perfect random numbers can be relaxed by utilizing random number generators that are $\eps_\text{rng}$-close  to perfect, in the composable security framework.\footnote{Random number generators are often designed to output uniformly random numbers. 
However, the QKD protocol may need to transform this output into nonuniform random values, as some steps such as signal preparation or basis choice are often based on nonuniform probability distributions. This can be done at essentially small cost, using the ``interval algorithm'' described in~\cite{han1997interval} or a variant thereof in~\cite{brown2020framework}.}

We assume that all classical communication between Alice and Bob takes place over authenticated channels, modeled formally in \cref{sec:reductionstatement}. Informally, we consider a setting where Eve, the adversary, controls the timing and delivery of all classical messages exchanged between Alice and Bob. When Alice sends a message, Eve may delay its delivery by an arbitrary amount of time. Upon delivery, the receiver either obtains the exact message that was sent or receives a special authentication-abort symbol $\authabort$, indicating that authentication has failed. Importantly, the authentication mechanism ensures that Eve cannot modify or forge messages without detection, nor can she cause them to be delivered earlier than they were sent. In other words, she may only delay or block messages, but any attempt to alter their content or timing prematurely necessarily results in an authentication abort.\footnote{Note that in this work we consider a scenario in which the probability of forged messages is zero. While no physical channel perfectly satisfies this property, we believe that an appropriate construction that is $\varepsilon_\mathrm{auth}$-close to this property, within the composable framework, is possible. At the same time,  we acknowledge that no explicit construction currently exists in the literature --- see Ref.~\cite[Section 2.2.1]{inprep_authentication}. Once such a construction is obtained, one can invoke the standard composability property to add $\varepsilon_\mathrm{auth}$ to the overall QKD security epsilon $\epssecure$ obtained in this work. Alternatively, one could also ignore composability, and attempt to straightforwardly modify the arguments in Ref.~\cite{inprep_authentication} for a scenario that includes a small probability of successful message forgery 
.} A modification of this setting, where the communication occurs over unauthenticated classical channels, and the transcript is authenticated only at the end of the protocol, is discussed in \cref{remark:delayedauthentication}.

While a rigorous treatment of authentication is an essential component of QKD security proofs, we postpone its detailed analysis to \cref{sec:reductionstatement}. There, following the results of Ref.~\cite{inprep_authentication}, we show that the security analysis of the QKD protocol can, without loss of generality, be carried out under the honest authentication setting, meaning that authentication never aborts and all messages are transmitted faithfully, including their timing.

In what follows, when specifying the protocol in \cref{subsec:abstractprotdesc}, we describe all of Alice’s and Bob’s operations in full detail. However, to simplify notation and focus on the core structure of the protocol, when specifying the corresponding classical registers and quantum states, we restrict our attention to the case where authentication behaves honestly, since it is that setting for which one performs the bulk of the QKD security analysis in \cref{sec:proof}.  In this setting, Alice and Bob agree on all public announcements, which are represented using a single register that is accessible to all parties (including Eve), rather than introducing separate registers to represent each party’s individual copy of the public information.  We stress that neither our protocol specification nor our security analysis requires the assumption that authentication behaves honestly, since we can rigorously reduce the general case to this setting in \cref{sec:reductionstatement}. 
\subsection{Generic QKD Protocol Description} \label{subsec:abstractprotdesc}

The generic protocol description is provided in this subsection, while the detailed protocol specification for decoy-state BB84 is described in \cref{subsec:detailprotdesc}. During the execution of the protocol, the parties exchange the following  classical communication which is described in the protocol steps below (see also \cref{sec:notation}):
\[
\CP^n_1, \flagkey,\flagEC, \CEC, \HEV, \CEV, \flagEV, \HPA, \Cauth.
\]
The final output of the protocol consists of the secret keys \( K_A \) and \( K_B \), generated by Alice and Bob, respectively. All parameters required to instantiate an instance of this \nameref{prot:abstractqkdprotocol} are given in \cref{table:abstractparameters}.

\begin{table}[h]
    \centering
    \renewcommand{\arraystretch}{1.3} 
    \begin{tabularx}{\textwidth}{>{\raggedright\arraybackslash}p{4cm} X}
        \toprule
        \textbf{Symbol} & \textbf{Meaning} \\
        \midrule
      $n \in \mathbb{N}$& Total number of rounds of the QKD protocol. \\
      $\sigma^{(j)}_k \in \dop{=}(A')$ & The $k$th signal state sent by Alice, in round $j$.\\
      $p^{(j)}_k \in [0,1]$ & Probability with which Alice sends $k$th signal state in round $j$. \\
      $M^{(j)}_k \in \Pos(B_j)$  & The POVM element corresponding to for outcome $k$ of Bob's measurement in the $j$th round.  \\
$\announcementfunction^{(j)} : \mathcal{X} \otimes \mathcal{Y} \rightarrow \mathcal{\CP}$ & Function mapping Alice and Bob's local data ($X_j, Y_j)$ to public announcements, for the $j$th round.  \\
    $\keymapfunction^{(j)}(\cdot,\cdot) : \mathcal{X} \times \mathcal{\CP} \rightarrow \mathcal{S} $ & Function implementing the mapping Alice's local data stored in $X_j$ to $S_j$, based on public announcements stored in $\CP_j$, for round $j$.\\
    $\leak(\cdot) : \mathcal{\CP}^n \rightarrow \mathbb{N}$ & Function that determines the number of transcripts of error-correction protocol, as a function of public announcements $\CP_1^n$. \\
      $\lkey(\cdot) : \mathcal{\CP}^n \rightarrow \mathbb{N}$ & Function that determines the number of bits of output key, as a function of public announcements $\CP_1^n$. \\
      $\epsEV \in [0,1]$ & Epsilon for error-verification. Determines output length of hash family used in error-verification, and the final correctness parameter. \\
      $\epsPA \in [0,1]$ & Epsilon for privacy amplification. The exact input and output lengths are determined by the protocol during runtime. \\ 
      $\hashfamily{l_\mathrm{in}}{l_\mathrm{out}}$ & Universal$_2$ hash family from $l_\mathrm{in}$ bits to $l_\mathrm{out}$ bits.  Used for error-verification. The exact input and output lengths are determined by the protocol during runtime. \\
$\idealhashfamily{l_\mathrm{in}}{l_\mathrm{out}}$  & Ideal universal$_2$ hash family from $l_\mathrm{in}$ bits to $l_\mathrm{out}$ bits. Used for privacy-amplification. \\
        \bottomrule
    \end{tabularx}
    \caption{ Parameters required to define an instance of \nameref{prot:abstractqkdprotocol}.}
\label{table:abstractparameters}
\end{table}

In the following protocol, it is to be understood that if authentication results in \authabort, then the receiving party aborts the protocol—i.e., replaces the key register with $\bot$ and sends \abort for all subsequent messages to the other party. Moreover, in the steps below, Alice and Bob exchange descriptions of functions (such as the error-correction maps or hash functions) that are to be applied to their respective strings. If one of them were to receive a function or message incompatible with the expected input—for example, a hash function whose input/output lengths do not match the agreed parameters—they would also abort.\footnote{Such inconsistencies can only arise through a successful forgery by Eve, which is excluded in the honest authentication setting of the communication model we adopt. Nonetheless, we include these protocol steps for completeness, since they can occur with small probability in practice.} 

Note that we do \emph{not} assume that Alice and Bob have synchronized clocks for our security analysis. All timings that we refer to in this description refer to the global or true time in which these events occur, and not to Alice's and Bob's local time (see also \cref{remark:delayedauthentication}). Thus, we do not require Alice or Bob to have access to these global time values. They are simply introduced here since they are required in our theoretical proof (see \cref{sec:reductionstatement}).

\begin{prot}[Generic QKD Protocol] \label{prot:abstractqkdprotocol} 
\leavevmode \\
    \begin{enumerate}
 \item For rounds $j$ from $1$ to $n$, Alice and Bob perform the following operations:
   
	\begin{enumerate}
		\item \label{protStep:statePrep} \textbf{State preparation and transmission:} In round $j$, Alice performs the following steps. 
        \begin{enumerate}
            \item  Alice prepares one of \(d_A\) possible signal states \(\{\sigma^{(j)}_k\}_{k}\), according to the probability distribution $p^{(j)}_{k}$, where $\sigma^{(j)}_k\in \dop{=} (A^\prime_j)$.\footnote{In this protocol specification, and in our proof, Alice is allowed send different states with different probabilities in the different rounds $j$, as long as the distributions are independent across rounds.} This step requires the use of local randomness. (Note that we assume that the adversary is uncorrelated with the state preparation; in particular, they do not hold any purification of the signal state, nor is any such purification made available to the adversary at any point during or after the protocol.)  
            \item She stores the label $k$ for her choice of the signal state in a classical register \(X_j\). The register $X_j$ has the alphabet $\mathcal{X}$.
            \item Alice sends the signal state to Bob via an insecure quantum channel. \item We let $t^\mathrm{A}_j$ be the (global) time this state leaves Alice's lab.  
        \end{enumerate} 
  
		\item \label{protStep:meas} \textbf{Measurements:} In round $j$, Bob performs the following steps:
        \begin{enumerate}
            \item  Bob measures his received state using a POVM  \(\{M_k^{(B_j)}\}_{k=1 \dots d_B}\), obtains one of $d_B$ possible outcomes, and stores his results in a classical register \(Y_j\) (which has alphabet $\mathcal{Y}$). Depending on the exact detection setup used, this step may require the use of local randomness.\footnote{Note that we also allow Bob's measurement POVMs to depend on the rounds $j$.}
            \item We let $t^\mathrm{B}_j$ be the (global) time this measurement is completed. 
        \end{enumerate}

     \item   \label{step:announce} \textbf{Public announcement:} Alice and Bob engage in public announcements in parallel to their signal transmission and measurement steps described above. 
        Alice and Bob perform the following steps for each round $j$:
        \begin{enumerate}

    \item Bob sends a fixed message to Alice indicating that he has completed the measurement for the $j$th round, after he has done so.\footnote{Note that  this is a fixed message which is already known to Eve, since she knows the time at which Bob measures. Thus, this message leaks no information to the adversary, and we can exclude it from the public announcement registers. } 
    \item Alice and Bob engage in further interactive public communication, which depends on the classical registers $X_j, Y_j$ as well as the values announced so far. The first message in this exchange is sent by Alice at time $t^\mathrm{ann}_j$, but only after two conditions are met: (i) she has already transmitted the $j$th quantum state to Bob, and (ii) she has received Bob’s message confirming that his measurement for that round is complete. This ordering guarantees that announcements never precede the corresponding quantum transmission or measurement. Consequently, under the honest authentication setting, the relation $t^\mathrm{ann}_j > t^\mathrm{A}_j, t^\mathrm{B}_j$ is always satisfied.
    \item This classical communication can be two-way and may differ from round to round. We use $\CP_j$ (which has alphabet $\mathcal{\CP}$) to denote the classical register storing all public announcements in the $j$th round.  For clarity, we stress that $\CP_j$ is determined solely by the Alice's and Bob's classical data of that round, namely the registers $X_j$ and $Y_j$, and can be computed by some function $\announcementfunction^{(j)}$ acting on $X_j,Y_j$.\footnote{For the security analysis, we only require that $\CP_j$ should be obtained by applying a CPTP map on $X_j,Y_j$. Thus, we can also allow $\announcementfunction^{(j)}$ to be a stochastic map. Of course, since these public announcements are carried out by two separate parties, each of which starts with access to $X_j$ and $Y_j$ respectively, this imposes constraints on the types of announcement functions $\announcementfunction^{(j)}$ that can be actually implemented in the protocol. } 

 \begin{remark} \label{remark:groupingannouncements} We stress that we do \emph{not} require Alice and Bob to complete the announcements for round $j$, before proceeding to signal preparation and measurement for the next round. We only require that the announcements for round $j$ occur at some time after signal preparation and measurement for that round. This feature of allowing Alice and Bob to perform public announcements \emph{before} all signals are sent are received, is referred to as \term{on-the-fly} announcements.  In particular,  Alice and Bob may announce their classical registers either immediately after each round or in grouped form, where several rounds are announced together in a block of size $\block$ rounds. In the latter, $\CP_j$ for all rounds $(k_\mathrm{block}-1)\block < j \leq k_\mathrm{block}\block$ is announced  simultaneously, at some time after the signal preparation and measurement of the $k_\mathrm{block}\block$th round. Grouping announcements in this way can reduce the overall consumption of authentication keys (see \cref{subsubsec:publicannouncements}).   As shown in \cref{sec:proof}, the security analysis remains valid under either choice. 
    \end{remark}
   
        \end{enumerate}
       	\item \label{step:sift} \textbf{Sifting and key map:} Alice uses knowledge of previous public announcements (and local randomness, if needed), to map her  register $X_1^n$ to the \term{pre-amplification string} $\PAstring$. 
        \begin{enumerate}
            \item Alice maps her local data $X_j$ to the classical registers $S_j$ (which has alphabet $\mathcal{S}$), using public announcements $\CP_j$. This is described by a function $\keymapfunction^{(j)}$ acting on $X_j,\CP_j$.\footnote{As with $\announcementfunction$, we only require this mapping to be CPTP. Thus we can allow $\keymapfunction$ to be a stochastic map.  }
           \item She then applies a deterministic rule, based solely on $\CP_j$, to discard certain rounds from $S_j$. This produces the (potentially shortened) pre-amplification string, which is stored in the register $\PAstring$.\footnote{Thus, the register $\PAstring$ takes values from the set of all possible strings of length less than or equal to $n$, and composed of symbols from the alphabet of the register $S_j$ that are \textit{not} discarded. In most common scenarios, $S_j$ takes values in $\{0,1,\singleRoundBot\}$ (based on $\CP_j$,$X_j$), and the last outcome refers to the outcome that is discarded. Thus, $\PAstring$ stores a binary string of length up to $n$ bits.} In this work, we consider protocols where  Alice’s remapping and discarding are defined so that whenever a value of $S_j$ will be discarded (based on $\CP_j$), $S_j$ is set to a fixed placeholder symbol $\singleRoundBot$. Since Alice determines the mapping to $S_j$, she can directly encode the rounds to be discarded in this way.\footnote{In principle, this means that Alice could implement the discarding step by inspecting the sequence $S_1^n$ alone, without separately referring to the announcements $\CP_1^n$.}
           \item  This step can be performed at any time, as long as Alice has the necessary registers $(X_j, \CP_j)$ to perform the required operations.  
        \end{enumerate}

  After all the announcements and sifting and key map operations are completed, the state of the protocol is given by 
\begin{equation}
    \rho_{ \PAstring S_1^n X_1^n Y_1^n  \CP_1^n   E_n  }.
    \end{equation} 
 where $E_n$ denotes Eve's quantum side information at the end of her attack on the $n$ rounds (which may contain a copy of the announcements $\CP_1^n$). Alice moves on to the next step in the protocol only after all public announcements are completed, i.e, after she has sent her last message and received the last message for the public announcements corresponding to round~$n$. Note that since Alice and Bob only perform announcements after they signal preparation and measurement for that round, this implies that Alice and only moves on to the next step after all signal preparation and measurements are completed.\footnote{In our protocol, it is Alice who sends the next message containing (i) the length of the output key to be produced and (ii) a parameter related to error correction, to Bob. If desired, this step could instead be initiated by Bob via a message to Alice; in that case, Bob would analogously proceed only after sending (and receiving) the final message associated with the public announcements.} 

\begin{remark} \label{rem:diffclassicalregisters}
We make the following clarification regarding the treatment of classical registers in this work.
During the state preparation, transmission, and measurement phases, all public announcements are recorded in the registers $\CP_j$. For each such announcement, we provide Eve with an explicit copy, denoted $\CPhat_j$ (introduced later in our security analysis), which she may choose to store, process, or discard. Consequently, all public announcements made during these phases are already included in Eve’s final quantum system $E_n$ at the end of the quantum communication stage. Thus, we do not necessarily need to explicitly specify that Eve has access to $\CP_1^n$ (although doing so does not cause any difference to the security analysis, since Eve is allowed to copy all announcements and keep them in $E_n$ anyway). 

In the remaining steps of the protocol, only classical communication takes place. We model these by introducing  classical registers that are also accessible to Eve. If Eve wishes, she may use these classical registers together with her existing quantum system $E_n$ to generate an updated system $E'_n$. However, all
operations performed by Alice and Bob during this stage act solely on their own classical
registers and thus commute with any operation Eve might apply to her systems. Consequently, we may, without loss of generality, postpone all of Eve’s remaining operations
until after the protocol has completed. Moreover, since the trace distance is non-increasing
under CPTP maps, it follows that if the required security criterion holds for the state
\emph{before} Eve’s postponed operations, then it also holds for the state \emph{after} she
applies them (which corresponds to the actual state she obtains as a result of her attack). Hence it suffices to focus solely on analyzing the former state; in other words, we can suppose without loss of generality that Eve does not act on these classical registers at all (other than storing them). 
\end{remark}
      
	\end{enumerate}

\item \label{step:varlengthdecision} \textbf{Variable length decision:} Alice and Bob use predetermined functions of the public announcements $\CP_1^n$ to determine parameters for error correction and key generation. We use $\cobs$ to denote the value stored in the classical register $\CP_1^n$.
\begin{enumerate}
    \item Alice computes the value $\leak(\cobs) \in \mathbb{N}$ (a parameter that will be used during error correction) and stores it in $\flagEC$.
    \item Alice computes the length of the output key to be produced, given by $\lkey(\cobs) \in \mathbb{N}$, using a predetermined function $\lkey(\cdot)$, and stores this in the register $\flagkey$.
    \item Alice sends $\flagkey, \flagEC$ to Bob using the authenticated classical channel.

\end{enumerate}

The state in the protocol, conditioned on observing a specific value in the $\CP_1^n$ registers, is given by
\begin{equation}
    \rho_{ \PAstring S_1^n X_1^n Y_1^n  \CP_1^n  \flagkey \flagEC E_n | \Omega(\cobs) },
    \end{equation}
    where $\Omega(\cobs)$ denotes the event that  the value $\cobs$ is observed in the $\CP_1^n$ registers. Note that the functions $\lkey(\cdot)$  and $\leak(\cdot)$ are related, since $\lkey(\cdot)$ must account for the amount of information leaked during error correction, which is quantified by $\leak(\cdot)$. The relationship is established by the security proof. 

\item[]  \textit{ \textbf{Abort Condition: } If $\flagkey$ stores $\lkey(\cobs)=0$, Alice and Bob abort the protocol. That is, they replace their key registers with $\bot$s and send $\abort$ for all future communication.}\footnote{In this case one would want Alice and Bob to avoid wasting authentication keys on messages since they intend to abort in the end anyway and instead simply jump to the end of the protocol. While this should not pose any real issue, we do not attempt to formalize this detail here.}

		\item \label{step:EC} \textbf{Error correction:} Alice and Bob perform an error-correction subprotocol that results in Bob outputting a guess for Alice's pre-amplification string, by performing the following steps:
        \begin{enumerate}
            \item Alice and Bob read the value of $\leak(\cobs)$ from the classical register $\flagEC$.
            \item They run an error-correction subprotocol that  is designed such that the total number of possible communication transcripts is at most  $2^{\leak(\cobs)}$.\footnote{For one-way error-correction protocols that leak a fixed number of bits $\leakfixed$, the number of possible transcripts is $2^{\leakfixed}$.} 
            \item The error correction procedure can have any number of communication rounds. Alice's steps can depend on her local information $\PAstring,S_1^n, X_1^n$, prior public announcements $\CP_1^n$, and announcements during the error-correction procedure itself. Bob's steps can depend on his local information $Y_1^n$, public announcements $\CP_1^n$, and announcements during the error-correction procedure.
            \item The communication transcript is stored in the classical register $\CEC$. 
            \item At the end of this procedure, Bob outputs a classical string stored in \(\PAstring_B\), intended as his guess of Alice’s pre-amplification string stored in \(\PAstring\). The protocol ensures that $\PAstring$ and $\PAstring_B$ are of the same length.\footnote{Since Alice's procedure of discarding rounds from $S_i^n$ is based on public announcements $\CP_i$, Bob can compute this length himself, and ensure that his guess is of the same length.}  
            \item From this point on, we ignore the registers $S_1^n, X_1^n, Y_1^n$. This is justified because these registers are either explicitly deleted by Alice and Bob after their use, or they no longer play any role in the subsequent steps of the protocol. In either case, they are never leaked to the adversary.
        \end{enumerate}
    		
The state in the protocol, conditioned on observing a specific value in the $\CP_1^n$ registers, is given by
\begin{equation}
    \rho_{ \PAstring \PAstring_B \CP_1^n  \CEC \flagkey \flagEC  E_n | \Omega(\cobs) }.
    \end{equation}

         \item \label{step:EV} \textbf{Error verification:}  Alice and Bob perform universal$_2$ hashing and compare the hash values, by performing the following: 
        \begin{enumerate}
            \item Alice and Bob look at the length $\lprePA$ of  the string in $\PAstring$ ($\PAstring_B$ for Bob). This determines the universal$_2$ hash family from the set of $\lprePA$-length strings to $\ceil{\log(1/\epsEV)}$ bits, which we denote using $\hashfamily{\lprePA}{\ceil{\log(1/\epsEV)}}$.\footnote{More generally, one could also use a $\delta$-almost-universal family. In this case, $\epsEV = \delta$, and $\ceil{\log(1/\epsEV)}$ is replaced by the number of the bits in the output of the chosen hash family.} 
            \item Alice uses her local randomness to choose a hash function from the chosen universal$_2$ hash family. She announces the choice of the function chosen in the classical register $\HEV$.
            \item Alice announces the result of the hash function applied to $\PAstring$, in the classical register $\CEV$. 
            \item Bob uses the announcement $\HEV$ to choose the same hash function as Alice, and applies it to his guess $\PAstring_\mathrm{B}$. Bob compares his computed hash value with Alice's announced value in $\CEV$, and outputs the binary result of the match in the register $\flagEV$. Bob sends $1$ in $\flagEV$ to indicate a match, and $0$ for a mismatch. 
        \end{enumerate}
       
\item[]        \textit{\textbf{Abort Condition:} If \(\flagEV\) indicates a mismatch, Alice and Bob  abort the protocol. That is, they replace their key registers with $\bot$s and send $\abort$ for all future communication.}
The state in the protocol, conditioned on $\Omega(\cobs)$ and error-verification passing, is given by
\begin{equation}
    \rho_{ \PAstring \PAstring_B \CP_1^n  \CEC \CEV \flagkey \flagEC \flagEV \HEV E_n | \Omega(\cobs) \wedge \OmegaEV }.
    \end{equation}
where $\wedge$ denotes the logical `and' operator and $\OmegaEV$ denotes the event that error-verification passes, i.e $\flagEV$ indicates a match.
        
		\item \textbf{Privacy amplification:}  Alice and Bob perform \emph{ideal}  universal$_2$\footnote{Recall that ideal universal$_2$ (see \cref{def:2universal}) corresponds to the case where the collision probability of the hash is \emph{equal} to $1/|D'|$, where $D'$ is the codomain of the hash family. Note that this requirement can be relaxed to universal$_2$ by sacrificing one additional bit of output key, see \cref{remark:LHLequalityissue} for a discussion.} hashing on their pre-amplification strings to generate the output key, as follows:
        \begin{enumerate}
           \item Alice looks at the length $\lprePA$ of the string in $\PAstring$, and the register $\flagkey$ to determine the length $\lkey(\cobs)$  of the output key. This determines the ideal universal$_2$ hash family from the set of $\lprePA$-length strings to $\lkey(\cobs)$ bits, which we denote using $\idealhashfamily{\lprePA}{\lkey(\cobs)}$.  
           \item Alice uses her local randomness to choose a hash function from the chosen universal$_2$ hash family. She announces the choice of the function chosen in the classical register $\HPA$. Bob uses the announcement $\HPA$ to choose the same function as Alice.
           \item Alice (Bob) applies the hash function to the $\PAstring$ ($\PAstring_\mathrm{B}$) register, and stores the result in the $K_A$ ($K_B$) registers.
           \item From this point on in the protocol, we ignore the registers $\PAstring,\PAstring_B$, since they are longer play any role in the subsequent steps of the protocol and are never leaked to the adversary. 
        \end{enumerate}
  The state in the protocol, conditioned on $\Omega(\cobs) \wedge \OmegaEV$, is given by 
\begin{equation}
    \rho_{K_A K_B  \CP_1^n  \CEC \CEV \flagkey \flagEC \flagEV \HEV \HPA E_n  | \Omega(\cobs) \wedge \OmegaEV }.
    \end{equation}

    \item \textbf{Authentication post-processing.}
Alice and Bob exchange their final accept or abort decisions following the procedure described below. Recall that the protocol is defined such that, if either party receives an $\authabort$ during preceding communication, they replace their corresponding key register with the symbol $\bot$. 
    \begin{enumerate}
        \item Bob looks at the size of the string stored in $K_B$. If the $K_B$ stores $\bot$,  Bob uses the classical authenticated channel once to send an \abort message to Alice. Otherwise, Bob sends an \accept message to Alice. This is Bob's preliminary decision.
        \item  Alice looks at the received message and the size of the string stored in $K_A$. If Alice receives an \accept, and $K_A$ does not store $\bot$, she uses the classical authentication channel to send an \accept message to Bob. Otherwise, Alice sends an \abort message to Bob, and replaces $K_A$ with $\bot$. This is Alice's final decision.
        \item If Bob receives \accept, he does nothing. In any other case, he replaces $K_B$ with $\bot$. This is Bob's final decision.
        \item The transcript of this communication is stored in $\Cauth$. 
    \end{enumerate}
\end{enumerate}
\end{prot}
This concludes the description of the generic QKD protocol studied in this work. We highlight that this protocol has several important features: it accommodates on-the-fly announcements (see \cref{remark:groupingannouncements}) and allows the states sent, measurements performed, the announcement procedure as well as the sifting procedure, to vary across different rounds. We discuss these aspects in more detail later in \cref{subsec:discussionabstractproof}. Moreover, the framework is highly general, encompassing a wide range of protocols.

\begin{remark}
Note for implementations: 
The protocol description given above follows a fixed ordering of local operations by Alice and Bob for convenience. However, this ordering may be varied without affecting the security of the protocol. In particular, Alice and Bob may (i) perform local operations in a different order, (ii) discard or delete registers that are no longer needed, or (iii) carry out certain steps earlier or later than specified, as long as such changes do not alter the order in which classical or quantum messages are sent and received between them. Any protocol obtained in this way is equivalent for the purposes of the security proof. For instance, Alice may begin constructing the $\PAstring$ register as soon as she receives $\CP_j$, and may discard $X_j$ immediately after applying the key-mapping and discard procedure for round $j$.
\end{remark}

We prove the security of \nameref{prot:abstractqkdprotocol}, in terms of the parameters used to describe it (see \cref{table:abstractparameters}),  in \cref{sec:proof}. In the next subsection, we provide additional details which, along with \nameref{prot:abstractqkdprotocol},  yield a concrete realization of the decoy-state BB84 protocol.

\subsection{Instantiated Decoy-state Protocol Description} \label{subsec:detailprotdesc}

We now instantiate the \nameref{prot:abstractqkdprotocol} as a polarization-encoded, decoy-state BB84 protocol with a passive detection setup. In doing so, we introduce several additional parameters that must be specified to obtain a concrete realization of the protocol. Some of these parameters (which are included in various Tables in this section) have already appeared at a more abstract level, but are repeated here for clarity and to emphasize their operational meaning in the physical implementation. 

At this stage, our goal is not to provide a complete, fully specified protocol description from scratch. Instead, we supplement the abstract (\cref{prot:abstractqkdprotocol}) with additional details. Thus the details provided here, along with the description of the protocol in the previous section, and choices for the free parameters, provides a complete, fully specified protocol description of the decoy-state BB84 QKD protocol. 

\subsubsection{State preparation and transmission}

\begin{table}[h]
    \centering
    \renewcommand{\arraystretch}{1.3} 
    \begin{tabularx}{\textwidth}{%
        >{\raggedright\arraybackslash}p{1.5cm}  
        >{\raggedright\arraybackslash}X       
        >{\raggedright\arraybackslash}p{8cm}  
    }
        \toprule
        \textbf{Symbol} & \textbf{Meaning} & \textbf{Instantiation} \\
        \midrule
        $\gamma$ & Probability that the round is a test round & Free parameter\\
        $p_{a,\mu \vert \testgenflag}$ & Probability of picking polarization $a$ and intensity $\mu$, conditioned on round being test or generation round, as determined by $\testgenflag$ & Free parameter\\
     $\sigma_{(a,\mu,z)}$ & State sent by Alice when polarization $a$ and intensity $\mu$ is picked for $z = $ test or gen & Fully phase-randomized coherent state: $\sum_{N=0}^\infty e^{-\mu}\frac{\mu^N}{N!}\ketbra{N}_a$\\
        \bottomrule
    \end{tabularx}
    \caption{Instantiation of Step \ref{protStep:statePrep} of the \nameref{prot:abstractqkdprotocol}.} \label{tab:statePrep}
\end{table}
We now specify the signal states $\{\sigma^{(j)}_k\}_k$ from the generic protocol description, together with the preparation probabilities $p^{(j)}_k$. Since we consider a protocol which has the same signal preparation in each round, we drop the superscript $j$. In the \nameref{prot:abstractqkdprotocol}, all possible states prepared by Alice are indexed with a single label $k$. For the decoy-state setting, however, it is more natural to separate this label into three components: the polarization choice $a$, the intensity setting $\mu$, and the test/generation flag $\testgenflag \in \{\text{test},\text{gen}\}$. Thus, every signal state is labeled as $k = (a, \mu, \testgenflag)$. 

The choice of the signal state occurs as follows. First, Alice decides whether the round is supposed to be key generation round, or a test round, choosing $\testgenflag=\text{gen}$ with  probability $1-\gamma$, and $\testgenflag=\text{test}$ with probability $\gamma$. Conditioned on this choice, she samples the polarization $a$ and intensity $\mu$ according to the probability distribution $p_{a,\mu | \testgenflag}$.

The signal preparation is as follows. There are four possible polarization choices, ${H, V, A, D}$. These are partitioned into the rectilinear basis ${H, V}$ and the diagonal basis ${A, D}$; the chosen subset is referred to as Alice’s basis choice $b_A 
\in \{\Zbasis,\Xbasis\}$. As described in  the \nameref{prot:abstractqkdprotocol}, Alice records her signal-state choice for round $j$ in her register $X_j$. Concretely, this means that the tuple $(a,\mu,\testgenflag)$ is stored in $X_j$. Thus the set of possible labels $k$ coincides with the alphabet $\mathcal{X}$ of the register $X_j$.

We assume that Alice prepares fully phase-randomized coherent states, and that the polarization encoding is perfect. We further assume that any multi-mode information such as the pulse spectrum is independent of the photon number and polarization\footnote{ See \cref{sec:imperfections} for more details on incorporating imperfections into Alice's state preparation.}. Thus, her prepared states for choices of polarization $a$ and intensity $\mu$ are given by
\begin{equation} \label{eq:AliceSignalStatesDescription}
    \sigma_{(a,\mu,\testgenflag)} = \sum_{N=0}^\infty e^{-\mu}\frac{\mu^N}{N!}\ketbra{N}_a.
\end{equation}
Here, $\ketbra{N}_a$ represents an $N$-photon state where all the photons have polarization $a$. Note that since the signal state only depends on $a,\mu$, Alice can prepare the same state in test and key generation rounds. We let all other variables be free parameters as described in \cref{tab:statePrep}.

\subsubsection{Measurements} \label{subsec:detProtMeas}

\begin{table}[h]
    \centering
    \renewcommand{\arraystretch}{1.3} 
    \begin{tabularx}{\textwidth}{%
        >{\raggedright\arraybackslash}p{1.5cm}  
        >{\raggedright\arraybackslash}X       
        >{\raggedright\arraybackslash}p{8cm}  
    }
        \toprule
        \textbf{Symbol} & \textbf{Meaning} & \textbf{Instantiation} \\
        \midrule
        $\beamsplit \in [0,1]$ & Bob's passive basis choice beam-splitting ratio & Free parameter\\
        \(\{M_k^B\}\) & Measurement made by Bob & POVM with $2^4$ elements corresponding to all click patterns for the passive detection setup.\\
        \bottomrule
    \end{tabularx}
    \caption{Instantiation of Step \ref{protStep:meas} of the \nameref{prot:abstractqkdprotocol}} \label{tab:meas}
\end{table}

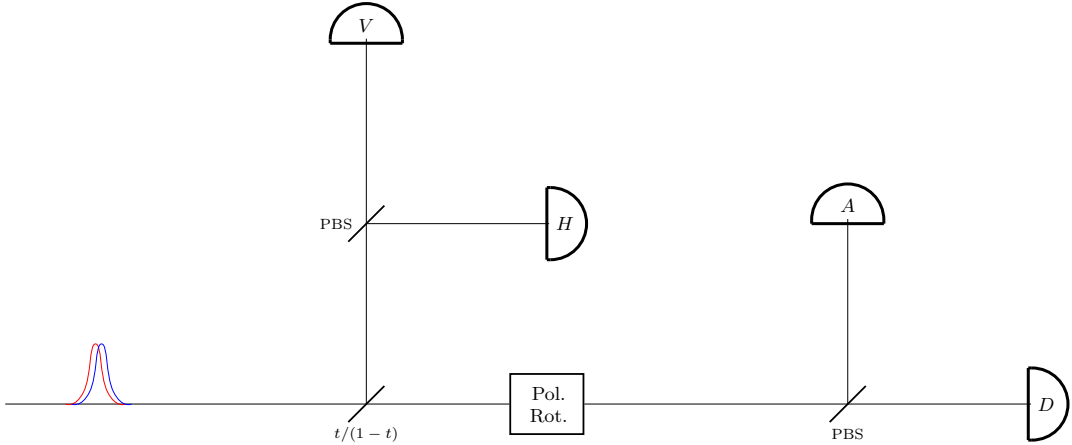
\begin{figure*}
    \centering
    \scalebox{0.8}{\begin{tikzpicture}

\pic{beamsplitter={5050,$\beamsplit/(1-\beamsplit)$,(0,-0.5)}};
\pic[above of = 5050, yshift = 2cm]{beamsplitter={HV,PBS,(-0.5,0)}};
\pic [right of = HV,xshift = 2cm,scale = 0.6]{detector={H, $H$, black}};
\pic [above of = HV,yshift = 2cm,scale = 0.6,rotate=90]{detector={V, $V$, black}};
\draw (-6,0) -- (5050);
\draw (5050) -- (HV);
\draw (HV) -- ([xshift = -5cm]H);
\draw (HV) -- ([yshift = -5cm]V);

\node[process,right of = 5050,xshift = 2cm](PolRot){Pol.\\Rot.};
\pic[right of = PolRot, xshift = 4cm]{beamsplitter={AD,PBS,(0,-0.5)}};
\pic[above of = AD, yshift = 2cm, ,scale = 0.6, rotate = 90]{detector = {A,$A$, black}};
\pic[right of = AD, xshift = 2cm, scale = 0.6]{detector = {D,$D$, black}};
\draw (5050) -- (PolRot);
\draw (PolRot) -- (AD);
\draw (AD) -- ([xshift = 0cm]A);
\draw (AD) -- ([yshift = 0cm]D);

\pic at (-5,0) {pulse = {red}};
\pic  at (-4.9,0) {pulse = {blue}};

\end{tikzpicture}}
    \caption{Passive polarisation-encoded BB84 detection setup.} \label{fig:PassiveBB84}
\end{figure*}

Bob measures using a passive detection setup, with threshold detectors (see \cref{fig:PassiveBB84}). Again, we consider a protocol that performs the same measurement in each round, and therefore drop the superscript $j$. We allow Bob's passive basis choice beam-splitting ratio $\beamsplit$ to be freely chosen. We assume perfect lossless, single-mode threshold detectors, with no dark counts\footnote{For more details on accommodating detection setup imperfections in the security proof, see \cref{sec:imperfections}.}. Thus, Bob's output modes can be written in terms of his input modes $\hat{a}_H$ and $\hat{a}_V$ as follows
\begin{equation} \label{eq:outputModes}
    \begin{aligned}
        \hat{b}_H &= \sqrt{\beamsplit} \hat{a}_H,\\
        \hat{b}_V &= \sqrt{\beamsplit} \hat{a}_V,\\
        \hat{b}_D &= \sqrt{\frac{1-\beamsplit}{2}} (\hat{a}_H + \hat{a}_V),\\
        \hat{b}_A &= \sqrt{\frac{1-\beamsplit}{2}} (\hat{a}_H - \hat{a}_V).
    \end{aligned}
\end{equation}
\begin{remark}
   Linear optical transformations are generally represented by unitary matrices that map the input modes to the output modes. In this case, there are formally four input modes: two from each input arm of the $\beamsplit/(1-\beamsplit)$ beam splitter in \cref{fig:PassiveBB84}. Since two of these modes are always prepared in the vacuum state, we instead use the truncated, unnormalized mode operators introduced in \cref{eq:outputModes}. For a more detailed discussion in the context of a related detection setup, see Ref.~\cite[Appendix A]{narasimhachar_study_}.
\end{remark}
There are $2^4$ possible detection outcomes, each corresponding to a different combination of detectors clicking. For a particular detection outcome, let $\Sclick \subset \{H,\ V,\ D,\ A\}$ be the set of detectors that click.  We refer to the subset of detection outcomes corresponding to $\abs{\Sclick}=1$ as single-clicks. In this case, we denote Bob's basis as $b_B$ which specifies whether the click occurred in the $H/V$ detectors corresponding to the $\Zbasis$ basis, or in the $A/D$ detectors corresponding to the $\Xbasis$ basis. The outcome where no detector clicks ($\Sclick = \emptyset$) is referred to as the no-click outcome. We refer to the subset consisting of all outcomes where $\abs{\Sclick}>1$ as multi-clicks. Then the POVM elements for outcomes corresponding to at least one detector clicking ($\Sclick \neq \emptyset$) are given by, 
\begin{equation}
    M_{\Sclick}^{(B)} = \sum_{\{N_j\geq1\}_{j\in\Sclick}} \frac{1}{\Pi_{j\in\Sclick} N_j!} \Pi_{j\in\Sclick}\left({(\hat{b}_j^\dag)}^{N_j}\right)\ketbra{\mathrm{vac}}\Pi_{j\in\Sclick}\left(\hat{b}_j^{N_j}\right),
\end{equation}
where $\ketbra{\mathrm{vac}}$ is the vacuum state. The POVM element for the no-detector click outcome is given by
\begin{equation}
    M_{\emptyset}^B = \ketbra{\mathrm{vac}}.
\end{equation}
For a simple example, consider the $\Sclick = \{H,\ D\}$. The POVM element corresponding to this particular double-click outcome is
\begin{equation}
    M_{\{H,\ D\}}^B = \sum_{N_H, N_D = 1}^\infty \frac{1}{N_H!N_D!} {(\hat{b}^\dag_H)}^{N_H}{(\hat{b}^\dag_V)}^{N_V}\ketbra{\mathrm{vac}}\hat{b}_H^{N_H}\hat{b}_V^{N_V}.
\end{equation}
The detailed structure of these POVM elements is irrelevant to our proof itself, and is only relevant for evaluating the final key rate expression.\footnote{As we shall show, the evaluation of the key rate expression does not require the detailed structure of these POVM elements for higher photon-numbers either. Specifically, for key rate computations, a description the POVM elements in the single-photon (across all modes) subspace is often sufficient. This corresponds to choosing the photon number cutoff $\nFSS=1$ when using the flag-state squasher (explained later in \cref{sec:optics}).} For the security proof, we rely only on the fact that they are jointly block‑diagonal, with each block corresponding to the total photon number across all modes. 
Note that Bob stores his exact measurement outcome in the $Y_j$ registers. Thus, the alphabet of $Y_j$ is
$$
\mathcal{Y} = \{ \emptyset,\, \{H\},\, \{V\},\, \{A\},\, \{D\},\, \{H,V\}, \{H,A\}, \dots, \{H,V,A,D\} \}.
$$

\subsubsection{Public announcement} \label{subsubsec:publicannouncements}

\begin{table}[ht!]
    \centering
    \renewcommand{\arraystretch}{1.3} 
    \begin{tabularx}{\textwidth}{%
        >{\raggedright\arraybackslash}p{1.5cm}  
        >{\raggedright\arraybackslash}X       
        >{\raggedright\arraybackslash}p{8cm}  
    }
        \toprule
        \textbf{Symbol} & \textbf{Meaning} & \textbf{Instantiation} \\
        \midrule
        $\block$ & Size of public announcement block & Free parameter\\
     $\announcementfunction^{(j)} $ & Function mapping Alice and Bob's local data ($X_j, Y_j)$ to public announcements, for the $j$th round. & Free parameter \\
     $\keymapfunction^{(j)} $ & Function mapping Alice local data ($X_j$) and public announcements $(\CP_j)$ to sifted data ($S_j$), for the $j$th round. & Free parameter \\
        \bottomrule
    \end{tabularx}
    \caption{Instantiation of Step \cref{step:announce,step:sift} of the \nameref{prot:abstractqkdprotocol}.} \label{tab:announce}
\end{table}

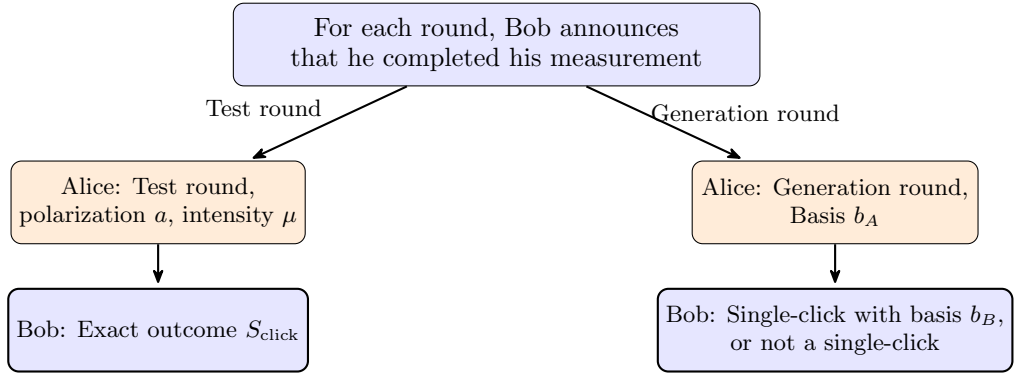
\begin{figure}[h]
    \centering
    \begin{tikzpicture}[
    level 1/.style={sibling distance=9cm, level distance=2.1cm},
    level 2/.style={sibling distance=4.8cm, level distance=1.7cm},
    edge from parent/.style={draw, thick, -{Stealth[round]}, shorten >=2pt},
    every node/.style={font=\small, align=center, rounded corners, minimum height=1.1cm},
    bob/.style={draw, fill=blue!10, minimum width=3.7cm},
    alice/.style={draw, fill=orange!15, minimum width=3.8cm}
]
\node[bob, font=\normalsize, minimum width=7cm, xshift=-2cm] (root) {For each round, Bob announces \\ that he completed his measurement}
        child {
            node[alice] (test) {Alice: Test round, \\polarization $a$, intensity $\mu$}
                child {
                    node[bob] (testbob) {Bob: Exact outcome $\Sclick$}
                }
            edge from parent node[pos=0.9, above=1pt] {Test round}
        }
        child {
            node[alice] (gen) {Alice: Generation round, \\Basis $b_A$}
                child {
                    node[bob] (genbob) {Bob: Single-click with basis $b_B$, \\ or not a single-click}
                }
            edge from parent node[pos=1, above=2pt] {Generation round}
    };

\end{tikzpicture}
    \caption{Schematic of public announcements stored in register $\CP_j$. Blue refers to Bob's announcements, and orange refers to Alice's announcements. See \cref{eq:fannounce} for a formal description.}
    \label{fig:CPidata}
\end{figure}

For each round \(j \in \{1,\dots,n\}\), Alice and Bob perform the following announcements at some time \(t^{\mathrm{ann}}_j\), \emph{after} Alice has sent the state for that round and Bob has completed his measurement.\footnote{As explained in \nameref{prot:abstractqkdprotocol}, this ordering is enforced in our protocol by having Bob first send a message to Alice indicating that he has completed his measurement. Alice proceeds with the announcements for round \(j\) only after also confirming that she has sent the state for that round.}. Note that since the announcement procedure we describe is the same for every round $j$, we drop the superscript $j$. We first describe the announcements in words (see \cref{fig:CPidata}), and later specify them formally in an explicit equation (\cref{eq:fannounce}).

Alice first announces the following information:
\begin{enumerate}
\item Whether the round was a test or a key-generation round.
\item In test rounds: her chosen polarization $a$ and intensity setting $\mu$.
\item In key-generation rounds: her chosen basis $b_A$.
\end{enumerate}

After Alice’s announcement, Bob discloses the following, depending on the round type:
\begin{enumerate}
\item If Alice announces a test round, Bob announces his exact measurement outcome $\Sclick$.
\item If Alice announces a key generation round, Bob announces whether a single-click was observed or not. If he observed a single-click, he further announces his basis  $b_B$.\footnote{Unlike Alice, who makes a choice that determines her basis, Bob's basis is determined by the measurement outcome he gets.}
\end{enumerate}

The function $\announcementfunction\left( (a,\mu,z),y\right)$ (where we again drop the $j$ superscript) determining the announcements is given by:

\newcommand{\nosingleclick}{\bot^\mathrm{not-sc}}
\newcommand{\notannounced}{\bot^\mathrm{not-ann}}
\begin{align} \label{eq:fannounce}
  \announcementfunction\left( (a,\mu,z),y\right) \defvar \begin{cases}
                (a,\mu,\mathrm{test},y) & \text{ if } z=\mathrm{test}\;\; \\
    (\Zbasis,\notannounced , \mathrm{gen},\Zbasis) & \text{ if } z=\mathrm{gen}\; \wedge \; a \in \{H,V\} \; \wedge \; y \in \{H,V\}  \\    (\Zbasis,\notannounced ,\mathrm{gen},\Xbasis) & \text{ if } z=\mathrm{gen}\; \wedge \; a \in \{H,V\} \; \wedge \; y \in \{A,D\}  \\   (\Xbasis,\notannounced ,\mathrm{gen},\Zbasis) & \text{ if } z=\mathrm{gen}\; \wedge \; a \in \{A,D\} \; \wedge \; y \in \{H,V\}  \\   (\Xbasis,\notannounced ,\mathrm{gen},\Xbasis) & \text{ if } z=\mathrm{gen}\; \wedge \; a \in \{A,D\} \; \wedge \; y \in \{A,D\}  \\ 
    (\Zbasis, \notannounced, \mathrm{gen}, \nosingleclick) & \text{ if } z=\mathrm{gen}\; \wedge \; a \in \{H,V\} \; \wedge \; y \notin \{H,V,A,D\} \\
      (\Xbasis, \notannounced, \mathrm{gen}, \nosingleclick) & \text{ if } z=\mathrm{gen}\; \wedge \; a \in\{A,D\} \; \wedge \; y \notin \{H,V,A,D\}
            \end{cases},
\end{align}
where $\notannounced$ indicates that the intensity is not announced, and $\nosingleclick$ denotes that Bob did not obtain a single-click outcome. All the announced data for round $j$ is stored in the register $\CP_j$. We denote the announcements stored in $\CP_1^n$ as $\cobs$.

Note that we only require that the announcements for each round $j$ occur after the signal preparation and measurement for that round. However, if each round goes through a separate announcement, then this is too expensive. In practical situations, the protocol proceeds by announcing data in blocks of $\block$ rounds. The size of the block $\block$ is a free parameter that depends on implementation constraints: larger blocks consume less authentication key, whereas smaller blocks reduces the memory required to store the protocol data.\footnote{For $\block=1$, we do not expect to obtain a net positive secret key rate, since the secret key consumed for authentication exceeds the amount of secret key generated by the QKD protocol. Thus, practical values of $\block$ must be sufficiently larger than $1$.} However, the security analysis is valid for all choices of $\block$, as we show in \cref{sec:proof}.

\subsubsection{Sifting and key map} \label{subsubsec:siftingkeymap}
For each round $j$, Alice constructs her secret register $S_j$ from her measurement outcomes $X_j$ and the announcements $\CP_j$ as follows (\cref{eq:fkeymap}). All test rounds are mapped to $\singleRoundBot$. All generation rounds where Bob announces that he did not receive a single-click are mapped to $\singleRoundBot$. For the remaining generation rounds, Alice checks if her basis choice matches Bob's basis; if not, the round is mapped to $\singleRoundBot$. Otherwise, she stores her bit value corresponding to her polarization choice $a$. This process is represented by the function $\keymapfunction$ (where we drop the superscript $j$), which is given by :
\begin{align} \label{eq:fkeymap}
    \keymapfunction\left( (a,\mu,z), (p_1,p_2,p_3,p_4) \right) \defvar \begin{cases}
                0 & \text{ if } p_3 = z = \mathrm{gen} \; \wedge \; p_1 = p_4 = \Zbasis \; \wedge \; a = H \\
                 1 & \text{ if } p_3 = z = \mathrm{gen} \; \wedge \; p_1 = p_4 = \Zbasis \; \wedge \; a = V \\
                   0 & \text{ if } p_3 = z = \mathrm{gen} \; \wedge \; p_1 = p_4 = \Xbasis \; \wedge \; a = A \\
                     1 & \text{ if } p_3 = z = \mathrm{gen} \; \wedge \; p_1 = p_4 = \Xbasis \; \wedge \; a = D \\
                     \singleRoundBot & \text{otherwise}.
            \end{cases}
\end{align}

To go from $S_1^n$ to $\PAstring$, Alice discards all rounds $j$ where $\CP_j$ does not correspond to $\mathrm{gen}$,  basis matching, and single-click. In other words, all rounds where $S_j$ stores a  $\singleRoundBot$  are dropped from the string $S_1^n$ to form Alice's pre-amplification string $\PAstring$. Note that since Bob has access to the same public announcements as Alice, he discards the same set of rounds from $Y_1^n$ to obtain a corresponding shortened string of the same length as $\PAstring$, which we denote by $\bm{Y}$. This procedure is described formally in the error-correction stage of the protocol, discussed later in \cref{subsubsec:errorcorrection}.

\subsubsection{Variable length decision}

\begin{table}[h]
    \centering
    \renewcommand{\arraystretch}{1.3} 
    \begin{tabularx}{\textwidth}{%
        >{\raggedright\arraybackslash}p{3.5cm}  
        >{\raggedright\arraybackslash}X       
        >{\raggedright\arraybackslash}p{8cm}  
    }
        \toprule
        \textbf{Symbol} & \textbf{Meaning} & \textbf{Instantiation} \\
        \midrule
        $\lkey(\cobs)$ & Function determining output key length & Determined by security proof. \\
        $\leak(\cobs)$ &  Function determining number of transcripts of error-correction protocol & Free parameter\footnote{Note that choosing this value too large forces the output key length to be $0$, yielding a useless protocol.} \\
     $f_\text{synd}^{\lprePA,\leak(\cobs)}(\cdot) $ & Function computing syndrome for error-correction, announced $\CEC$ & Should have number of output bits equal to $\leak(\cobs)$\\
    $f_\text{corr}^{\lprePA,\leak(\cobs)}(\cdot,\cdot) $ & Function implementing error-correction based on Bob's local data and public-information  & Free parameter \\
    $\keymapfunction^\mathrm{Bob} : \mathcal{Y} \times \mathcal{\CP} \rightarrow \mathcal{S}$ & Function implementing Bob's keymap. The output goes through a discarding step. &  Outputs a $\singleRoundBot$ on exactly the same set of rounds as Alice's key map function\\ 
        \bottomrule
    \end{tabularx}
    \caption{Instantiation of all remaining steps of the \nameref{prot:abstractqkdprotocol}.} \label{tab:announce}
\end{table}

The concrete version of this step is identical to that described in the generic protocol. We choose an error-correction protocol such that $\leak(\cobs)$ quantifies the information leaked during error correction when $\cobs$ is observed in the public announcements $\CP_1^n$. In particular, this choice fixes the number of possible transcripts of the error-correction protocol to $2^{\leak(\cobs)}$. The computation of the length of the output key $\lkey(\cobs)$ is then determined by the security proof developed in this work, which is stated in \cref{theorem:decoystatebb84securitystatement}.

\subsubsection{Error correction and error verification} \label{subsubsec:errorcorrection}
\paragraph*{Error correction:} Alice and Bob perform one-way forward error correction via the following steps. Note that neither \nameref{prot:abstractqkdprotocol} nor the security analysis in this work require any such restriction, which is introduced merely to specify a more concrete instantiation of the QKD protocol.
\begin{enumerate}
    \item Alice computes the length of the string stored in $\PAstring$, given by $\lprePA$.  Alice reads the value of $\leak(\cobs)$ from the $\flagEC$ register. This specifies a function
\[
f_\text{synd}^{\lprePA,\leak(\cobs)}(\cdot) : \{0,1\}^{\lprePA} \to \{0,1\}^{\leak(\cobs)},
\]
which takes as input Alice’s pre-amplification string $\PAstring $ of length $\lprePA$ and outputs a syndrome of length $\leak(\cobs)$.\footnote{For practical implementations, this function can be chosen to be the parity-check matrix of a suitable error-correcting code. While it may not be feasible to choose separate parity check matrices for every possible combination of $\lprePA$ and $\leak(\cobs)$, one can always pick only a handful of them, and include appropriate padding within the functions $f_\text{synd}^{\lprePA,\leak(\cobs)}(\cdot)$}  This function corresponds to the error-correction protocol’s syndrome computation, i.e., it computes the error syndrome associated with the string in $\PAstring$.  
    \item Alice announces the result of applying the chosen function to $\PAstring$ through the register $\CEC$. Thus, $\CEC$ contains $f_\text{synd}^{\lprePA,\leak(\cobs)}(\PAstring)$.
    \item Bob implements a key map on his side, given by $\keymapfunction^\mathrm{Bob}$:
    \begin{align} \label{eq:fkeymapbob}
    \keymapfunction^\mathrm{Bob}\left( y, (p_1,p_2,p_3,p_4) \right) \defvar \begin{cases}
                0 & \text{ if }  p_3 = \mathrm{gen} \; \wedge \; p_1 = p_4 = \Zbasis \; \wedge \;  y = H \; \\
                 1 & \text{ if } p_3  = \mathrm{gen} \; \wedge \; p_1 = p_4 = \Zbasis \; \wedge \; a = V \\
                   0 & \text{ if } p_3  = \mathrm{gen} \; \wedge \; p_1 = p_4 = \Xbasis \; \wedge \; y = A \\
                     1 & \text{ if } p_3  = \mathrm{gen} \; \wedge \; p_1 = p_4 = \Xbasis \; \wedge \; y = D \\
                     \singleRoundBot & \text{otherwise}.
            \end{cases}
\end{align}
Bob then discards all rounds for which the key map outputs $\singleRoundBot$. In other words, in generation rounds where the bases match, Bob maps $Y_i$ to $1$ if his detector click corresponds to $H$ or $D$, and to $0$ if it corresponds to $V$ or $A$. In all other rounds, $Y_i$ is mapped to $\bot$ and later removed. This procedure results in a binary string $\bm{Y}$ of the same length as the binary string $\PAstring$.

    \item Bob computes the length of the string stored in $\bm{Y}$ to obtain $\lprePA$, and reads the value of $\leak(\cobs)$ from the $\flagEC$ register. This specifies a function
\[
f_\text{corr}^{\lprePA,\leak(\cobs)}(\cdot,\cdot) : \{0,1\}^{\lprePA} \times \{0,1\}^{\leak(\cobs)} \to \{0,1\}^{\lprePA},
\]
which takes as input Bob’s sifted string $\bm{Y}$ of length $\lprePA$ together with Alice’s syndrome of length $\leak(\cobs)$, and outputs a corrected string of length $\lprePA$. This function represents the error-decoding and correction step of the error-correction protocol.\footnote{Our proof can be straightforwardly extended to scenarios where error-correction occurs via stochastic maps or randomized algorithms, instead of deterministic functions. One simply needs to obtain a bound on the number of possible transcripts of the (randomized) algorithms.}
    \item Bob applies this function on $\bm{Y},\CEC$ to get $\PAstring_B$, i.e $\PAstring_B = f_\text{corr}^{\lprePA,\leak(\cobs)}(\bm{Y},\CEC)$.
\end{enumerate}

\paragraph*{Error verification:} Alice and Bob perform  error verification by the following steps:
\begin{enumerate}
    \item Alice and Bob use compute the value  $\lprePA$ from their registers $\PAstring$ and $\PAstring_B$, respectively. This determines the universal$_2$ hash family (see \cref{def:2universal}) mapping $\lprePA$ bits to $\log\left(\big\lceil \tfrac{1}{\epsEV} \big\rceil\right)$ bits, which we denote by $\hashfamily{\lprePA}{\log\left(\big\lceil \tfrac{1}{\epsEV} \big\rceil\right)}$.
    \item Alice uses local randomness to pick a hash  function $f^\mathrm{EV}_\mathrm{hash}$ from the chosen hash family $\hashfamily{\lprePA}{\log\left( \ceil{\frac{1}{\epsEV}}\right)}$, applies it to $\PAstring$ to compute $f^\mathrm{EV}_\mathrm{hash}(\PAstring)$, and announces $f^\mathrm{EV}_\mathrm{hash}(\PAstring)$ in the register $\CEV$ and the random seed used for picking the hash function in $\HEV$
    \item Bob receives $\HEV$ and picks the same hash function $f_\mathrm{hash}$ as Alice, and applies it to compute $f_\mathrm{hash}(\PAstring_B)$. Bob announces binary bit $\flagEV$, indicating a match if values stored in $f^\mathrm{EV}_\mathrm{hash}(\PAstring_B)$ and $ \CEV$ match, and a mismatch otherwise.
    \item As specified in \nameref{prot:abstractqkdprotocol}, both Alice and Bob abort the protocol if $\flagEV$ stores $0$, indicating a mismatch.
\end{enumerate}

\subsubsection{Privacy amplification}
Alice and Bob perform  privacy amplification by the following steps:
\begin{enumerate}
    \item Alice and Bob compute $\lprePA$ from their registers $\PAstring,\PAstring_B$ respectively. Alice and Bob read the length of the output key to be produced $\lkey(\cobs)$ from $\flagkey$. This fixes the \emph{ideal} universal$_2$ hash family $\idealhashfamily{\lprePA}{ \lkey(\cobs) }$ from $\lprePA$ bits to $\lkey(\cobs)$ bits.
    \item Alice uses local randomness to pick a hash  function $f^\mathrm{PA}_\mathrm{hash}$ from the chosen hash family $\idealhashfamily{\lprePA}{ \lkey(\cobs) }$, applies it to $\PAstring$ to compute $f_\mathrm{hash}(\PAstring)$. She stores the result of her hash $f^\mathrm{PA}_\mathrm{hash}(\PAstring)$ in the key register $K_A$, and announces the random seed used in the register $\HPA$.
    \item Bob receives $\HPA$ and picks the same hash function $f^\mathrm{PA}_\mathrm{hash}$ as Alice, and applies it to compute $f^\text{PA}_\mathrm{hash}(\PAstring_B)$. Bob stores the result of applying the hash in the key register $K_B$.
\end{enumerate}

\subsubsection{Authentication post-processing}
This step is already fully specified in \nameref{prot:abstractqkdprotocol}. We now have an instance of a decoy-state BB84 protocol that is specified  in sufficient detail to allow the protocol to be implemented and its security analysis to be carried out. Of course, we have omitted certain additional details (for example, the precise error-correction protocol), since these do not affect the security analysis beyond their contribution to the parameter $\leak(\cobs)$. Moreover, we have not yet specified the function $\lkey(\cobs)$ that determines the length of the output key, as this quantity is fixed by the security proof.

\section{Reduction to scenario with honest authentication } \label{sec:reductionstatement}
We now begin the security analysis of \nameref{prot:abstractqkdprotocol}. Recall that the QKD security definition (\cref{def:qkdsecurityasymmetric}) is expressed in terms of the set of possible output states that a given protocol may produce. This set --- and therefore the security properties of the protocol --- depends on the authentication assumptions under which the protocol is analyzed. Our goal is to reduce the security analysis to the setting where the authenticated channel can be assumed to behaves honestly, using the results of Ref.~\cite{inprep_authentication}. By honest, we mean that all messages that are sent are received correctly at some time after they were sent.  We begin by describing the classical communication model within which we work. Note that this model is only required for reduction to the honest authentication setting, and is not utilized anywhere else in this work. 

We note that there exist constructions of authenticated classical channels that closely approximate the functionality described by our model below. One example is Wegman-Carter Authentication \cite{wegman_new_1981} with two sets of shared secret keys, one for sending and one for receiving. We acknowledge however that to our knowledge, there is no explicit construction of such a resource in the literature on composable security; still, we believe such a construction to be achievable via similar arguments as in Ref.~\cite{portmann_key_2014}, and leave it as a point to be resolved in future work (see also \cite[Section 2.2.1]{inprep_authentication}).\footnote{Ref.~\cite{portmann_key_2014} constructs a resource that is very similar to the one described here, but does not explicitly model time.} 

We begin by setting up some notation to describe the sending and receiving of messages. For each message, we associate a register label and (global) time value associated to when it is sent or received (see \cref{fig:classicalcommmodel}).We stress that this time value refers to the \emph{true} time at which the sending or receiving event occurs and is independent of any timekeeping performed by Alice or Bob.\footnote{We assume that relativistic effects can be neglected.} In particular, we do not assume that Alice and Bob share synchronized clocks, nor do we require them to record or know the timing of messages during the protocol. The time values introduced here are purely conceptual: they exist for the purpose of theoretical reasoning, but are neither accessible to nor used by Alice or Bob.
Furthermore, we emphasize that no assumptions are made about the alignment between the sender’s and receiver’s message ordering; these may differ in the presence of an active adversary.  Thus, the various time values introduced here are used only for the purpose of theoretical analysis and refer to the global time when these events occur. (This setting can be modified: see \cref{remark:delayedauthentication}.)

\begin{itemize}
    \item \textbf{Alice sending: } Alice sends messages  in registers $C^{(i)}_{A \rightarrow E}$, where the index $i$ denotes the ordering of messages from her perspective. We let $t^{(i)}_{A \rightarrow E}$ denote the time at which this message leaves Alice's lab.

     \item \textbf{Alice receiving: } Alice receives messages  $C^{(i)}_{E \rightarrow A}$, where the index $i$ denotes the ordering of messages from her perspective. We let $t^{(i)}_{E \rightarrow A}$ denote the time at which this message is received by Alice. 
    \item \textbf{Bob sending:} Bob sends messages in registers $C^{(i)}_{B \rightarrow E}$, where the index $i$ denotes the ordering of messages from Bob’s perspective. We let $t^{(i)}_{B \rightarrow E}$ denote the time at which this message leaves Bob's lab. 

    \item \textbf{Bob receiving: } Bob receives messages  $C^{(i)}_{E \rightarrow B}$, where the index  $i$ denotes the ordering of messages from his perspective. We let $t^{(i)}_{E \rightarrow B}$ denote the time at which this message is received by Bob. 
   
\end{itemize}
Thus, each sent (received) message is labeled from the sender’s (receiver’s) point of view. We assume that the classical authenticated channel  between the two honest parties has the following properties:
\begin{enumerate}
    \item \textbf{Timing:} If the $i$th message is received \emph{before} the $i$th message was sent, then the received message is the special symbol $\authabort$. Formally,
    \begin{equation}
        \begin{aligned}
        &t^{(i)}_{A\rightarrow E} > t^{(i)}_{E \rightarrow B} \quad \implies \quad C^{(i)}_{E\rightarrow B} \text{ stores } \authabort, \quad \forall i.  \\
        &t^{(i)}_{B\rightarrow E} > t^{(i)}_{E \rightarrow A} \quad \implies \quad C^{(i)}_{E\rightarrow A} \text{ stores }\authabort, \quad \forall i.  
        \end{aligned}
    \end{equation}

    \item \textbf{Modifying messages:} If $i$th message is received \textit{after} the $i$th message was sent, then the received message is either a copy of the sent message or it is an $\authabort$. Formally,
     \begin{equation}
        \begin{aligned}
        &t^{(i)}_{A\rightarrow E} \leq t^{(i)}_{E \rightarrow B} \quad \implies \quad \text{Either $C^{(i)}_{E\rightarrow B}$ and $C^{(i)}_{A\rightarrow E}$ store identical messages, or $C^{(i)}_{E\rightarrow B}$ stores  \authabort}, \quad \forall i.  \\
        &t^{(i)}_{B\rightarrow E} \leq t^{(i)}_{E \rightarrow A} \quad \implies \quad \text{Either $C^{(i)}_{E\rightarrow A}$ and $C^{(i)}_{B\rightarrow E}$ store identical messages, or $C^{(i)}_{E\rightarrow A}$ stores  \authabort}, \quad \forall i.  \\
        \end{aligned}
    \end{equation}
\end{enumerate}
For a given protocol $\QKDprotocol$, we denote by $\worldreal(\QKDprotocol)$ the set of all output states that can arise when the authenticated classical communication behaves as described above, where messages may be delayed and tampered with, and  $\authabort$s may occur.
We denote by $\worldhonest(\QKDprotocol)$ the set of output states obtained when Eve is \emph{not allowed} to perform any action that would cause an $\authabort$. In this honest authentication setting, all messages are delivered correctly, in the same order in which they were sent, and each message is received at some time after it was sent.

\newcommand{\rowsep}{1.3}            
\tikzset{
  participant/.style={font=\small\bfseries},
  lifeline/.style={gray,dashed},
  message/.style={-Latex,thick},    
  attackbox/.style={draw,rounded corners,fill=blue!8,inner sep=4pt,font=\footnotesize}
}

\newcommand{\nextrow}[2]{\coordinate (#1) at ($(#2)+(0,-\rowsep)$);}
\begin{figure}[h!]
  \centering
  \begin{tikzpicture}[every node/.style={inner sep=1pt}]
    \node[participant] (Alice) {Alice};
    \node[participant] (Eve)   [right=5cm of Alice] {Eve};
    \node[participant] (Bob)   [right=5cm of Eve]   {Bob};

    \draw[lifeline] (Alice) -- coordinate (AliceEnd) ++(0,-8);
    \draw[lifeline] (Eve)   -- ++(0,-8);
    \draw[lifeline] (Bob)   -- coordinate (BobEnd) ++(0,-8);

    \coordinate (A0) at ($(Alice)+(0,-\rowsep)$);
    \coordinate (B0) at ($(Bob)+(0,-\rowsep)$);

    \node[attackbox,align=center,minimum height=7.2cm,anchor=north] (Ebox)
      at ($(Eve)+(0,-0.7)$) {Eve
      may delay, drop, add \\
      or replace messages \\
      (subject to the specified details \\
      of the communication model)};

\node at ($ ($(Alice)!0.5!(A0)$) + ( 2,-0.5)$) {$\vdots$};
\node at ($ ($(Bob)!0.5!(B0)$)   + (-2,-0.5)$) {$\vdots$};

    \nextrow{A1}{A0}
    \nextrow{B1}{B0}

    \draw[message] (A1) -- node[midway,above] {$C_{A\to E}^{(5)}$} (Ebox.west |- A1);
    \draw[message] (Ebox.east |- A1) -- node[midway,above] {$C_{E\to B}^{(5)}$} (B1);

    \nextrow{A2}{A1}
    \nextrow{B2}{B1}

    \draw[message] (B2) -- node[midway,above] {$C_{B\to E}^{(14)}$} (Ebox.east |- B2);

    \nextrow{A3}{A2}
    \nextrow{B3}{B2}

    \draw[message] (Ebox.east |- A3) -- node[midway,above] {$C_{E\to B}^{(6)}$} (B3);

    \nextrow{A4}{A3}
    \nextrow{B4}{B3}

    \draw[message] (B4) -- node[midway,above] {$C_{B\to E}^{(15)}$} (Ebox.east |- B4);
    \draw[message] (Ebox.west |- B4) -- node[midway,above] {$C_{E\to A}^{(14)}$} (A4);

    \node at ($(A4)+(2,-0.5*\rowsep)$) {$\vdots$};
    \node at ($(B4)+(-2,-0.5*\rowsep)$) {$\vdots$};

\draw[->, thick, dashed]
  ($(Alice)+(-2,0)$) -- ($(AliceEnd)+(-2,-4)$)
  node[midway, left] {time};

  \end{tikzpicture}
\caption{Authenticated  classical communication model in QKD. Messages pass through Eve, who may delay, drop, or substitute them with $\authabort$, subject to the constraints described in \cref{sec:reductionstatement}. Time flows from top to bottom in the figure, which illustrates an example scenario: in earlier parts of the protocol (not shown in the figure), 4 messages have been sent from Alice to Bob, and 13 messages from Bob to Alice.
Eve does not interfere with Alice’s 5th message to Bob. However, she chooses to delay Bob’s 14th message. (Presumably, Alice does not send a new message during this period because she is waiting to receive one.) During the delay, Eve receives Bob’s 15th message and also delivers the 6th message to Bob. According to our communication model, this implies that $C^{(6)}_{E \rightarrow B}$ must be $\authabort$. Figure and caption from Ref.~\cite{inprep_authentication}.}
\label{fig:classicalcommmodel}
\end{figure}
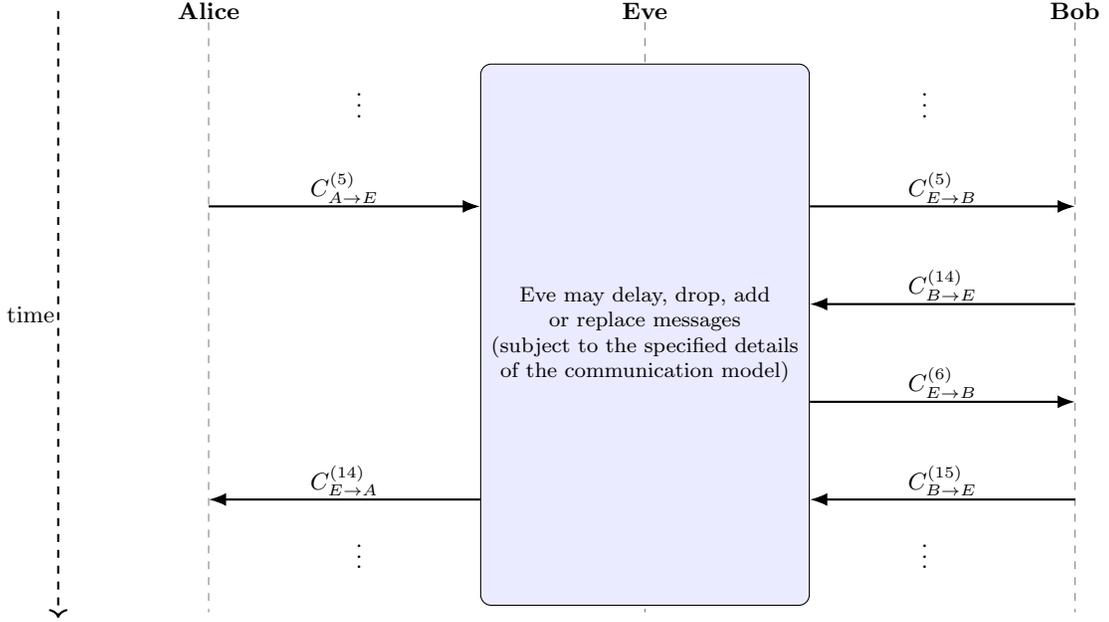

We are now ready to state the theorem that reduces the security analysis of QKD protocols to the setting in which authentication behaves honestly.

\begin{theorem}[Reduction to the honest authentication setting]
\label{theorem:reductionstatement}
Let $\QKDprotocol$ denote the full protocol described in \nameref{prot:abstractqkdprotocol}, and let $\coreQKDprotocol$ denote the same protocol excluding its final step that performs authentication post-processing operations.
We refer to this final step as $\APPprotocol$, so that the complete protocol can be written as
\[
\QKDprotocol = \APPprotocol \circ \coreQKDprotocol,
\]
meaning that $\QKDprotocol$ is obtained by running $\APPprotocol$ on the output of $\coreQKDprotocol$.

Let $\rho^\mathrm{real}_{K_A K_B \CfinalQKD \EfinalQKD}$ denote the output state at the end of $\coreQKDprotocol$, where $\CfinalQKD$ stores all public communication during the protocol (which is assumed to be accessible to Eve), and Eve has access to registers $\CfinalQKD \EfinalQKD$ at the end of $\coreQKDprotocol$.\footnote{Note that any scenario in which Eve chooses to ignore or forget part of the public classical communication can be treated as one where she first records all communication and then traces out whatever she wishes at the end of the QKD protocol. Moreover, since our analysis already allows Eve to retain all public communication, keeping the public‐communication register explicit and accessible to her is without loss of generality.} Similarly, let $\rho^\mathrm{real,final}_{K_A K_B \Cfinal \Efinal}$ denote the final output state at the end of $\QKDprotocol$, where $\Cfinal$ also stores the additional classical communication undertaken during $\APPprotocol$ (which can potentially be used by Eve to update her registers $\EfinalQKD$ to $\Efinal$), and Eve has access to the registers $\Cfinal \Efinal$ at the end of $\QKDprotocol$. We define the corresponding ideal output states via the action of the map $\idealmap$ (see~\cref{sec:securitydefinition}) as
\begin{equation}
\begin{aligned}
\rho^\mathrm{ideal}_{K_A K_B \CfinalQKD \EfinalQKD} &= \idealmap\left[ \rho^\mathrm{real}_{K_A K_B \CfinalQKD \EfinalQKD} \right], \\
\rho^\mathrm{ideal,final}_{K_A K_B \Cfinal \Efinal} &= \idealmap\left[ \rho^\mathrm{real,final}_{K_A K_B \Cfinal \Efinal} \right].
\end{aligned}
\end{equation}

Then, $\epssecure$-security of $\coreQKDprotocol$ for all output states in $\worldhonest(\coreQKDprotocol)$ implies $\epssecure$-security of $\QKDprotocol$ for all output states in $\worldreal(\QKDprotocol)$. That is,
\begin{equation}
    \begin{aligned}
          \tracedist{ \rho^\mathrm{real}_{K_A K_B \CfinalQKD \EfinalQKD} -  \rho^\mathrm{ideal}_{K_A K_B \CfinalQKD \EfinalQKD}} &\leq \epssecure \qquad \forall \rho^\mathrm{real}_{K_A K_B \CfinalQKD \EfinalQKD} \in \worldhonest(\coreQKDprotocol) \\
          &\Downarrow \\
     \tracedist{ \rho^\mathrm{real,final}_{K_A K_B \Cfinal \Efinal} -  \rho^\mathrm{ideal,final}_{K_A K_B \Cfinal \Efinal}} &\leq \epssecure \qquad \forall \rho^\mathrm{real}_{K_A K_B \Cfinal \Efinal} \in \worldreal(\QKDprotocol).
    \end{aligned}
\end{equation}
\end{theorem}

The proof follows directly from \cite[Theorem 2.1]{inprep_authentication}. The {\realistic} authentication setting, and honest authentication setting used here is identical to the one considered in that work. Moreover, the construction of the post-processing protocol $\APPprotocol$ in this work satisfies all the conditions required to apply that theorem, which allows for arbitrary QKD protocols $\coreQKDprotocol$.

Intuitively, \cite[Theorem 2.1]{inprep_authentication} holds because $\APPprotocol$ is designed to ensure that both parties abort whenever an $\authabort$ is received during the execution of $\coreQKDprotocol$. In such cases, the resulting real and ideal states are identical, and the trace distance between them is therefore zero. Consequently, it suffices to bound the trace distance for the case where no $\authabort$ occurs during $\coreQKDprotocol$, that is, when message contents and timings are not tampered with. This reduces the problem to analyzing the security of the protocol under the honest-authentication assumption.

Thus, from this point onward, we focus on proving that the \nameref{prot:abstractqkdprotocol} (excluding its final authentication post-processing step) satisfies the security definition under the assumption of honest authentication. In this setting,  Alice and Bob’s key lengths always coincide, and it suffices to use \cref{def:qkdsecuritysymmetric} as the security definition for the QKD protocol.

\begin{remark} \label{remark:delayedauthentication}

Note that \cite[Theorem 4.1]{inprep_authentication} provides an analogous reduction for a different setting as well, namely one where Alice and Bob use unauthenticated classical communication during \(\coreQKDprotocol\), and authenticated classical communication only during \(\APPprotocol\). In that case, \(\APPprotocol\) must be modified so that the entire communication transcript is authenticated using two messages, one from Alice to Bob and one from Bob to Alice. One may therefore use this result to obtain a modified version of \cref{prot:abstractqkdprotocol} that employs unauthenticated communication for all steps except the final one, which is adjusted as prescribed by \cite{inprep_authentication}. We do not state this result formally, but merely note that this approach is possible. However, it requires the additional assumption that Alice and Bob possess synchronized clocks and retain timestamp information until the end of the protocol for transcript comparison.
\end{remark}

\section{Background information theory} \label{sec:backgroundinfotheory}

In this section, we collect some definitions and background from information theory that will be used throughout this work. In particular, we state the relevant properties of {\Renyi} entropies that will be needed in our analysis. We note that the exact definitions of  {\Renyi} divergences and entropies are not used directly in this work, but are included here for completeness.  

\begin{definition}\label{def:freq}
(Frequency distributions) For a string $z_1^n\in\mathcal{Z}^n$ on some alphabet $\mathcal{Z}$, $\freq_{z_1^n}$ denotes the following probability distribution on $\mathcal{Z}$:
\begin{align}
\freq_{z_1^n}(z) \defvar \frac{\text{number of occurrences of $z$ in $z_1^n$}}{n} .
\end{align}
\end{definition}

\begin{definition}\label{def:purify}
For registers $Q,Q'$ with $\dim(Q) \leq \dim(Q')$, a \term{purifying function for $Q$ onto $Q'$} is a function $\pf: \dop{\leq}(Q) \to \dop{\leq}(QQ')$ such that for any state $\rho_Q$, the state $\pf(\rho_Q)$ is a purification of $\rho_Q$ onto the register $Q'$, i.e.~a (possibly subnormalized) rank-$1$ operator such that $\tr_{Q'}[{\pf(\rho_Q)}] = \rho_Q$.
\end{definition}
Note that a purifying function is \emph{not} a channel (i.e.~CPTP map), for instance, because it is necessarily nonlinear. As a concrete example, we denote $\pfc: \dop{\leq}(Q) \to \dop{\leq}(QQ')$ to be a purifying function that outputs the ``canonical purification'' (with respect to some arbitrary choice of basis), i.e.,
\begin{align}\label{eq:pfexample}
\pfc(\rho_Q) = \dim(Q) \left(\sqrt{\rho_Q} \otimes \id_{Q'}\right) \ketbra{\Phi^+}{\Phi^+} \left(\sqrt{\rho_Q} \otimes \id_{Q'}\right),
\end{align}
where $\ket{\Phi^+}$ is the normalized maximally entangled state across $QQ'$ (up to the support of $Q$) with respect to that choice of basis.

\begin{definition}\label{def:sandwiched divergence}
({\Renyi} divergence~\cite{Muller_Lennert_2013,WWY14})
For any $\rho\in\dop{=}(A)$ and $\sigma\in\Pos (A)$ with $\alpha\in(0,1)\cup (1,\infty)$, the (sandwiched) \Renyi\ divergence between $\rho$, $\sigma$ is defined as:
\begin{align}
    \label{eq:sand_renyi_div}
    \widetilde{D}_\alpha(\rho\Vert\sigma)=\begin{cases}
    \frac{1}{\alpha-1}\log\Tr{ \left(\sigma^{\frac{1-\alpha}{2\alpha}}\rho\sigma^{\frac{1-\alpha}{2\alpha}}\right)^\alpha} &\left(\alpha < 1\ \land\ \rho\not\perp\sigma\right)\vee \left(\supp(\rho)\subseteq\supp(\sigma)\right) \\ 
    +\infty & \text{otherwise},
    \end{cases}  
\end{align}
where for $\alpha>1$ the $\sigma^{\frac{1-\alpha}{2\alpha}}$ terms are defined via the Moore-Penrose pseudoinverse if $\sigma$ is not full-support~\cite{tomamichel_quantum_2016}.
The above definition can be extended to $\alpha \in \{0,1,\infty\}$ by taking the respective limits.
For the $\alpha=1$ case, it reduces to the Umegaki divergence:
\begin{align}
    \label{eq:umegaki_div}
    D(\rho\Vert\sigma)=\begin{cases}
        \Tr{\rho\log\rho-\rho\log\sigma} & \supp(\rho)\subseteq\supp(\sigma)\\
    +\infty & \text{otherwise}. 
    \end{cases}
\end{align}
For any two classical probability distributions $\mbf{p},\mbf{q}$ on a common alphabet, we also define the {\Renyi} divergence $D_\alpha(\mbf{p}\Vert\mbf{q})$ analogously, e.g.~by viewing the distributions as diagonal density matrices in the above formulas; in the $\alpha=1$ case this gives the Kullback–Leibler (KL) divergence.
\end{definition}

\begin{definition}\label{def:sandwiched entropy}
({\Renyi} conditional entropies)
For any bipartite state $\rho\in
\dop{=}(AB)
$, and $\alpha\in[0,\infty]$, we define the following {\Renyi} conditional entropies:
\begin{align}
    \label{eq:cond_renyi}
    &\HalphaDown(A|B)_\rho=-\widetilde{D}_\alpha(\rho_{AB}\Vert\id_A\otimes\rho_B)\notag\\
    &\Halpha(A|B)_\rho=\sup_{\sigma_B\in\dop{=}(B)}-\widetilde{D}_\alpha(\rho_{AB}\Vert\id_A\otimes\sigma_B)\notag \\
\end{align}
For $\alpha=1$, all of the above values coincide and are equal to the von Neumann conditional entropy. In this work, we always have $\alpha \in (1,2)$ throughout our security analysis, which is the regime where all the relevant statements stated here hold. 
\end{definition}

\begin{definition}($f$-weighted {\Renyi} entropies \cite[Definition~4.1]{arqand_marginal_2025}  \cite[Definition~4.1]{arqand_generalized_2024})
	\label{def:QES}
	Let $\rho \in \dop{=}(\CP Q Q')$ be a state where $\CP$ is classical with alphabets $\alphCP$. A \term{tradeoff function on $\CP$} is simply a function $f:\alphCP \to \mathbb{R}$; equivalently, we may denote it as a real-valued tuple $\mbf{f}
	\in \mathbb{R}^{|\alphCP|}$ where each term in the tuple specifies the value $f(\cP)$. Given a tradeoff function $f$ and a value $\alpha\in(0,1)\cup (1,\infty)$, we define two versions of the \term{$f$-weighted entropy of order $\alpha$} for $\rho$ as
    \begin{align}
	\label{eq:fweightedonlyH}
	\Halpha[f](Q|\CP Q')_{\rho} &\defvar
	\frac{\alpha}{1-\alpha} \log \left( \sum_{\cP} \rho(\cP) \, 2^{\left(\frac{1-\alpha}{\alpha}\right) \left(\Halpha(Q|Q')_{\rho_{|\cP}} - f(\cP) \right) } \right), \\
    \label{eq:fweightedonlyH}
	\HalphaDown[f](Q|\CP Q')_{\rho} &\defvar
	\frac{1}{1-\alpha} \log \left( \sum_{\cP} \rho(\cP) \, 2^{\left(1-\alpha\right) \left(\HalphaDown(Q|Q')_{\rho_{|\cP}} - f(\cP) \right) } \right).
\end{align}
\end{definition}
Note that the $f$-weighted {\Renyi} entropy reduces to the usual {\Renyi} entropy when the function $f$ is set to be zero for all inputs \cite[Proposition 5.1]{tomamichel_quantum_2016}.  To gain some intuition of this quantity,
we can define the notion of ``log-mean-exponential'' of a random variable with respect to a base $b \in (0, \infty)$ as:
\begin{align}\label{eq:lme}
\underset{P_X}{\operatorname{lme}_b} \left[g(X)\right] \defvar \log_b \left( \sum_{x} P_X(x) \, b^{g(x)} \right).
\end{align}
With this, Eq.~\eqref{eq:fweightedonlyH} can be written as
\begin{align}\label{eq:interpretaslogmeanexp}
\Halpha[f](Q|\CP Q')_{\rho} = \underset{\bsym{\rho}_{\CP}}{\operatorname{lme}_b} \left[\Halpha[f](Q|Q')_{\rho_{|\CP}} - f(\CP) \right] , \quad\text{where } b = 2^{1-\alpha} \in (0,\infty).
\end{align}
This means non-negativity of  $\Halpha[f](Q|\CP Q')_\rho\geq 0$ is equivalent to having $f(\CP)$ lower bound $\Halpha(Q|Q')_{\rho_{|\CP}}$ in a log-mean-exponential sense. We will mostly be working with $\Halpha[f]$ in this work, and will only require the use of $\HalphaDown[f]$ when we discuss numerics in \cref{sec:numerics}.
\begin{lemma}\label{lemma:DPI}
	(Data-processing \cite[Theorem~1]{frank2013monotonicity}) Let $\rho \in \dop{=}(Q Q')$, take any $\alpha\in [\frac{1}{2},1)\cup(1,\infty]$. Then for any channel $\mathcal{E} \in \CPTP(Q',Q'')$, 
	\begin{align}
		\Halpha(Q|Q'')_{\mathcal{E}[\rho]} \geq \Halpha(Q|Q')_{\rho}.
	\end{align}
	If $\mathcal{E}$ is an isometry, then we have equality in the above bound.
\end{lemma}

\begin{lemma}\label{lemma:DPIfweighted}
	(Data-processing \cite[Lemma~4.4]{arqand_marginal_2025}) Let $\rho \in \dop{=}(\CP Q Q')$ be classical on $\CP$, let $f$ be a tradeoff function on $\CP$, and take any $\alpha\in [\frac{1}{2},1)\cup(1,\infty]$. Then for any channel $\mathcal{E} \in \CPTP(Q',Q'')$, 
	\begin{align}
		\Halpha[f](Q|\CP Q'')_{\mathcal{E}[\rho]} \geq \Halpha[f](Q|\CP Q')_{\rho}.
	\end{align}
	If $\mathcal{E}$ is an isometry, then we have equality in the above bound.
\end{lemma}
\begin{lemma} \label{lemma:dpinonconditioningregister}
    (Data-processing on non-conditioning register) Let $\rho \in \dop{=}(\CP Q Q')$ be classical on $\CP$, let $f$ be a tradeoff function on $\CP$, and take any $\alpha\in [\frac{1}{2},1)\cup(1,\infty]$. Then for any unital channel $\mathcal{F} \in \CPTP(Q,Q'')$, i.e, $\mathcal{F}$ maps the identity operator $\id_Q$ to $\id_{Q'}$, we have 
    \begin{align}
		\Halpha[f](Q''|\CP Q')_{\mathcal{F}[\rho]} \geq \Halpha[f](Q|\CP Q')_{\rho}.
	\end{align}
	If $\mathcal{F}$ is an isometry, then we have equality in the above bound.
\begin{proof}
    Let $\alpha\in\left(\frac{1}{2},1\right]$, then we have
    \begin{align}
        &\Halpha(Q|Q')_{\rho_{|\cP}}\leq\Halpha(Q''|Q')_{\mathcal{F}[\rho_{|\cP}]}\qquad ;\forall\cP\in\CP\nonumber\\
        &\Longrightarrow \rho(\cP)2^{\left(\frac{1-\alpha}{\alpha}\right)\left(\Halpha(Q|Q')_{\rho_{|\cP}}-f(\cP)\right)} \leq \rho(\cP)2^{\left(\frac{1-\alpha}{\alpha}\right)\left(\Halpha(Q''|Q')_{\mathcal{F}[\rho_{|\cP}]}-f(\cP)\right)},
    \end{align}
    where the first line is an application of data processing for unital channels~\cite[Corollary~5.1]{tomamichel_quantum_2016}. Note that the inequality is saturated if $\mathcal{F}$ is an isometry. Summing over all values of $\cP$, followed by taking log and multiplying both sides by $\frac{\alpha}{1-\alpha}$, yields the desired results. The proof for $\alpha>1$ follows similarly.
\end{proof}
\end{lemma}

\begin{lemma}\label{lemma:Monotonicity_f}
	(Monotonicity in $f$) Let $\rho \in \dop{=}(\CP Q Q')$ be classical on $\CP$, let $f,g$ be tradeoff functions on $\CP$ such that for all $\cP\in\CP$ we have $f(\cP)\leq g(\cP)$. Then, for any $\alpha\in [\frac{1}{2},1)\cup(1,\infty]$, we have
	\begin{align}
		\Halpha[g](Q|\CP Q')_{\rho} \leq \Halpha[f](Q|\CP Q')_{\rho}.
	\end{align}
\end{lemma}
\begin{proof}
   Let $\alpha > 1$. Then, we have
   \begin{align}
       &2^{\left(\frac{1-\alpha}{\alpha}\right) \left(\Halpha(Q|Q')_{\rho_{|\cP}} - f(\cP) \right) }\leq2^{\left(\frac{1-\alpha}{\alpha}\right) \left(\Halpha(Q|Q')_{\rho_{|\cP}} - g(\cP) \right) }\qquad  ;\forall\cP\in\CP\notag\\
       &\Longrightarrow\sum_{\cP}\rho(\cP) 2^{\left(\frac{1-\alpha}{\alpha}\right) \left(\Halpha(Q|Q')_{\rho_{|\cP}} - f(\cP) \right) }\leq\sum_{\cP}\rho(\cP)2^{\left(\frac{1-\alpha}{\alpha}\right) \left(\Halpha(Q|Q')_{\rho_{|\cP}} - g(\cP) \right) }.
   \end{align}
   Taking $\log$ and multiplying both sides by $\frac{\alpha}{1 - \alpha}$ yields the desired result. The $\alpha < 1$ case follows similarly.
\end{proof}

\begin{lemma}({\Renyi} Leftover-hashing Lemma~\cite[Theorem~8]{dupuis_privacy_2023})
\label{lemma:LHL}
Let $\rho_{AE}$ be a classical-quantum state, and $(\mathcal{F}_{\mathcal{A}\rightarrow\mathcal{Z}},p_f)$ be a \emph{ideal} family of universal$_2$ hash functions with $\mathcal{Z}=\{0,1\}^l$. Consider the state $\rho_{ZEF}$ that is obtained when $f$ is a function drawn from that family with probability $p_f$, and applied to the register $A$ to obtain the register $Z$, and the choice of the function is stored in $F$. Then, for $\alpha\in(1,2)$, we have
\begin{align}
    \label{eq:privacy_amp}
 \frac{1}{2}\left\|\rho_{ZEF}-\frac{1}{|\mathcal{Z}|}\id_z\otimes\rho_{EF} \right\|_1 \leq 2^{\frac{1-\alpha}{\alpha}\big(\Halpha(A|E)_\rho-l + 2\big)}.
\end{align}
\end{lemma}
\begin{remark}\label{remark:LHLequalityissue}
Note that one must be careful to distinguish between \emph{universal\(_2\)} hashing (defined by requiring the collision probability to be $\leq 2^{-l}$) and \emph{ideal universal\(_2\)} hashing (where the collision probability is exactly $2^{-l}$), a subtle difference that is often missed in the literature. In particular, the {\Renyi} leftover hash lemma as stated in Ref.~\cite{dupuis_privacy_2023} requires \emph{ideal} universal\(_2\) hashing. However, this requirement can be relaxed to ordinary universal\(_2\) hashing, at the cost of losing only one bit of key length~\cite{kamin_phd_2025}. For this work, we restrict our analysis and statements to protocols that utilize ideal universal$_2$ hashing in the privacy amplification step. Note that Toeplitz hashing is ideal universal$_2$ \cite{mansour_computational_1993}.
\end{remark}
\begin{lemma}\label{lemma:conditioning}
    (\cite[Lemma~B.5]{dupuis_entropy_2020})
    Let $\rho_{ABC} \in \dop{=}(ABC)$ be classical on $C$, such that $\rho_{ABC} = \sum_c p_c \ketbra{c}{c} \otimes \rho_{AB|c}$ for some probability distribution $\{p_c\}$ and normalized conditional states $\rho_{AB_{|c}}$. Then, for each $c$ and any $\alpha \in (1, \infty)$, we have:
\begin{align}\label{eq:conditioning}
\Halpha(A|B)_{\rho_{|c}}\geq \Halpha(A|B)_\rho-\frac{\alpha}{\alpha-1}\log\left(\frac{1}{p_c}\right)
\end{align}
\end{lemma}

\begin{lemma}\label{lemma:EC_cost}
    (\cite[Proposition~2.9]{LWD16})
    Let $\rho_{ABC} \in \dop{=}(ABC)$ be classical on $C$. Then, for any $\alpha\in(0,\infty)$, we have:
    \begin{align}
        \label{eq:EC_cost}
        \Halpha(A|BC)_\rho\geq \Halpha(A|B)_\rho-\log 
        |C|
    \end{align}
\end{lemma}
\begin{lemma}
    \label{lemma:pur_to_ext}
    Let $\rho_A\in\dop{=}(A)$ be a density operator. Furthermore, let $\rho_{AB}\in\dop{=}(AB)$ and $\rho_{AA'}\in\dop{=}(AA')$ be any extension and purification of $\rho_A$, respectively. Then
    \begin{align}
        \exists\,\mathcal{N}\in\CPTP(A',B)\;\operatorname{s.t.}\;\rho_{AB}=\mathcal{N}[\rho_{AA'}].
    \end{align}
    \begin{proof}
        Let $\rho_{ABR}$ be any purification of $\rho_{AB}$with $\dim(A') \leq \dim( BR)$, then by the isometric equivalence of purifications \cite[Proposition 2.29]{watrous_theory_2018}, there exists an isometry $V:A'\to BR$ such that
        \begin{align}
            \rho_{ABR}=V\rho_{AA'}V^\dagger.
        \end{align}
        Therefore, $\rho_{AB}=\Tr_R\left[V\rho_{AA'}V^\dagger\right]$. The claim thus follows by setting $\mathcal{N}$ to be the channel $\mathcal{N} (\cdot) = \Tr_R \circ V (\cdot) V^\dagger$.
    \end{proof}
\end{lemma}

We also reproduce below the core result of the MEAT framework \cite{arqand_marginal_2025}. The $A$ registers are renumbered to reflect their role in this work, and we include only the statements that are required for our analysis.

\begin{theorem}(MEAT \cite[Theorem 4.1a]{arqand_marginal_2025}) \label{theorem:MEAT}
	For each $j\in\{1,2,\cdots,n\}$, take a state $\sigma^{(j)}\in\dop{=}(A_{j})$, and a channel $\mathcal{M}_j\in\CPTP(A_{j}E_{j-1},S_j\CP_jE_j)$, such that $\CP_j$ are classical. Let $\rho$ be a state of the form $\rho_{S_1^n\CP_1^nE_n}=\mathcal{M}_n\circ\cdots\circ\mathcal{M}_1[\omega_{A_1^{n}E_0}]$ for some $\omega\in\dop{=}(A_1^{n}E_0)$, such that $\omega_{A_1^{n}}=\sigma_{A_1}^{(1)}\otimes\cdots\otimes\sigma_{A_{n}}^{(n)}$. For each $j$, suppose that for every value $\cP_1^{j-1}$, we have a tradeoff function $f_{|\cP_1^{j-1}}$ on registers $\CP_j$. Define
    \begin{equation}
        \begin{aligned}
            \kappa_{\cP_1^{j-1}} &\defvar \inf_{\nu\in\Sigma_j} \Halpha[ f_{|\cP_1^{j-1}}](S_j| \CP_j E_j \widetilde{E})_{\nu}, \\
             \Sigma_j &\defvar \left\{
\mathcal{M}_j\left[\omega_{A_{j}E_{j-1}\widetilde{E}}\right] 
\;\middle|\;
\omega \in \dop{=}(A_{j}E_{j-1}\widetilde{E}) \;\text{s.t.}\; \omega_{A_{j}}=\sigma_{A_{j}}^{(j)}
\right\},
        \end{aligned}
    \end{equation}
  with $\widetilde{E}$ being a register of large enough dimension to serve as a purifying register for any of the $A_{j}E_{j}$ registers.   Define the following ``normalized'' tradeoff function on $\CP_1^n$:
	\begin{align}\label{eq:fullQESnorm_simp}
		\hat{f}_\mathrm{full}( \cP_1^n) \defvar \sum_{j=1}^n \hat{f}_{|\cP_1^{j-1}}(\cP_j), \quad\text{where}\quad \hat{f}_{|\cP_1^{j-1}}(\cP_j) \defvar f_{|\cP_1^{j-1}}(\cP_j) + \kappa_{\cP_1^{j-1}}.
	\end{align}
	Then for any $\alpha\in (1,\infty]$ we have
	\begin{align}\label{eq:chainQESnorm_simp}
	H[\hat{f}_\mathrm{full}](S_1^n | \CP_1^n E_n)_\rho \geq 0.
	\end{align}  

\end{theorem}

\section{Proof sketch} \label{sec:sketch}
In this section, we provide a brief proof sketch of QKD security using MEAT \cite{arqand_marginal_2025}, before presenting the full proof in \cref{sec:proof}. By our discussion in \cref{sec:reductionstatement}, we know it suffices to consider only the ``honest authentication'' scenario, where all messages are received correctly in the correct order, and authentication never aborts (see \cref{theorem:reductionstatement}). We hence focus on such a scenario throughout this overview.

For the purposes of this sketch, let us first restrict ourselves to the simpler case of a fixed-length protocol, which makes a single overall accept/abort decision and outputs a key of some predetermined {fixed} length $\lfixed$ whenever it accepts. Thus, we restrict \nameref{prot:abstractqkdprotocol} from \cref{subsec:abstractprotdesc} as follows: 
\begin{itemize}
\item The variable-length decision step in the protocol is restricted to a simple \term{acceptance test} form, in which Alice only computes the observed frequency distribution on the classical string $\cobs$, and checks whether this frequency distribution lies inside some $\Sacc$ (chosen before the protocol), which is a predetermined set of frequency vectors often called an \term{acceptance set}. If it does, she sets the classical register $\flagkey$ to the value $\lfixed$ (we shall informally refer to this as the acceptance test ``passing''); otherwise she sets it to the value $0$. She then sends $\flagkey$ to Bob as in the original protocol. 

\item We require that there exists a \emph{fixed} constant value $\leakfixed \in \mathbb{N}$, such that the number of possible transcripts in the error correction step (conditioned on the acceptance test passing) is at most $2^{\leakfixed}$. With this, we can take the register $\flagEC$ to always be set to this fixed value $\leakfixed$, so it is effectively trivial and we can ignore it in our  analysis.

\item At the privacy amplification step, Alice and Bob instead check only whether $\flagkey$ was set to $\lfixed$ and $\flagEV$ indicates a match (let us denote these two events as $\OmegaAT$ and $\OmegaEV$ respectively). If both these events hold, Alice and Bob accept the protocol overall, and hash to final keys of length $\lfixed$; otherwise they abort the protocol. Note that this means the event $\Omega_\mathrm{acc}$ of the protocol accepting overall is exactly the joint event 
\begin{align}
\Omega_\mathrm{acc} = \OmegaAT \land \OmegaEV,
\end{align}
and thus we may instead denote it as $\OmegaAT \land \OmegaEV$ in contexts where that is more convenient.
\end{itemize}

The first step in many QKD security proofs consists of observing that in order to prove $\epssecure$-security (\cref{def:qkdsecuritysymmetric}), it suffices to prove two slightly simpler conditions, known as correctness and secrecy. Namely, in the case of a fixed-length protocol, we say it is $\epscorr$-correct if, for all output states, the following inequality is satisfied:
\begin{equation}
    \Pr(K_A \neq K_B \wedge \Omega_\mathrm{acc}) \leq \epscorr,
\end{equation}
and we say it is $\epssecret$-secret if, for all output states, the following inequality is satisfied:
\begin{equation}\label{eq:secrecyfixedlengthgeneral}
    \tracedist{ \rho^\text{real}_{K_A   \Esecdef} -  \rho^\text{ideal}_{K_A   \Esecdef}}  = \Pr(\Omega_\mathrm{acc})  \tracedist{ \rho^\text{real}_{K_A   \Esecdef | \Omega_\mathrm{acc}} -  \rho^\text{ideal}_{K_A  \Esecdef| \Omega_\mathrm{acc}}} \leq \epssecret.
\end{equation}
It is known that if a protocol is $\epscorr$-correct and $\epssecret$-secret, then it is $(\epscorr+\epssecret)$-secure --- see Ref.~\cite{portmann_security_2022} for fixed-length protocols. For the more general case of variable-length protocols, one can similarly introduce definitions of correctness and secrecy with analogous properties, as discussed in Ref.~\cite{tupkary_security_2024} or \cref{lemma:correctandsecret} later. 

With this in mind, we now sketch how to prove correctness and secrecy for such protocols, which can thus be combined to obtain the required security statement. 

\subsection{Correctness}

To prove correctness, by definition we need to upper bound the probability that the final keys differ \emph{and} the protocol accepts. This is fairly straightforward: first, observe that since this event can only occur if the pre-amplification strings differ and the error verification step accepts, it suffices to upper bound the probability of that event instead. In turn, this (joint) probability can be upper bounded in terms of a conditional probability, as follows (writing $\Omegadiff$ to denote the event of the pre-amplification strings differing):
\begin{align}
\\Pr[\Omegadiff \land \OmegaEV] 
&= \Pr[\OmegaEV|\Omegadiff] \Pr[\Omegadiff] \nonumber\\
&\leq \Pr[\OmegaEV|\Omegadiff],
\end{align}
where the first line is immediate from the definition of conditional probability, and the second line holds since $\Pr[\Omegadiff] \leq 1$. Finally, we note that $\Pr[\OmegaEV|\Omegadiff]$ is upper bounded by $\epsEV$ from the fundamental property of the universal$_2$ hashing procedure in error verification (recalling the hash length is $\ceil{ \log(\frac{1}{\epsEV})}$), and thus the protocol is $\epsEV$-correct.\footnote{We emphasize that, as noted in~\cite{tupkary2025qkdsecurityproofsdecoystate}, the proof does \emph{not} involve trying to upper bound the probability that the pre-amplification strings differ conditioned on error verification accepting, i.e.~$\Pr[\Omegadiff|\OmegaEV]$ (the ``opposite'' conditional probability from the proof described here). In particular, it is in general impossible to find a nontrivial upper bound on $\Pr[\Omegadiff|\OmegaEV]$ against all possible attacks, as discussed in that work.}

\newcommand{\hupalpha}{h^\uparrow_\alpha}
\subsection{Secrecy}

We first remark that strictly speaking, many steps in the secrecy proof have only been fully proven for finite-dimensional systems, though they mostly do not have explicit dependence on the system dimensions. Therefore, applying such an analysis to e.g.~optical QKD protocols technically requires some argument that it suffices to consider finite-dimensional systems in some suitable fashion. However, we do not further discuss this  technical point within this sketch, deferring it to the full security proof (see in particular \cref{app:infinitetofinite,sec:optics,subsec:evedimensions}).

Putting that aside, we now outline the overall structure of how to show that the fixed-length protocol is $\epsPA$-secret as long as the output key length (when the protocol accepts) is chosen to be some constant $\lfixed$ satisfying
\begin{align}
\lfixed &\leq n \hupalpha - \leakfixed - \ceil{ \log(\frac{1}{\epsEV})} - \frac{\alpha}{\alpha-1}\log\frac{1}{\epsPA} + 2, \label{eq:lfixed}
\end{align}
where $\alpha$ can be chosen to be any arbitrary value in the interval $(1,2]$, and $\hupalpha$ is a value that is defined in terms of a convex optimization we describe below in \cref{eq:halpha}. For now, we simply note that $\hupalpha$ has the property that it is decreasing with respect to $\alpha$, and thus the choice of $\alpha$ ``trades off'' between two terms in this expression --- namely, as we take $\alpha \to 1$, the $n \hupalpha$ term increases but the $-\frac{\alpha}{\alpha-1}\log\frac{1}{\epsPA}$ term becomes increasingly negative. When using this bound in practice, the value of $\alpha$ should be tuned to obtain the best possible finite-size key length\footnote{Note that since \emph{any} choice of $\alpha\in(1,2]$ yields a valid choice of $\lfixed$, this optimization of $\alpha$ is only relevant in the sense of obtaining good finite-size performance --- the resulting key rates are always ``secure'', i.e.~not an over-estimate of the true secure key rate.} for a given $n$; qualitatively, the best $\alpha$ value approaches $1$ as $n$ increases, since $n \hupalpha$ becomes more dominant over $-\frac{\alpha}{\alpha-1}\log\frac{1}{\epsPA}$. In the asymptotic $n\to\infty$ limit, by taking $\alpha\to1$ at a suitable rate, it can be shown~\cite[Section~5.2]{arqand_generalized_2024} that the resulting key rate converges to the standard ``Devetak-Winter formula''~\cite{devetak2005distillation} in the literature, though we do not discuss these details regarding asymptotic behaviour in this work.

Let $\rho_{|\OmegaAT \land \OmegaEV}$ be the state just before privacy amplification, conditioned on accepting overall. 
In this state, Eve has access to the registers $\CEC \CEV \flagkey \flagEV \HEV E_n$ (see \cref{table:abstractprodesc}), and as discussed in \cref{rem:diffclassicalregisters}, we suppose without loss of generality that Eve did not act on those registers during the classical phase of the protocol. Note that we ensure that Eve has access to the public announcements $\CP_1^n$ as well, which are included within the $E_n$ register; also, we have ignored the register $\flagEC$ because of our restriction to protocols that always trivially set it to a fixed value.  
With this in mind, we can rewrite the fixed-length secrecy condition (\cref{eq:secrecyfixedlengthgeneral}) as
\begin{align}\label{eq:secrecyfixedlengthdetailed}
\Pr[\OmegaEV\land\OmegaAT] \tracedist{\rho_{K_A \CEC \CEV \flagkey \flagEV \HPA \HEV E_n |\OmegaEV\land\OmegaAT }  - \tau^{\lfixed}_{K_A} \otimes \rho_{\CEC \CEV \flagkey \flagEV \HPA \HEV E_n |\OmegaEV\land\OmegaAT } } \leq \epssecret.
\end{align}
Comparing this secrecy condition against the Leftover Hashing Lemma (\cref{lemma:LHL}), it is thus natural that our proof proceeds via bounding the entropy $\Halpha(S_1^n |\CEC \CEV \flagkey \flagEV \HEV E_n)_{\rho_{|\OmegaAT \land \OmegaEV}}$ (a subtle technicality is that the pre-amplification string $\PAstring$ may differ from $S_1^n$, a point that we return to the end of this section).
To do so, we first remove some of the conditioning registers and conditioning events. Specifically, we have
\begin{align}
\Halpha(S_1^n |\CEC \CEV \flagkey \flagEV \HEV E_n)_{\rho_{|\OmegaAT \land \OmegaEV}}
&= \Halpha(S_1^n |\CEC \CEV \HEV E_n)_{\rho_{|\OmegaAT \land \OmegaEV}} \nonumber\\
&\geq 
\Halpha(S_1^n |\CEC \CEV \HEV E_n)_{\rho_{|\OmegaAT}} - \frac{\alpha}{\alpha-1}\log\frac{1}{\Pr[\OmegaEV|\OmegaAT]}\nonumber\\
&= \Halpha(S_1^n |\HEV E_n)_{\rho_{|\OmegaAT}} - \leakfixed - \ceil{ \log(\frac{1}{\epsEV})} - \frac{\alpha}{\alpha-1}\log\frac{1}{\Pr[\OmegaEV|\OmegaAT]}\nonumber\\
&\geq \Halpha(S_1^n | E_n)_{\rho_{|\OmegaAT}} - \leakfixed - \ceil{ \log(\frac{1}{\epsEV})} - \frac{\alpha}{\alpha-1}\log\frac{1}{\Pr[\OmegaEV|\OmegaAT]}.
\end{align}
In the above, the first line holds since the registers $\flagkey \flagEV$ have a deterministic value in the conditional state $\rho_{|\OmegaAT \land \OmegaEV}$. The second line is by \cref{lemma:conditioning}. The third line holds by the chain rule in \cref{lemma:EC_cost}.
The fourth line is by using the fact that we can consider $\rho_{S_1^n \HEV E_n|\OmegaAT}$ to be the (appropriate marginal of the) state \emph{immediately} after $\HEV$ is generated (conditioned on $\OmegaAT$) --- in this state, $\HEV$ is independent of all other registers and we can thus remove it using the data-processing inequality in both directions (\cref{lemma:DPI}). 

Hence our task is now reduced to  bounding $\Halpha(S_1^n | E_n)_{\rho_{|\OmegaAT}}$. The technical tool that allows us to do so is
the MEAT (in the form of~\cite[Theorem~4.2a]{arqand_marginal_2025}), which states that for the value $\hupalpha$ we describe later in \cref{eq:halpha}, we have
\begin{align}\label{eq:MEATfixedlength}
\Halpha(S_1^n | E_n)_{\rho_{|\OmegaAT}} \geq 
n \hupalpha - \frac{\alpha}{\alpha-1}\log\frac{1}{\Pr[\OmegaAT]}.
\end{align}
Note that to justify the validity of this bound, we need to formulate a model for the protocol in terms of abstract channels compatible with the MEAT statement. We defer this discussion, along with the description of $\hupalpha$, to \cref{subsec:sketchMEAT} later in this proof sketch.

Putting the above bounds together, we get
\begin{align}
&\Halpha(S_1^n |\CEC \CEV \flagkey \flagEV \HEV E_n)_{\rho_{|\OmegaAT \land \OmegaEV}} \nonumber\\
\geq& n \hupalpha - \leakfixed - \ceil{ \log(\frac{1}{\epsEV})} - \frac{\alpha}{\alpha-1}\log\frac{1}{\Pr[\OmegaEV|\OmegaAT]} - \frac{\alpha}{\alpha-1}\log\frac{1}{\Pr[\OmegaAT]} \nonumber\\
=& n \hupalpha - \leakfixed - \ceil{ \log(\frac{1}{\epsEV})} - \frac{\alpha}{\alpha-1}\log\frac{1}{\Pr[\OmegaEV\land\OmegaAT]} .
\label{eq:finalentbound}
\end{align}
Then as long as we choose $\lfixed$ to be some value satisfying \cref{eq:lfixed},
we can show that the secrecy definition (in \cref{eq:secrecyfixedlengthdetailed}) holds with secrecy parameter $\epsPA$:
\begin{align}
&\Pr(\OmegaEV\land\OmegaAT)\tracedist{\rho_{K_A \CEC \CEV \flagkey \flagEV \HPA \HEV E_n |\OmegaEV\land\OmegaAT }  - \tau^{\lfixed}_{K_A} \otimes \rho_{\CEC \CEV \flagkey \flagEV \HPA \HEV E_n |\OmegaEV\land\OmegaAT } } \nonumber\\
\leq & \Pr(\OmegaEV\land\OmegaAT)  2^{\frac{1-\alpha}{\alpha}\left(\Halpha(S_1^n |\CEC \CEV \flagkey \flagEV \HEV E_n)_{\rho_{|\OmegaAT \land \OmegaEV}} - \lfixed + 2\right)} \nonumber\\
\leq& \Pr(\OmegaEV\land\OmegaAT) 2^{\log\frac{1}{ \Pr(\OmegaEV\land\OmegaAT) } - \log\frac{1}{\epsPA}} \qquad \text{ by substituting \cref{eq:finalentbound,eq:lfixed}} \nonumber\\
=&  \epsPA,
\end{align} 
as desired.
In the above, the second line is essentially based on the Leftover Hashing Lemma (\cref{lemma:LHL}) but is technically not a direct application of it, because that lemma by itself does not accommodate the fact that our \nameref{prot:abstractqkdprotocol} allows for $\PAstring$ to differ from $S_1^n$ (see also \cite[Section 5.6]{tupkary2025qkdsecurityproofsdecoystate}), and for the privacy amplification hash family to depend on $\CP_1^n$. A complete justification of that second line needs to consider both of these points --- this was rigorously addressed in~\cite[Lemmas 4--5]{tupkary_security_2024} (under the \nameref{prot:abstractqkdprotocol} structure, which implicitly places some minor constraints on how $\PAstring$ and the hash family are chosen), and we also provide a complete analysis later in \cref{lemma:equalityentropy} and \cref{theorem:entropytovarlength} of our full security proof, via a slightly different sequence of steps from that work.

\subsection{Bounding the {\Renyi} entropy}\label{subsec:sketchMEAT}
\newcommand{\channeloutputset}{\Sigma}
\newcommand{\onerndchann}{\mathcal{M}}
\newcommand{\Alicechann}{\mathcal{M}^{A}}
\newcommand{\Bobchann}{\mathcal{M}^{B}}
\newcommand{\processchann}{\mathcal{M}^\mathrm{process}}

Let us first outline how to justify that the MEAT can be validly invoked, to obtain the bound in \cref{eq:MEATfixedlength}. The first step is to apply the source-replacement argument, which is a standard technique for converting a PM protocol to an equivalent entanglement-based picture. 
We provide a detailed explanation in \cref{subsec:expressingprotocolassequence}; for this sketch here, we will simply lay out the broad ideas. First note that without loss of generality, we can instead consider a scenario where Alice sends out the signal states for \emph{all} rounds at the very start of the protocol, since giving Eve earlier access to these states does not reduce her class of possible attacks. In this scenario, immediately after Alice sends out these signal states, we have a global state of the form $\bigotimes_{j=1}^n \sigma^{(j)}_{X_j A'_j}$ with the $X_j$ registers being classical.
The source-replacement technique now makes the key observation that this state could equivalently have been produced by starting with a pure state of the form $\bigotimes_{j=1}^n \ket{\sigma^{(j)}}_{A_j A'_j}$, then for each $j$, performing a suitable measurement channel $\Alicechann_j \in \CPTP(A_j,X_j)$. The $A_j$ registers can include ``shield systems'' that contain purifications of signal states if they are mixed; this is a  technical point that we describe in more detail in \cref{subsec:expressingprotocolassequence}. 

For ease of explanation, let us now restrict this proof sketch only to protocols where the public announcements of \emph{all} the $\CP_j$ registers only take place after all state preparations and measurements have been completed (though still before other steps such as error correction and sifting).\footnote{In other words, we exclude protocols with on-the-fly announcements ---  analyzing such protocols requires a fairly detailed argument and we do not attempt to cover it within this sketch, deferring it to the full security proof (see in particular \cref{subsec:expressingprotocolassequence}, though some later parts of the proof are also relevant in analyzing the single-round quantities).}
In such a protocol, Eve's actions on the signal states before Bob's measurements cannot depend on the public announcements, and can thus be modelled as applying some arbitrary attack channel $\mathcal{A} \in \CPTP( (A')_1^n,B_1^n \bar{E})$ across all the signal states to produce some global state $\rho_{X_1^n B_1^n \bar{E}}$, then forwarding $B_1^n$ to Bob and keeping $\bar{E}$ as Eve's quantum side-information. Crucially, this channel $\mathcal{A}$ commutes with the measurements $\Alicechann_j$ defined above in the source-replaced picture, which implies this state $\rho_{X_1^n B_1^n \bar{E}}$ could equivalently have been produced by \emph{starting} with a suitable initial state $\rho_{A_1^n B_1^n \bar{E}} $ and \emph{then} applying the measurements $\Alicechann_j$; more explicitly, by taking $\rho_{A_1^n B_1^n \bar{E}} \defvar \mathcal{A} \left[\bigotimes_{j=1}^n \ket{\sigma^{(j)}}_{A_j A'_j}\right]$ (for some unknown $\mathcal{A}$) we can write 
\begin{align}
\rho_{X_1^n B_1^n \bar{E}} = \left(\bigotimes_{j=1}^n \Alicechann_j\right) [\rho_{A_1^n B_1^n \bar{E}}], \text{ where $\rho_{A_1^n B_1^n \bar{E}}$ satisfies $\rho_{A_1^n } = \bigotimes_{j=1}^n \sigma^{(j)}_{A_j}$}.
\end{align}
We refer to the fact that $\rho_{A_1^n } = \bigotimes_{j=1}^n \sigma^{(j)}_{A_j}$ in the above equation as the \term{marginal constraint} on $A_1^n$. Note that the marginal constraint on $A_1^n$ holds regardless of Eve's choice of attack, because the channel $\mathcal{A}$ does not act on the $A_1^n$ registers. This marginal constraint is a subtle but critical property that is needed for tight key rate bounds later. 

Finally, note that for each round under the \nameref{prot:abstractqkdprotocol} structure, Bob's measurement on $B_j$ can be described by some measurement channel $\Bobchann_j \in \CPTP(B_j, Y_j)$, and the subsequent processing of the single-round values $X_j Y_j$ into $S_j \CP_j$ can be described by yet another channel $\processchann_j \in \CPTP(X_j Y_j, S_j \CP_j)$ (we can exclude  $X_j Y_j$ from the output of this channel as they can be discarded after this processing).
Putting these points together, we see that the state on $S_1^n \CP_1^n \bar{E}$ immediately after $\CP_1^n$ are announced has the following form, defining $\onerndchann_j \defvar \processchann_j \circ (\Alicechann_j \otimes \Bobchann_j)$:
\begin{align}\label{eq:tensoroutputstate}
\rho_{S_1^n \CP_1^n \bar{E}} =
\left(\bigotimes_{j=1}^n \onerndchann_j\right) [\rho_{A_1^n B_1^n \bar{E}}], \text{ where $\rho_{A_1^n B_1^n \bar{E}}$ satisfies $\rho_{A_1^n } = \bigotimes_{j=1}^n \sigma^{(j)}_{A_j}$}.
\end{align}
This is the state for which we wish to write the entropy bound in \cref{eq:MEATfixedlength}, identifying $E_n \equiv \CPhat_1^n \bar{E}$ where $\CPhat_1^n$ is a classical copy\footnote{Here we implicitly extend the state $\rho_{S_1^n \CP_1^n \bar{E}}$ from \cref{eq:tensoroutputstate} with this classical copy of $\CP_1^n$; introducing this copy is mainly a technical formality for consistency with the later conventions in our full security proof.} of $\CP_1^n$, since 
Eve has access to those registers. 
States of this form are precisely what is covered by the MEAT (in the form presented in~\cite[Corollary~4.2]{arqand_marginal_2025}), which gives exactly the claimed bound in \cref{eq:MEATfixedlength}, in terms of the value $\hupalpha$ --- which we now finally turn to discussing. 

For the remainder of this sketch, we focus on protocols where Alice's state preparation,  Bob's measurement, and the processing $\processchann_j$ in the original PM picture is the same in every round.\footnote{We emphasize that this is not assuming that Eve's \emph{attack} is IID; we are only taking the trusted operations by Alice and Bob to be IID.} (This mirrors the analysis from Ref.~\cite{kamin_renyi_2025}.) Then all the single-round channels $\onerndchann_j$ described above in the source-replaced picture are isomorphic to each other, and similarly for the marginal states $\sigma^{(j)}_{A_j}$; hence for brevity, we will now drop the subscript $j$ in the subsequent discussion of single rounds. With this notation, $\hupalpha$ has the following value~\cite[Eq.~(137) in Corollary~4.2]
{arqand_marginal_2025}, letting $\widehat{E}$ be a register of large enough dimension to purify $AB$:
\begin{align}\label{eq:halpha}
\hupalpha &= \inf_{\nu\in\channeloutputset} \inf_{\mbf{q}\in\convSacc}  \left(\frac{\alpha}{\alpha-1}D\left(\mbf{q}\Vert\bsym{\nu}_{\CP} \right)+\sum_{\cP}q(\cP)\Halpha(S| \widehat{E})_{\nu_{|\cP}}\right) ,
\end{align}
where $\convSacc$ denotes the convex hull of the acceptance set $\Sacc$,
and $\channeloutputset \defvar \{\onerndchann[\omega_{AB\widehat{E}}] \;|\; \omega \in \dop{=}(AB\widehat{E}) \text{ s.t. } \omega_A = \sigma_A \}$ can be qualitatively viewed as the set of all states that could be generated by a single protocol round, keeping in mind Alice's marginal state in the source-replaced picture. As briefly noted above, having this marginal constraint ensures that the set $\channeloutputset$ does not become ``too large'', in the sense of including many states that could not have been produced via the prepare-and-measure process in the protocol.
For more general protocols than the class we have restricted to in this sketch, one still arrives at an optimization involving an essentially similar choice of $\channeloutputset $; however, the explanation is rather technical and we defer it to \cref{subsec:reformulatingattackmin}.\footnote{Moreover, this may not be exactly the correct set to consider when some particular types of squashing maps are used. However, we also defer this technical discussion to the full security proof, for instance in \cref{remark:sourcemaplemma}.}

Informally, we can gain some intuition for this optimization by viewing it as being similar in spirit to finding the minimal single-round entropy $\Halpha(S| \CP \widehat{E})$ over  \emph{exactly} the set of states ``compatible with'' the acceptance test, i.e.
\begin{align}\label{eq:exactopt}
\begin{aligned}
\inf_{\nu\in\channeloutputset} &\qquad \Halpha(S| \CP \widehat{E})_\nu, \\
    \text{ s.t.} &\qquad \bsym{\nu}_{\CP} \in \convSacc.
\end{aligned}
\end{align}
Qualitatively, the difference between the optimizations in \cref{eq:halpha} and \cref{eq:exactopt} is that in place of the ``hard constraint'' $\bsym{\nu}_{\CP} \in \convSacc$  in the latter, we have a ``soft'' version of that constraint in the former, as quantified by the term $\frac{\alpha}{\alpha-1}D\left(\mbf{q}\Vert\bsym{\nu}_{\CP} \right)$  that acts as a ``penalty'' whenever $\bsym{\nu}_{\CP} \notin \convSacc$. In other words, in the \cref{eq:halpha} optimization, the state $\nu$ does not need to strictly obey a $\bsym{\nu}_{\CP} \in \convSacc$ constraint, but is still ``penalized'' for violating that constraint by increasing the objective value by $\frac{\alpha}{\alpha-1}D\left(\mbf{q}\Vert\bsym{\nu}_{\CP} \right)$ (which would be positive since $\mbf{q}$ \emph{is} constrained to lie inside $\convSacc$). Hence we can intuitively view the former as a relaxed version of the latter, with the latter being what we might have informally expected in an IID scenario while ignoring finite-size effects.\footnote{For this informal intuition, we put aside the fact that \cref{eq:halpha} has a ``weighted average'' $\sum_{\cP}q(\cP)\Halpha(S| \widehat{E})_{\nu_{|\cP}}$ rather than the entropy $\Halpha(S| \CP \widehat{E})_\nu$ in \cref{eq:exactopt}; these quantities have similar behaviours in some ways as $\alpha$ becomes close to $1$.} Also, as noted at the start of this section, as $n\to\infty$ we typically take $\alpha\to1$, in which case the ``penalty'' term increases and thus enforces the constraint ``more stringently'', and moreover the {\Renyi} entropy $\Halpha$ converges towards the von Neumann entropy.

We stress that while the above intuitive comparison to \cref{eq:exactopt} was informal, the point of the MEAT is that the \cref{eq:halpha} formula (substituted into \cref{eq:MEATfixedlength}) yields a \emph{rigorous} lower bound on the entropy of the final state, which has already accounted for all finite-size and non-IID effects. Hence one can rigorously prove security of the fixed-length protocol by evaluating or lower-bounding the optimization in \cref{eq:halpha}.

\subsection{Differences for the variable-length case}

For the more general scope of variable-length protocols, the proof of correctness remains identical, but secrecy is somewhat more challenging to prove. This is basically because it no longer suffices to only analyze the {\Renyi} entropy conditioned on a single overall accept event $\OmegaAT \land \OmegaEV$ --- rather, since the length of the final key can potentially depend on the observed value on the $\CP_1^n$ registers, we need a more fine-grained analysis that considers the {\Renyi} entropy conditioned on the different values those registers can take. At a very informal level, the rough idea of the proof is that we instead aim to construct a function $\fhatfull(\CP_1^n)$ of the value on those registers, which serves as a ``statistical estimator'' of the final {\Renyi} entropy conditioned on that value. Since this ``statistical estimator'' $\fhatfull(\CP_1^n)$ can be physically computed in the protocol (from the observed value on the public announcements $\CP_1^n$), Alice and Bob can then hash to a final key length chosen based on $\fhatfull(\CP_1^n)$ (and various other values such as $\leak(\CP_1^n)$), in such a way that we can apply the Leftover Hashing Lemma to show that the variable-length secrecy definition (\cref{def:qkdsecuritysymmetric}) is satisfied. 

The main technical tool that allows us to construct $\fhatfull$ is the MEAT statement that we present as \cref{theorem:MEATQKDfirst} later in the full security proof. Qualitatively, it specifies ``single-round functions''\footnote{Here we put aside the most general scope of that statement, which in fact allow these functions to depend on past rounds as well.} $\fhat$ that can be evaluated on the $\CP_j$ values observed from single rounds of the protocol, in such a way that summing these single-round values gives a function $\fhatfull$ with the properties sketched above. Evaluating these single-round functions requires bounding an optimization that is similar in some ways to \cref{eq:halpha} for the fixed-length version, though we defer the detailed explanation to the full security proof later.

An important conceptual difference between the fixed-length security proof and the above sketch for the variable-length version is that unlike the \cref{eq:finalentbound} bound, the ``statistical estimator'' $\fhatfull(\CP_1^n)$ does not have the property of \emph{always} being a lower bound on the final {\Renyi} entropy, for every value $\CP_1^n$ that could occur. Rather, it can be informally said to bound the final {\Renyi} entropy ``on average'' (thus our choice of the term ``statistical estimator'' for it), albeit in a modified sense of averaging that we explain further in \cref{subsec:proofgenericstuff} later. Fortunately, this sense of ``averaged'' bound is sufficient for us to \emph{rigorously} prove that the variable-length secrecy definition holds --- informally, this is because that definition is also essentially an ``averaged'' statement over the different possible key lengths. In the remainder of this manuscript, we formalize the qualitative properties discussed above, thereby obtaining a rigorous security proof for protocols of the form described in this work. 

\section{Proving security of QKD Protocol using MEAT} \label{sec:proof}

In this section, we embark upon proving the security of the \nameref{prot:abstractqkdprotocol}. Recall that due to our analysis of authentication  from \cref{theorem:reductionstatement},  we are only concerned with proving the security of the QKD protocol (before the final authentication postprocessing step) in the setting where authentication can be assumed to be honest. We will first begin by expressing the \nameref{prot:abstractqkdprotocol} as a sequence of CPTP maps, before analyzing the sequence of maps using MEAT \cite[Theorem 4.1a]{arqand_marginal_2025} (\cref{theorem:MEAT}).

\subsection{Expressing the protocol as a sequence of channels} \label{subsec:expressingprotocolassequence}
The analysis in this section is identical to the analysis found in \cite[Section 5]{arqand_marginal_2025}. The goal of this subsection is to reduce the QKD protocol we are interested in analyzing to one which can be expressed as a sequence of channels as described in \cref{lemma:stateevolution,fig:MEAT_Attack}, in preparation for the application of MEAT.

\subsubsection{Modifying timings of the QKD protocol}

Let us consider the \nameref{prot:abstractqkdprotocol} described in \cref{subsec:abstractprotdesc}, under the honest authentication setting. 
In this case, the protocol can be equivalently expressed as the following \nameref{prot:pmprotocoll}, which will be the object of our subsequent analysis. Note that since we will only be modifying certain timings and signal-preparation steps of this protocol, only the relevant parameters are listed explicitly below.

\begin{prot}[Prepare-and-measure protocol]\label{prot:pmprotocoll} 

\leavevmode \\

\noindent \textbf{Parameters:}

\begin{tabularx}{0.9\linewidth}{r c X}
\(n \in \mathbb{N}_0\) 			    &:& 	Total number of rounds \\
\(\{\sigma^{(j)}_{X_jA'_j}\}_{j=1}^n\) 	&:& 	Classical-quantum states prepared by Alice \\
\(\{t_j^A,t_j^B,t^{\mathrm{ann}}_j\}_{j=1}^n\) 						&:& 	Timings of various steps (see below)
\end{tabularx}

\noindent We assume that the times listed above satisfy the following conditions (see \cref{remark:justifytiming}) :
\begin{itemize}
\item For each $j \in \{1,2,\dots,n\}$, we have $t^A_j,t^B_j<t^{\mathrm{ann}}_j$. 
\item Each of the sequences $\{t^A_j\}_{j=1}^n$, $\{t^B_j\}_{j=1}^n$, and $\{t^{\mathrm{ann}}_j\}_{j=1}^n$ is monotonically increasing.
\end{itemize}
We do not impose \emph{any} other restrictions on those timings; for instance, the sequences $\{t^A_j\}_{j=1}^n$, $\{t^B_j\}_{j=1}^n$, and $\{t^{\mathrm{ann}}_j\}_{j=1}^n$ can be ``interleaved'' with each other in an arbitrary fashion as long as the first condition is satisfied. In particular, we do \emph{not} require the public announcements to have  blocksize $\block=1$; larger blocksizes are permitted precisely because such interleaving is allowed.

\noindent \textbf{Protocol steps:}
\begin{enumerate}
\item Alice and Bob perform the following steps for every $j \in \{1,2,\dots,n\}$. \label{step:singlerounds}
\begin{enumerate}
\item \textbf{State preparation and transmission:} At time \(t_j^A\), independently for each round, Alice prepares some state \(\sigma^{(j)}_{k} \in S_{=} (A'_j) \) with probability $p^{(j)}_k$, and stores $k$ in the classical register \(X_j\).
She then sends out \({A'_j}\) via a public quantum channel, after which Eve may interact freely with it. This is represented via the preparation of the state $\sigma^{(j)}_{X_j A'_j} = \sum_{k=1}^{d_A} p^{(j)}_k \ketbra{k}_{X_j} \otimes \sigma^{(j)}_k$.

\item \textbf{Measurements:} 
At time $t^B_j$, Bob receives some quantum register from Eve and measures it using some POVM, storing the outcome in a register \(Y_j\). 

\item \label{step:announce} \textbf{Public announcement:} At time \(t_j^\mathrm{ann}\), Alice and Bob commence announcements for round $j$, based only on the values $X_j Y_j$. We denote all such public information as a classical register $\CP_j$.  Formally, $\CP_j$ can be viewed as the output of a CPTP map acting on the registers $X_j Y_j$.

\item \label{step:sift} \textbf{Sifting and key map:} 
At any time after \(t_j^\mathrm{ann}\), Alice computes a classical value \(S_j\) based only on her raw data \(X_j\) and the public announcements \(\CP_j\). This value \(S_j\) will later be used to generate her final key via privacy amplification. 

\end{enumerate}
\item \label{step:classical_processing} 
Alice and Bob apply various further classical procedures such as discarding values from $S_1^n$, variable-length decision, error correction, error verification, and privacy amplification as described in \nameref{prot:abstractqkdprotocol}.
\end{enumerate}
\end{prot}

\begin{remark} \label{remark:justifytiming}
Note that these conditions on the timings are extremely minimal. The first condition states that the announcements in round $j$ only happen after Alice has sent the state in that round and Bob has measured it. This is enforced in our protocol by design: Alice only starts announcing after she has received Bob's announcement, which in turn only occurs after Bob has finished measuring.  Moreover, under the honest-authentication assumption, Eve cannot cause Alice to announce prematurely by impersonating Bob. The second condition is simply a basic time-ordering condition on the sequences of preparations, measurements, and announcements \emph{individually}, without otherwise constraining them with respect to each other. These can be enforced by each party locally. In particular, these conditions do \emph{not} limit the repetition rate as in the existing generalized EAT-based proofs \cite{metger_security_2023}, which require the protocol to ensure that Eve has access to only one signal at a time (i.e, Alice can only send the state for the next round after Bob has measured the current round).
\end{remark}

For the security proof, as we will see later, we are interested in analyzing the state at the end of Step \cref{step:singlerounds} in \nameref{prot:pmprotocoll}. However, the structure of the above protocol is not yet compatible with the MEAT. Hence we first perform some modifications to the above protocol, which modify the timings $t^\mathrm{A}_j,t^\mathrm{B}_j, t^\mathrm{ann}_j$ of various steps to new values  $\tilde{t}^\mathrm{A}_j,\tilde{t}^\mathrm{B}_j, \tilde{t}^\mathrm{ann}_j$ satisfying:
\begin{equation} \label{eq:newtimes}
    \tilde{t}_1^\mathrm{A} = \dots = \tilde{t}^\mathrm{A}_n < \tilde{t}^\mathrm{B}_1 < \tilde{t}^\mathrm{ann}_1 < \tilde{t}^\mathrm{B}_2 < \tilde{t}^\mathrm{ann}_2 < \dots < \tilde{t}^\mathrm{ann}_n.
\end{equation}

To perform the above modification, we 
\begin{itemize}
\item Do not modify Bob's measurement timings, i.e, we set $\tilde{t}_j^\mathrm{B} = t^\mathrm{B}_j$.
    \item Move forward all of Alice's preparations such that they happen before Bob's first measurement.
    \item Move forward all announcements for round $j$ such that they happen right after Bob's measurement for that round, i.e, $\tilde{t}^\mathrm{B}_j < \tilde{t}^\mathrm{ann}_j < \tilde{t}^\mathrm{B}_{j+1}$. In this step, we ensure that \textit{all} announcements for round $j$ happen at $\tilde{t}^\mathrm{ann}_j$, 
\end{itemize}
Thus, in the modified protocol, Alice prepares and sends out all states at time $\tilde{t}^\mathrm{A}_1=\tilde{t}^\mathrm{A}_n$. Eve is allowed to perform any operation she wants on these states. After all states have been sent, Bob begins receiving and measuring states. After he measures the first state, Alice and Bob perform all announcements for round $j$. Bob then measures the second state and so on.

We stress that we have not lost any generality in considering the modified protocol. This is because moving the timings of Alice's preparations and public announcements to an earlier time does not reduce Eve's capabilities in any way: any action that Eve could have done in \nameref{prot:pmprotocoll} is also possible in the modified protocol, since at any point she always has access to more registers than in the modified protocol,  and she is free to ignore them until the time she would have interacted with them in the original protocol. This argument lets us write the following lemma.

\begin{lemma}[Modified Protocol] \label{lemma:modifiedprotocol}
Consider \nameref{prot:abstractqkdprotocol} with timings $t^\mathrm{A}_j,t^\mathrm{B}_j,t^\mathrm{ann}_j$ such that $t^\mathrm{A}_j,t^\mathrm{B}_j < t^\mathrm{ann}_j$, and  each of the sequences $\{t^A_j\}_{j=1}^n$, $\{t^B_j\}_{j=1}^n$, and $\{t^{\mathrm{ann}}_j\}_{j=1}^n$ is monotonically increasing. Consider the modified protocol that is identical to \nameref{prot:abstractqkdprotocol} except that the timings $\tilde{t}^\mathrm{A}_j,\tilde{t}^\mathrm{B}_j,\tilde{t}^\mathrm{ann}_j$ satisfy \cref{eq:newtimes}. Any Alice-Bob-Eve state that can be obtained in the original protocol, can also be obtained in the modified protocol.
\end{lemma}
\begin{proof}
    As described above.
\end{proof}
Therefore, for the purposes of proving security, we can now focus on analyzing the modified version. In the modified version, Alice prepares and sends out all states before Bob's first measurement, i.e, she prepares the global state 
\begin{equation} \label{eq:aliceprepearesstatesflobal}
\begin{aligned}
 \sigma_{X_1^n (A')_1^n} &=   \bigotimes_{j=1}^n \sigma^{(j)}_{X_j A'_j}, \\
   \sigma^{(j)}_{X_j A'_j} &= \sum_{k=1}^{d_A} p^{(j)}_k \ketbra{k}_{X_j} \otimes (\sigma^{(j)}_{k})_{A'_j},
\end{aligned}
\end{equation}
 before sending the $(A')_1^n$ systems to Eve. The evolution of the state in the modified protocol can now be described using the following lemma. Note that it is this protocol with modified timings for which we prove certain results in later sections. For this reason, we also formalize the notion of this protocol in terms of its internal maps and states (\cref{def:PMQKD}), and we express the relevant security requirement directly in terms of the maps that arise in the state evolution (\cref{def:qkdsecurityChannelVersion}).
 
\begin{lemma}[State evolution for modified protocol] \label{lemma:stateevolutionmodifiedprotocol}
Consider the \nameref{prot:abstractqkdprotocol} with timings $t^\mathrm{A}_j,t^\mathrm{B}_j,t^\mathrm{ann}_j$ such that $t^\mathrm{A}_j,t^\mathrm{B}_j < t^\mathrm{ann}_j$, and  each of the sequences $\{t^A_j\}_{j=1}^n$, $\{t^B_j\}_{j=1}^n$, and $\{t^{\mathrm{ann}}_j\}_{j=1}^n$ is monotonically increasing. Consider the modified protocol that is identical to the \nameref{prot:abstractqkdprotocol} except that the timings $\tilde{t}^\mathrm{A}_j,\tilde{t}^\mathrm{B}_j,\tilde{t}^\mathrm{ann}_j$ satisfy \cref{eq:newtimes}. Then, the evolution of the state through the modified protocol can be described as follows:
\begin{enumerate}
\item Alice prepares the global state $ \sigma_{X_1^n (A')_1^n}$  as in \cref{eq:aliceprepearesstatesflobal}. She sends $(A')_1^n$ to Eve. Eve thus starts with $E_0 = (A')_1^n$.\footnote{Of course, Eve may always start with additional auxiliary systems. However, since these systems are completely uncorrelated (i.e., in a trivial tensor-product form) with all the other registers in the QKD protocol, they can simply be added by Eve whenever needed, for instance when she implements her first attack channel $\attack{1}$.}
 
She keeps $X_1^n$ to herself, and sends the $(A')_1^n$ systems to Eve, which we relabel to be $E_0$.   
\item For each round $j \in \{1,\dots,n\}$, the following maps are applied:
\begin{enumerate}
    \item In each round $j$, Eve implements her attack $\attack{j} \in \CPTP(E_{j-1},B_j E^\prime_j)$.  \item The $B_j$ system is forwarded to Bob. Bob performs measurements on the $B_j$ system, and stores the outcome in the $Y_j$ register (Alice already has the $X_j$ register). They perform public announcements for the round $j$, which are stored in the register $\CP_j$. A copy  of the public announcements is given to Eve in the register $\CPhat_j$. Alice and Bob perform sifting to generate the $S_j$ registers. This entire process can be described by a map $\QKDGmapfullbeforeSR{j} \in \CPTP(X_j B_j , S_j  X_j Y_j \CP_j \CPhat_j)$.
    \item The $E'_j$ register and the $\CPhat_j$ register is combined into a unified $E_j$ register, which is passed into Eve's attack channel for the next round $\attack{j+1}$.\footnote{When $j=n$ is the last round, there is no $\attack{j+1}$ that is applied.}
\end{enumerate}
Thus, the state after the application of these maps is given by
\begin{equation}
      \rho_{S_1^n X_1^n Y_1^n \CP_1^n E_n } = \left( \QKDGmapfullbeforeSR{n} \circ \attack{n} \dots \QKDGmapfullbeforeSR{1} \circ \attack{1} \right) \left[ \sigma_{X_1^n (A')_1^n} \right] 
\end{equation}
\item Alice performs the final sifting step, generating the register $\PAstring$ from $S_1^n$ and $\CP_1^n$.
\item Alice and Bob perform steps such as the variable-length decision, error-correction, error-verification and privacy amplification. The final sifting operation, and the above procedures are all classical operations that are described by a map $\QKDpostprocessingmap\in \CPTP(X_1^n Y_1^n S_1^n \CP_1^n, K_A K_B \flagkey \flagEC \CEC \HEV \flagEV \HPA)$.
\end{enumerate}
\end{lemma}
\begin{proof}
    The proof follows from the description of the \nameref{prot:abstractqkdprotocol} with the modified timings. 
\end{proof}

\begin{definition}\label{def:PMQKD}
    At this stage, it is convenient to use the tuple $\left\{\{\QKDGmapfullbeforeSR{j}\}_{j=1}^n,\QKDpostprocessingmap, \sigma_{X_1^n (A')_1^n} \right\}$ to define an instance of the time-modified \nameref{prot:abstractqkdprotocol}, where $\QKDGmapfullbeforeSR{j},\QKDpostprocessingmap,\sigma_{X_1^n (A')_1^n}$ are defined in \cref{lemma:stateevolutionmodifiedprotocol}.
\end{definition} 
We can also restate the security definition \cref{def:qkdsecurityasymmetric} in terms of this sequence of channels.
\begin{definition}[QKD security as sequence of channels] \label{def:qkdsecurityChannelVersion}
Let $\left\{\{\QKDGmapfullbeforeSR{j}\}_{j=1}^n,\QKDpostprocessingmap, \sigma_{X_1^n (A')_1^n} \right\}$ define a QKD protocol that is described through the series of maps from \cref{lemma:stateevolutionmodifiedprotocol}. Let $\idealmap$ be the map that produces the ideal output state as defined in \cref{subsec:securitydefinition}.
Then, the QKD protocol is $\epssecure$-secure if, for all Eve's attack channels $\attack{j}\in \CPTP(E_{j-1},B_jE'_j)$, the following inequality is satisfied:
\begin{align}
    \left\|
        \left(
            \QKDpostprocessingmap \circ \QKDGmapfullbeforeSR{n} \circ \attack{n} \circ \cdots \circ \QKDGmapfullbeforeSR{1}\circ\attack{1}
            - \idealmap \circ \QKDpostprocessingmap \circ \QKDGmapfullbeforeSR{n} \circ \attack{n} \circ \cdots \circ \QKDGmapfullbeforeSR{1}\circ\attack{1}
        \right)
        \left[\sigma_{X_1^n (A'')_1^n}\right]
    \right\|_1 \leq \epssecure.
\end{align}
Here, without loss of generality, we can assume that $E_j,E'_j = \Hinf$ is infinite-dimensional; embedding any smaller spaces in $\Hinf$ if necessary.
\end{definition} 

\subsubsection{Applying the source-replacement scheme}

 In the above description, the starting state is one where Alice holds the classical register $X_1^n$  storing the labels for the states she sent in the protocol. This state then goes through the sequence of channels described in the above lemma. While the MEAT can be be applied to this sequence as well, doing so yields trivial results. Intuitively, this is because the application of the MEAT results in a single-round optimization that involves granting Eve access to an arbitrary purification of the system under Alice's control, which is a common feature in QKD security analysis. In this scenario, that would allow Eve to hold a purifying system for $X_j$ which is simply a copy of $X_j$. This issue is straightforwardly remedied by the use of the source-replacement scheme \cite{bennett_quantum_1992,curty_entanglement_2004}, which equivalently describes Alice's initial operation as preparing an entangled states followed by performing measurements on it. This is obtained in the following lemma.

\begin{lemma}[Shield System and Source Replacement] \label{lemma:shieldandsourcereplacement}
Consider a protocol where Alice first prepares the global state
\begin{equation}\label{eq:AlicePreMeasuredClassicalState}
\begin{aligned}
 \sigma_{X_1^n (A')_1^n} &=   \bigotimes_{j=1}^n \sigma^{(j)}_{X_j A'_j} \\
   \sigma^{(j)}_{X_j A'_j} &= \sum_{k=1}^{d_A} p^{(j)}_k \ketbra{k}_{X_j} \otimes (\sigma^{(j)}_{k})_{A'_j} &
\end{aligned}
\end{equation}
where $\{\ket{k}\}_{X_j}$ denotes the 
classical basis of 
$X_j$. She keeps $X_1^n$ to herself, and sends the $(A')_1^n$ systems to Eve.  
Then, the state $\sigma_{X_1^n (A')_1^n}$ can equivalently be obtained by
\begin{enumerate}
    \item Preparing the global state
   \begin{equation} \label{eq:globalsourcereplacedstate}
       \begin{aligned}
      \sourcesymbol_{\Ameas_1^n \Ashield_1^n (A')_1^n} &=\bigotimes_{j=1}^n \sourcesymbol^{(j)}_{\Ameas_j \Ashield_j A'_j}\\
          \sourcesymbol^{(j)}_{\Ameas_j \Ashield_j A'_j} &=\sum_ {k,i=1}^{d_A}\sqrt{p^{(j)}_ip^{(j)}_k} \ketbra{i}{k}_{\Ameas} \otimes \ketbra{\sigma^{(j)}_{i}}{\sigma^{(j)}_k}_{\Ashield_j A'_j}
       \end{aligned}
   \end{equation}
  where $\ket{\sigma_k^{(j)}}_{\Ashield_j A'_j}$ denotes a purification of the state $(\sourcesymbol_k^{(j)})_{A'_j}$, $\{\ket{k}\}_{k =1,\dots,d_A}$ is an orthornomal basis for $\Ameas$, and $\Ashield_j$ is referred to as the shield system.
   \item Sending the state $(A')_1^n$ to Eve,
   \item For each $j \in \{1,\dots, n\}$, measuring the $\Ameas_j \Ashield_j$ systems using the POVM $\{\ketbra{k}_{\Ameas_j}\otimes\mathbb{I}_{\Ashield_j}\}_{k\in\{1,\dots,d_A\}}$, and storing the result in $X_j$.
\end{enumerate}
\end{lemma}
\begin{proof}
For any round $j$, first consider the single round states $\sigma_{X_j A'_j}$ and $\sourcesymbol_{\Ameas_j \Ashield_j A'_j}$ as defined in \cref{eq:AlicePreMeasuredClassicalState,eq:globalsourcereplacedstate}. Simple algebra tell us that measuring $\sourcesymbol_{\Ameas_j \Ashield_j A'_j}$ with the POVM $\{\ketbra{k}_{\Ameas_j}\otimes\mathbb{I}_{\Ashield_j}\}_{k\in\{1,\dots,d_A\}}$, and storing the result in $X_j$ would result exactly in $\sigma_{X_j A'_j}$.
The $n$-round case follows similarly by considering the $n$-round states and $n$-round IID POVM instead.
\end{proof}

Note that  $\ket{\sigma_k^{(j)}}_{\Ashield_j A'_j}$ can be \emph{any} purification of $(\sigma_k^{(j)})_{A'_j }$.Crucially, we now observe that after the usage of the source replacement scheme to describe signal preparation using \cref{lemma:shieldandsourcereplacement}, Alice's operations are now equivalent to her preparing a global pure state $\sourcesymbol_{\Ameas_1^n \Ashield_1^n (A')_1^n}$, and then performing a suitable measurement on the $\Ameas_j$ registers to produce classical data in the $X_j$ systems. This has the critical property that Alice's measurements on the $\Ameas_j$ systems commute with all other operations that do not depend on the $X_j$ registers. In particular, Alice's measurements commutes with Eve's attack, and thus we can delay her measurements and assume that she does her measurements at the same time as Bob. This allows us to focus our analysis to that of the following virtual entanglement-based protocol, which produces the \textit{same exact} final state as the modified protocol from \cref{lemma:modifiedprotocol} (or the evolution of the state described in \cref{lemma:stateevolutionmodifiedprotocol}). This will be the protocol we analyze in subsequent sections. A security proof for this protocol implies the security proof of \nameref{prot:abstractqkdprotocol} via \cref{lemma:shieldandsourcereplacement,lemma:modifiedprotocol,lemma:stateevolutionmodifiedprotocol}.

\begin{remark} \label{remark:timecommutealice}
    Note that in the arguments above, we modify the timings of Alice's operations \emph{twice} - first by moving them earlier in time, and then by moving some of them later in time. Specifically, we initially argue that without loss of generality, one can assume that all state preparations are performed at the start of the protocol. This modification gives more power to Eve and thus only strengthens our security claims. Next, we apply the source-replacement technique, replacing Alice's state sending operations with the preparation of a global entangled state, part of which is sent to Eve ($(A')_1^n$), part of which is kept with Alice but not measured  ($\Ashield_1^n$), and part of which is kept with Alice and measured ($\Ameas_1^n$). Finally, we argue that since Alice's measurements commute with all operations except the public announcements (which depend on $X_1^n$), those measurements can be postponed to  after Eve's attack and just before public announcements for that round, without affecting Eve's strategy. Note that Alice's global state preparation still occurs at the initial time.
\end{remark}

\begin{prot}[Virtual entanglement-based protocol].\label{prot:virtualprotocoll} 

\noindent \textbf{Parameters:}

\begin{tabularx}{0.9\linewidth}{r c X}
\(n \in \mathbb{N}_0\) 			    &:& 	Total number of rounds \\
\(\{\sourcesymbol^{(j)}_{\Ameas_j \Ashield_jA'_j}\}_{j=1}^n\) 	&:& 	Pure quantum states prepared by Alice \\
\(\{\tilde{t}_j^A=\tilde{t}_j^B,\tilde{t}^{\mathrm{ann}}_j\}_{j=1}^n\) 						&:& 	Timings of various steps (see below), satisfying $\tilde{t}^\mathrm{B}_1 < \tilde{t}^\mathrm{ann}_1 < \tilde{t}^\mathrm{B}_2 < \tilde{t}^\mathrm{ann}_2 < \dots < \tilde{t}^\mathrm{ann}_n.$
\end{tabularx}

\noindent \textbf{Protocol steps:}
\begin{enumerate}

\item \textbf{State preparation and transmission:} Alice prepares a global pure state $\sourcesymbol_{\Ameas_1^n \Ashield_1^n (A')_1^n} = \bigotimes_{j=1}^n \sourcesymbol^{(j)}_{\Ameas_j \Ashield_j A'_j}$. She sends out all of the $(A')_1^n$ registers to Eve. She keeps $\Ameas_1^n \Ashield_1^n$ with herself. 
\item Alice and Bob perform the following steps for every $j \in \{1,2,\dots,n\}$. \label{step:singleroundsvirtual}
\begin{enumerate}
\item \textbf{Measurements:} \label{step:announcevirtual} At time $\tilde{t}_j^\mathrm{B}$, Bob receives quantum register $B_j$ from Eve, and performs a measurement on it, storing the outcome in the register $Y_j$. At the same time\footnote{Since this is a purely virtual protocol used only within the security proof, there is no need for Alice and Bob to physically implement these operations simultaneously.}
, Alice performs a measurement in the computational basis on the $\Ameas_j$ system, and stores the measurement outcome in the $X_j$ register. 
\item \textbf{Public Announcements:} At time $\tilde{t}_j^\mathrm{ann}$, Alice and Bob perform some public announcements which may be interactive. We denote all public announcements for round $j$ with the register $\CP_j$. A copy of these announcements in the register $\CPhat_{j}$ is made available to Eve. 
\item \textbf{Sifting and key map:}\label{step:siftvirtual} At time $\tilde{t}_j^\mathrm{ann}$, Alice maps her private data $X_j$ and public announcements $\CP_j$ to a classical value $S_j$.
\end{enumerate}

\item \label{step:classical_processing} 
Alice and Bob apply various further classical procedures such as variable-length decision, error correction, error certification, and privacy amplification as described in \nameref{prot:abstractqkdprotocol}.
\end{enumerate}
\end{prot}	

It is convenient to view the above \nameref{prot:virtualprotocoll} as a sequence of maps, depicted in \cref{fig:MEAT_Attack} and described in the following \cref{lemma:stateevolution}.

\newcommand{\MyTikzFigure}{

\begin{tikzpicture}[>=Stealth,auto,
  box/.style  ={draw,minimum width=1cm,minimum height=1cm,
                align=center,fill=red!20},
  cpbox/.style ={draw,minimum width=1.4cm,minimum height=0.8cm,
                align=center,fill=gray!20},
  sig/.style  ={->,rounded corners=4pt},
  merge/.style={circle,fill,inner sep=1.2pt}
]


\node[box]   (N1) {$\attack{1}$};
\node[cpbox, below right=\offsetY and \offsetX of N1] (L1) {$\QKDGmap{1}$};

\node[box,   right=\colsep of N1] (N2) {$\attack{2}$};
\node[cpbox, below right=\offsetY and \offsetX of N2] (L2) {$\QKDGmap{2}$};

\node[right=\colsep of N2] (dots) {$\cdots$};
\node[box,   right= of dots] (Nn) {$\attack{n}$};
\node[cpbox, below right=\offsetY and \offsetX of Nn] (Ln) {$\QKDGmap{n}$};

\node[left =1.4cm of N1] (E0) {};
\node[right=4cm   of Nn] (En) {};

\coordinate (join1) at ($(L1.east |- N1) + (1.0,0)$);
\coordinate (join2) at ($(L2.east |- N2) + (1.0,0)$);
\coordinate (joinn) at ($(Ln.east |- Nn) + (1.0,0)$);

\draw[sig] (E0) node[above=2pt, xshift=10pt]{$E_{0}=(A')_1^n$} -- (N1.west);

\draw[sig] (N1.east) node[above=2pt, xshift=10pt] {$E'_1$} -- (join1);
\draw[sig] (L1.east) node[above=2pt, xshift=10pt]   {$\CPhat_{1}$}
           -- ++(0.6,0) |- (join1);
\node[merge] at (join1) {};
\draw[sig] (join1) -- node[above=2pt,xshift=1pt] {$E_{1}$} (N2.west);

\draw[sig] (N2.east) node[above=2pt, xshift=10pt] {$E'_2$} -- (join2);
\draw[sig] (L2.east) node[above=2pt, xshift=10pt]   {$\CPhat_{2}$}
           -- ++(0.6,0) |- (join2);
\node[merge] at (join2) {};
\draw[sig] (join2) -- node[above=2pt,xshift=1pt] {$E_{2}$} (dots.west);


\draw[sig] (Nn.east) node[above=2pt, xshift=10pt] {$E'_n$} -- (joinn);
\draw[sig] (Ln.east) node[above=2pt, xshift=10pt]   {$\CPhat_{n}$}
           -- ++(0.6,0) |- (joinn);
\node[merge] at (joinn) {};

\draw[sig] (dots.east) node[above=2pt,xshift = 3pt] {$E_{n-1}$} -- (Nn.west);
\draw[sig] (joinn) -- node[above=2pt,xshift=1pt] {$E_{n}$} (En);

\foreach \i/\N/\L in {1/N1/L1, 2/N2/L2, n/Nn/Ln}{
  \draw[sig] (\N.south) -- ++(0,-0.6)
             |- node[above right=3pt] {$B_{\i}$} (\L.west);
}

\foreach \i/\L in {1/L1, 2/L2, n/Ln}{
  \node[below=1.5cm of \L] (S\i) {$S_{\i}\widehat{C}_{\i}$};
  \draw[sig] (\L.south) -- (S\i);

  \node[above=2.5cm of \L] (A\i) {$A_{\i}$};
  \draw[sig,<-] (\L.north) -- (A\i);
}

\draw[red,thick,rounded corners]
  ($ (N1.north west) + (-3pt,3pt)$)  
  rectangle
  ($ (join1 |- L1.south) + (3pt,-6pt)$);   
\node[text=red] at ($($(N1.north west) + (0pt, 10pt)$)$) {\(\mathcal{M}_1\)};
\draw[red,thick,rounded corners]
  ($ (N2.north west) + (-3pt,3pt)$)
  rectangle
  ($ (join2 |- L2.south) + (3pt,-6pt)$);
\node[text=red] at ($($(N2.north west) + (0pt, 10pt)$)$) {\(\mathcal{M}_2\)};
\draw[red,thick,rounded corners]
  ($ (Nn.north west) + (-3pt,3pt)$)
  rectangle
  ($ (joinn |- Ln.south) + (3pt,-6pt)$);
\node[text=red] at ($($(Nn.north west) + (0pt, 10pt)$)$) {\(\mathcal{M}_n\)};
\end{tikzpicture}

} 

\begin{figure}
    \centering
    \scalebox{0.9}{\MyTikzFigure}
    \caption{The evolution of the state through Eve's attack channels $\{ \attack{j} \}$, and Alice and Bob's operations $\{\QKDGmap{j}\}$ for the \nameref{prot:virtualprotocoll}. The evolution of states is also described in \cref{lemma:stateevolution}. Note that the announcements $\CP_j$ are made available to Eve through an explicit copy $\CPhat_j$, which gets merged with $E'_j$ to form $E_j$.}
    \label{fig:MEAT_Attack}
\end{figure}

\begin{lemma}[Describing evolution of states for QKD protocol from \nameref{prot:virtualprotocoll}] \label{lemma:stateevolution}
The evolution of the state in the \nameref{prot:abstractqkdprotocol} occurs as follows:
\begin{enumerate}
    \item Alice prepares the state $\sourcesymbol_{\Ameas_1^n \Ashield_1^n (A')_1^n}$, and sends ${A'}_1^n$ to Eve. We treat this as $E_0$.  
    \item For each round $j$ from $1$ to $n$, the following maps are applied:
    \begin{enumerate}
    \item In each round $j$, Eve implements her attack $\attack{j} \in \CPTP(E_{j-1}, B_j E^\prime_j)$.
    \item The $B_j$ system is forwarded to Bob. Alice and Bob then perform measurements on the $\Ameas_j$ and $B_j$ systems, and store them in registers $X_j$ and $Y_j$, and perform public announcements $\CP_j$. A copy $\CPhat_j$ of the announcements $\CP_j$ is given to Eve. They perform sifting to generate the $S_j$ registers. This entire process can be described by a map $\QKDGmapfull{j} \in \CPTP(A_j B_j , S_j X_j Y_j \CP_j \CPhat_j)$, where $A_j = \Ameas_j \Ashield_j$ denotes the registers that do not leave Alice's lab. We let $\QKDGmap{j} \in \CPTP(A_j B_j , S_j \CP_j \CPhat{j})$ denote the map $\QKDGmap{j} \defvar \Tr_{X_j Y_j} \circ \QKDGmapfull{j}$, i.e.~$\QKDGmapfull{j}$ with its output registers restricted to the secret register and public announcements. 
    \item The $E^\prime_j$ register and the $\CPhat_j$ register is combined into a unified $E_j$ register, which is passed into Eve's attack channel for the next round $\attack{j+1}$.\footnote{When $j=n$, there is no $\attack{n+1}$ that is applied.}
    \end{enumerate}
    Thus, the state after the application of these maps is given by
    \begin{equation}\label{eq:stateevolved}
    \begin{aligned}
        \rho_{S_1^n X_1^n Y_1^n \CP_1^n E_n } &= \left( \QKDGmapfull{n} \circ \attack{n} \dots \QKDGmapfull{1} \circ \attack{1} \right) \left[  \sourcesymbol_{\Ameas_1^n (\Ashield)_1^n (A')_1^n} \right] \\
         \rho_{S_1^n \CP_1^n E_n } &= \left( \QKDGmap{n} \circ \attack{n} \dots \QKDGmap{1} \circ \attack{1} \right) \left[  \sourcesymbol_{\Ameas_1^n (\Ashield)_1^n (A')_1^n} \right].
         \end{aligned}       
    \end{equation}
    Furthermore, we have $\QKDGmapfull{j} = \QKDGmapfullbeforeSR{j} \circ \measChannel{\{\ketbra{k}_{\Ameas_j}  \otimes \id_{\Ashield} \}}$, where $\measChannel{\{\ketbra{k}_{\Ameas_j}\otimes \id_{\Ashield}\}}  \in \CPTP(\Ameas_j \Ashield_j, X_j)$ is a channel that measures the $\Ameas_j$ system using POVM $\{\ketbra{k}_{\Ameas_j} \otimes \id_{\Ashield} \}_k$ and stores the outcome in the $X_j$ system (see \cref{def:measurementchannels}), and  
    $\QKDGmapfullbeforeSR{j}$ does all remaining operations except the initial measurement by Alice, and is the same as in \cref{lemma:stateevolutionmodifiedprotocol}.
    \item Alice performs the final sifting step, generating the register $\PAstring$ from $S_1^n$ and $\CP_1^n$.
    \item Alice and Bob perform steps such as the variable-length decision, error-correction, error-verification and privacy amplification. The final sifting operation, and the above procedures are together described by a map $\QKDpostprocessingmap \in \CPTP(X_1^n Y_1^n S_1^n \CP_1^n, K_A K_B \flagkey \flagEC \CEC \HEV \flagEV \HPA)$.
\end{enumerate}
\end{lemma}
\begin{proof}
The proof follows directly from the description of \nameref{prot:virtualprotocoll}. Recall that this protocol is obtained from \nameref{prot:pmprotocoll} (with the modified timings specified in \cref{lemma:modifiedprotocol}) by replacing the state preparation step through the source-replacement scheme (\cref{lemma:shieldandsourcereplacement}). The state evolution of \nameref{prot:pmprotocoll} with the modified timings is given in \cref{lemma:stateevolutionmodifiedprotocol}. The required state evolution for \nameref{prot:virtualprotocoll} is then obtained by applying the same modification (arising from the source-replacement scheme) to the state evolution from \cref{lemma:stateevolutionmodifiedprotocol}.
    \end{proof}

\subsection{Dimensions of Eve's systems} \label{subsec:evedimensions}
Our goal is to prove that the security statement holds for \emph{all} possible attacks by Eve. Concretely, this requires showing that the security definition (\cref{def:qkdsecurityChannelVersion}) is satisfied for \textit{every} possible attack channel $\attack{j}$. A subtlety arises here: since  $\attack{j} \in \CPTP(E_{j-1}, B_j E'_j)$, our analysis must also account for the \emph{dimensions} of the auxiliary spaces $E'_j$ that Eve may introduce. In principle, these spaces could even be infinite-dimensional. Moreover, Alice’s state preparation ($\Ashield_j, A'_j$) and Bob’s measurements ($B_j$) may themselves involve infinite-dimensional systems. However, the MEAT theorem is formulated only under the assumption that all registers are finite-dimensional. To reconcile this mismatch, we proceed as follows:

\begin{enumerate}

\item \textbf{Dimensions of Alice and Bob:} We first assume that Alice prepares finite-dimensional states and Bob performs finite-dimensional measurements. The proof in this section is restricted to this finite-dimensional Alice and Bob setting. Later, in \cref{sec:optics}, we show how these restrictions can be lifted by introducing appropriate source maps and squashing maps. With this, the only remaining issue is the dimensionality of Eve’s systems.

  \item \textbf{Security against arbitrary but finite-dimensional Eve:} Next, we fix an arbitrary finite dimension $d_j$ for each $E'_j$. For this setting, we consider a sequence of attack channels $\{\attack{j}\}_j$ with $\attack{j} \in \CPTP(E'_{j-1} \CPhat_{j-1}, E'_j B_j)$ and establish security in a way that does not depend on the particular choice of $\{\attack{j}\}_j$. Critically, the resulting security statement we obtain is independent of the values of $d_j$. Hence, security holds against all attacks where Eve’s systems are finite-dimensional.

\item \textbf{Infinite-dimensional Eve: } Finally, we extend the argument to infinite-dimensional systems. Intuitively, since our security statement holds for arbitrary finite dimensions $d_j$ of Eve's systems, one can take the limit $\lim_{d_j \rightarrow \infty}$ to recover the result when Eve uses infinite-dimensional systems. This limiting argument is made rigorous in \cref{thm:finiteToInfiniteFixedA,remark:infiniteEve} of \cref{app:infinitetofinite}. 
\end{enumerate}
 
Thus, in the following subsections, we fix a finite dimension $d_j$ for each $E'_j$. Note that this issue of specifying Eve’s dimension is primarily a technical nuisance rather than a conceptual obstacle. Intuitively, in a protocol where state preparation, measurements, and announcements are all finite-dimensional, Eve cannot do better than implement some unitary operation in each round as part of her attack. In such a setting, she only ever requires finite-dimensional registers. For a fully rigorous analysis, one must formalize a suitable version of this intuition. In this work, we adopt one such method.

\subsection{Security proof for \nameref{prot:abstractqkdprotocol} (finite-dimensional case)}
\label{subsec:proofgenericstuff} 
Recall that we are in the setting where Alice prepares finite-dimensional states, Bob performs finite-dimensional measurements and each $E'_j$ has dimension $d_j$. We start by fixing the set of attack channels $\{\attack{j}\}_j$, and consider the fixed state obtained corresponding to this attack. We will first break up the required security statement into correctness and secrecy, as is often done in QKD security analyses.

\begin{lemma}[Correctness and Secrecy imply Security] \label{lemma:correctandsecret}

Let $\rho_{K_A K_B \CfinalQKD \EfinalQKD}$ be the output state of the QKD protocol, and let the ideal output state be defined as $ \rho^{\mathrm{ideal}}_{K_A K_B  \CfinalQKD \EfinalQKD| \Omega_{\mathrm{len}=l} } \defvar \idealmap \left(  \rho^{\mathrm{real}}_{K_A K_B  \CfinalQKD \EfinalQKD| \Omega_{\mathrm{len}=l}}  \right)$. Suppose that for some $\epscorr,\epssecret \in [0,1]$, the output state satisfies correctness, that is:
\begin{equation}
    \Pr(K_A \neq K_B) \leq \epscorr,
\end{equation}
and secrecy 
\begin{equation} \label{eq:secrecy}
    \sum_l \Pr(\Omega_{\mathrm{len}=l} ) \tracedist{ \rho^\mathrm{real}_{K_A  \CfinalQKD \EfinalQKD | \Omega_{\mathrm{len}=l} } - \rho^{\mathrm{ideal}}_{K_A  \CfinalQKD \EfinalQKD| \Omega_{\mathrm{len}=l} } } \leq \epssecret.
\end{equation}
Then the output state is $(\epscorr + \epssecret)$-secure (see \cref{def:qkdsecuritysymmetric}), i.e.

\begin{equation}
    \sum_l \Pr(\Omega_{\mathrm{len}=l} ) \tracedist{ \rho^\mathrm{real}_{K_AK_B  \CfinalQKD \EfinalQKD | \Omega_{\mathrm{len}=l} } - \rho^{\mathrm{ideal}}_{K_A K_B  \CfinalQKD \EfinalQKD| \Omega_{\mathrm{len}=l} } } \leq \epssecret + \epscorr.
\end{equation}

\end{lemma}
\begin{proof}
  This proof of the secrecy–correctness split of the security criterion for variable-length protocols is essentially identical to the one in \cite[Lemma 12]{tupkary_security_2024} (see also Ref.~\cite{ben-or_universal_2004}) and closely resembles the analogous proofs for fixed-length protocols \cite{renner_security_2005,portmann_security_2022}. We include it here for completeness.\footnote{Strictly speaking, security, secrecy, and correctness are typically defined to be properties of a \emph{protocol} rather than of the output  state. Referring to the “$\eps$-security” of a state is therefore a slight abuse of terminology. Nonetheless, the statement of this lemma is self-contained, mathematically well-defined, and holds for any output state for the QKD protocol.}
   Let us write  
   \begin{equation}
   \begin{aligned}
\rho^\mathrm{real}_{K_AK_B  \CfinalQKD \EfinalQKD | \Omega_{\mathrm{len}=l} } &=\sum_{k_A , k_B \in\{0,1\}^l } \Pr(k_A ,k_B| \Omega_{\mathrm{len}=l} ) \ketbra{k_A,k_B}_{K_A K_B} \otimes \rho^{(k_A,k_B)}_{\CfinalQKD \EfinalQKD | \Omega_{\mathrm{len}=l}} \\
\rho^\mathrm{correct}_{K_AK_B  \CfinalQKD \EfinalQKD | \Omega_{\mathrm{len}=l} } &=\sum_{k_A , k_B \in\{0,1\}^l } \Pr(k_A ,k_B| \Omega_{\mathrm{len}=l} ) \ketbra{k_A,k_A}_{K_A K_B} \otimes \rho^{(k_A,k_B)}_{\CfinalQKD \EfinalQKD | \Omega_{\mathrm{len}=l}} \\
\rho^\mathrm{ideal}_{K_AK_B  \CfinalQKD \EfinalQKD | \Omega_{\mathrm{len}=l} } &=\sum_{k_A  \in\{0,1\}^l } \frac{1}{2^k} \ketbra{k_A,k_A}_{K_A K_B} \otimes \rho^\mathrm{real}_{\CfinalQKD \EfinalQKD|\Omega_{\mathrm{len}=l}},
   \end{aligned}
   \end{equation}
where the ideal output state written above can be easily verified to be the output of the $\idealmap$ acting on the real output state.  Here the states $\rho^{(k_A,k_B)}_{\CfinalQKD \EfinalQKD | \Omega_{\mathrm{len}=l}}$ are simply Eve's conditional states when the key registers hold keys $k_A, k_B$ respectively, each of length $l$. The states labeled ``correct" are obtained by simply replacing Bob's key with Alice's key in the real output state.
Then, using the triangle inequality, the required security condition can be upper bounded as
\begin{equation} \label{eq:secrecycorrectsplit}
\begin{aligned}
 & \sum_l \Pr(\Omega_{\mathrm{len}=l} ) \tracedist{ \rho^\mathrm{real}_{K_AK_B  \CfinalQKD \EfinalQKD | \Omega_{\mathrm{len}=l} } - \rho^{\mathrm{ideal}}_{K_A K_B  \CfinalQKD \EfinalQKD| \Omega_{\mathrm{len}=l} } } \\
 &\leq  \sum_l \Pr(\Omega_{\mathrm{len}=l} ) \tracedist{ \rho^\mathrm{real}_{K_AK_B  \CfinalQKD \EfinalQKD | \Omega_{\mathrm{len}=l} } - \rho^{\mathrm{correct}}_{K_A K_B  \CfinalQKD \EfinalQKD| \Omega_{\mathrm{len}=l} } } \\
 &+ \sum_l \Pr(\Omega_{\mathrm{len}=l} ) \tracedist{ \rho^\mathrm{correct}_{K_AK_B  \CfinalQKD \EfinalQKD | \Omega_{\mathrm{len}=l} } - \rho^{\mathrm{ideal}}_{K_A K_B  \CfinalQKD \EfinalQKD| \Omega_{\mathrm{len}=l} } }
    \end{aligned}
\end{equation}
The first term in \cref{eq:secrecycorrectsplit} can be upper bounded by the correctness condition as follows:
\begin{equation}
    \begin{aligned}
        &\sum_l \Pr(\Omega_{\mathrm{len}=l} ) \tracedist{ \rho^\mathrm{real}_{K_AK_B  \CfinalQKD \EfinalQKD | \Omega_{\mathrm{len}=l} } - \rho^{\mathrm{correct}}_{K_A K_B  \CfinalQKD \EfinalQKD| \Omega_{\mathrm{len}=l} } } \\
        &\leq \sum_l \Pr(\Omega_{\mathrm{len}=l} ) \sum_{k_A , k_B \in\{0,1\}^l }  \Pr(k_A, k_B |  \Omega_{\mathrm{len}=l} ) \\
        &\times \tracedist{\ketbra{k_A,k_B}_{K_A K_B} \otimes \rho^{(k_A,k_B)}_{\CfinalQKD \EfinalQKD | \Omega_{\mathrm{len}=l}}  - \ketbra{k_A,k_A}_{K_A K_B} \otimes \rho^{(k_A,k_B)}_{\CfinalQKD \EfinalQKD | \Omega_{\mathrm{len}=l}} } \\
        & \leq \sum_l \Pr(\Omega_{\mathrm{len}=l} ) \sum_{k_A , k_B \in\{0,1\}^l, k_A \neq k_B }  \Pr(k_A, k_B |  \Omega_{\mathrm{len}=l} )  \\
        &= \Pr(K_A \neq K_B),
    \end{aligned}
\end{equation}
where we apply the triangle inequality for the first inequality. In the second inequality, we replace the one-norm with $0$ when $k_A = k_B$ and with $2$ otherwise. The final equality then follows from basic properties of probability.

The second term in \cref{eq:secrecycorrectsplit} is identical to the left-hand-side \cref{eq:secrecy}, since $K_A = K_B$ and hence $K_B$ can be traced out without affecting the one-norm.
 Thus, upper bounding it by $\epssecret$ is identical to  the secrecy requirement. 
This concludes the proof. 
\end{proof}

\begin{lemma}[Correctness] \label{lemma:correct}
The output state of the QKD protocol (defined via \cref{lemma:stateevolution}) is $\epsEV$-correct.
\end{lemma}
\begin{proof}
    This proof is again a simple consequence of the error-verification step from \nameref{prot:abstractqkdprotocol}, and the proof of correctness can be found in many works. We have
    \begin{equation}
        \begin{aligned}
            \Pr(K_A \neq K_B) &= \Pr(K_A \neq K_B \wedge \OmegaEV) + \Pr(K_A \neq K_B \wedge \OmegaEV^\complement) \\
            &=\Pr(K_A \neq K_B \wedge \OmegaEV) \\
            &\leq \Pr(\PAstring \neq \PAstring_B \wedge \OmegaEV) \\
            &= \Pr(\OmegaEV | \PAstring \neq \PAstring_B)\Pr(\PAstring \neq \PAstring_B) \\
            &\leq \Pr(\OmegaEV | \PAstring \neq \PAstring_B ) \\
            &\leq 2^{-\left(\ceil{\log(1/\epsEV)}  \right)} \\
            &\leq \epsEV,
        \end{aligned}
    \end{equation}
    where the first line follows from the properties of probability, and the second line follows from the fact that $\OmegaEV^\complement \implies K_A = K_B$, since the protocol aborts. The third line follows from that fact that $K_A \neq K_B \implies \PAstring\neq \PAstring_B$, i.e, the final keys can differ only if the pre-amplification strings differ. The fourth and fifth lines follow from the properties of probability.  The sixth line follows from the fact that error-verification involves checking universal$_2$ hashes, and the final line follows from simple algebra.
\end{proof}

Having satisfied the correctness requirement, let us now focus on the secrecy requirement. Let us first consider the event $\Ocp$, defined as the event that the public announcements $\CP_1^n$ took the value $\cobs$. Observe that in the protocol, the privacy amplification step is performed conditioned on the value $\cobs$ (and conditioned on error verification accepting), i.e.~it is applied on the state $\rho_{\PAstring  \CEC \CEV \flagkey \flagEC \flagEV \HEV E_n | \Ocp \wedge \OmegaEV}$. Hence we would be interested in lower bounding the {\Renyi} entropy of register $\PAstring$ in this state, which will be used in the application of the Leftover Hashing Lemma (\cref{lemma:LHL}). However, the formulation of MEAT \cite[Theorem 4.1a]{arqand_marginal_2025} that we use later in \cref{theorem:MEATQKDfirst} is best suited in analyzing \textit{sequential} processes, and thus allows us to lower bound the {\Renyi} entropy of register $S_1^n$ of the state   $\rho_{S_1^n  \CEC \CEV \flagkey \flagEC \flagEV \HEV E_n | \Ocp \wedge \OmegaEV}$. Since $S_1^n$ and $\PAstring$ can be transformed into one another by using the public announcements $\CP_1^n$, we expect the two entropies to be equal. This can be argued using the following lemma.

\begin{lemma}[Equality of entropies of $S_1^n$ and $\PAstring$] 
\label{lemma:equalityentropy}
For any event $\Ocp$, consider the state just after the key map and sifting steps in the QKD protocol (see \cref{lemma:stateevolution}), given by $\rho_{\PAstring S_1^n X_1^n Y_1^n \CP_1^n E_n}$. Then, the following statement holds:
\begin{equation}
\Halpha(\PAstring| E_n)_{\rho_{ | \Ocp }} = \Halpha(S_1^n| E_n)_{\rho_{ | \Ocp }}.
\end{equation}
Moreover, this equality continues to hold at any later point in the protocol prior to privacy amplification. In particular, 
\begin{equation}
\Halpha(\PAstring|\CEC \CEV \flagkey \flagEC \flagEV  \HEV E_n)_{\rho_{ | \Ocp \wedge \OmegaEV}} = \Halpha(S_1^n|\CEC \CEV \flagkey \flagEC \flagEV \HEV E_n)_{\rho_{ | \Ocp \wedge \OmegaEV}},
\end{equation}
where we consider a version of the protocol in which the register $S_1^n$ is not deleted after use in the above equation. 
\end{lemma}
\begin{proof}
Recall that the protocol generates $\PAstring$ from $S_1^n$ by applying a deterministic discard rule based on $\CP_1^n$. In particular, $S_1^n$ can be transformed into $\PAstring$ by discarding certain positions (as specified by $\CP_1^n$), and $\PAstring$ can be transformed back into $S_1^n$ by inserting $\singleRoundBot$ symbols at the appropriate locations (again determined by $\CP_1^n$).
Since we are considering states conditioned on specific values of $\CP_1^n$, the two states are related by isometries $V_{\PAstring \rightarrow S_1^n}$ and $V_{S_1^n \rightarrow \PAstring}$ (both of which depend on $\cobs$), such that
   \begin{equation}
\begin{aligned}
   V_{\PAstring \rightarrow S_1^n} \rho_{\PAstring  E_n | \Ocp \wedge \OmegaEV}V^\dagger_{\PAstring \rightarrow S_1^n} &= \rho_{S_1^n  E_n | \Ocp \wedge \OmegaEV} \\
 V_{S_1^n \rightarrow \PAstring}\rho_{S_1^n  E_n | \Ocp \wedge \OmegaEV} V^\dagger_{S_1^n \rightarrow \PAstring} &=\rho_{\PAstring  E_n | \Ocp \wedge \OmegaEV}.
    \end{aligned}
\end{equation}
Then, the required statement follows from the fact that {\Renyi} entropies are invariant under isometries on the first system (\cref{lemma:dpinonconditioningregister}).\footnote{Note that we only require the isometry to hold in one direction for this proof to work. } Note that the statement holds throughout the protocol, since the same isometries exist at every stage. 
\end{proof}

We will now state the following theorem, which states that if the $f$-weighted {\Renyi} entropy of the private string $S_1^n$ is positive, then the secrecy requirement can be satisfied by a suitable choice of the $l(\cdot)$ and $\leak(\cdot)$ functions. This suggests that the $f$-weighted {\Renyi} entropy is a particularly useful object of study for the purposes of QKD security analysis. The proof of the following theorem is essentially the same as the one developed in Ref.~\cite{inprep_vanhimbeeck_tight_2024} (reproduced with permission in Ref.~\cite{kamin_renyi_2025}), and that work should be cited as the source whenever possible.

\begin{theorem}[$f$-weighted entropy to variable-length secrecy] \label{theorem:entropytovarlength}
Consider the state just after the key map step in \nameref{prot:virtualprotocoll} that is obtained from the \nameref{prot:abstractqkdprotocol}. 
Consider any $\alpha \in (1,2)$, and let $\fhatfull$ be any tradeoff function on the classical registers $\CP_1^n$ such that
\begin{equation} \label{eq:conditiononentropy}
    \Halpha[\fhatfull](S_1^n | \CP_1^n E_n)_{\rho} \geq 0,
\end{equation}
and let $\lkey(\cobs)$ be given by
\begin{equation} \label{eq:lexpression} 
     \lkey(\cobs) = \max\left\{ 0 , \floor{ \fhatfullQKD(\cobs)  - \leak(\cobs) - \ceil{ \log(\frac{1}{\epsEV})} -  \frac{\alpha}{\alpha-1} \log(\frac{1}{\epsPA})  + 2 }   \right\},
\end{equation}
where $\fhatfullQKD$ `lower-bounds' $\fhatfull$, i.e, satisfies
\begin{equation} \label{eq:fqkdcondition} 
    \fhatfullQKD(\cobs) \leq \fhatfull(\cobs) \qquad \forall \cobs.
\end{equation}
Then, the state obtained by performing sifting, error-correction and privacy amplification on this state as described in \nameref{prot:abstractqkdprotocol} is $\epsPA$-secret. 
\end{theorem}
\begin{proof}
Recall that for proving $\epsPA$-secrecy, one has to show that that the distance between the real and the ideal Alice-Eve states is smaller than $\epsPA$, i.e,
\begin{equation}
    \sum_{l} \Pr(\Omega_{\mathrm{len}=l}) \tracedist{\rho_{K_A \CEC \CEV \flagkey \flagEC \flagEV \HPA \HEV E_n | \Omega_{\mathrm{len}=l} }  - \tau^{(l)}_{K_A} \otimes \rho_{ \CEC \CEV \flagkey \flagEC \flagEV \HPA \HEV E_n | \Omega_{\mathrm{len}=l}} } \leq \epsPA
\end{equation}
where we recall that $\tau^{(l_A)}_{K_A}$ was defined in \cref{eq:taukAkB} to be the ideal output key state. 
The required statement follows from the following series of inequalities (explained in the paragraph later):

\begin{align}
  & \sum_{l=0}^\infty \Pr(\Omega_{\mathrm{len}=l})
  \tracedist{
    \rho_{K_A \CEC \CEV \flagkey \flagEC \flagEV \HPA \HEV E_n |\Omega_{\mathrm{len}=l}}
    - \tau^{(l)}_{K_A} \otimes
    \rho_{\CEC \CEV \flagkey \flagEC \flagEV \HPA \HEV E_n | \Omega_{\mathrm{len}=l}}
  }
  \label{eq:conditiononkeylength:a} \\[0.5em]
  &\leq \sum_{\{\cobs \mid \lkey(\cobs) > 0\}}
  \Pr(\Ocp \wedge \OmegaEV)
  \tracedist{
    \rho_{K_A \CEC \CEV \flagkey \flagEC \flagEV \HPA \HEV E_n | \Ocp \wedge \OmegaEV}
    - \tau^{(l)}_{K_A} \otimes
    \rho_{\CEC \CEV \flagkey \flagEC \flagEV \HPA \HEV E_n | \Ocp \wedge \OmegaEV}
  }
  \label{eq:conditiononkeylength:b} \\[0.5em]
  &\leq \sum_{\{\cobs \mid \lkey(\cobs) > 0\}}
  \Pr(\Ocp \wedge \OmegaEV)
  2^{\frac{1-\alpha}{\alpha}\left(
    \Halpha(\PAstring | \CEC \CEV \flagEC \flagEV \flagkey \HEV E_n)_{\rho| \Ocp \wedge \OmegaEV}
    - \lkey(\cobs) + 2
  \right)}
  \label{eq:conditiononkeylength:c} \\[0.5em]
  &= \sum_{\{\cobs \mid \lkey(\cobs) > 0\}}
  \Pr(\Ocp \wedge \OmegaEV)
  2^{\frac{1-\alpha}{\alpha}\left(
    \Halpha(\PAstring | \CEC \CEV \HEV E_n)_{\rho| \Ocp \wedge \OmegaEV}
    - \lkey(\cobs) + 2
  \right)}
  \label{eq:conditiononkeylength:d} \\[0.5em]
  &\leq \sum_{\{\cobs \mid \lkey(\cobs) > 0\}}
  \Pr(\Ocp)
  2^{\frac{1-\alpha}{\alpha}\left(
    \Halpha(\PAstring | \CEC \CEV \HEV E_n)_{\rho|\Ocp}
    - \lkey(\cobs) + 2
  \right)}
  \label{eq:conditiononkeylength:e} \\[0.5em]
  &\leq \sum_{\{\cobs \mid \lkey(\cobs) > 0\}}
  \Pr(\Ocp)
  2^{\frac{1-\alpha}{\alpha}\left(
    \Halpha(\PAstring | \HEV E_n)_{\rho|\Ocp}
    - \leak(\cobs)
    - \ceil{\log\!\left(\frac{1}{\epsEV}\right)}
    - \lkey(\cobs) + 2
  \right)}
  \label{eq:conditiononkeylength:f} \\[0.5em]
  &\leq \sum_{\{\cobs \mid \lkey(\cobs) > 0\}}
  \Pr(\Ocp)
  2^{\frac{1-\alpha}{\alpha}\left(
    \Halpha(\PAstring | E_n)_{\rho|\Ocp}
    - \leak(\cobs)
    - \ceil{\log\!\left(\frac{1}{\epsEV}\right)}
    - \lkey(\cobs) + 2
  \right)}
  \label{eq:conditiononkeylength:g} \\[0.5em]
  &\leq \sum_{\{\cobs \mid \lkey(\cobs) > 0\}}
  \Pr(\Ocp)
  2^{\frac{1-\alpha}{\alpha}\left(
    \Halpha(\PAstring | E_n)_{\rho|\Ocp}
    - \fhatfull(\cobs)
    + \frac{\alpha}{\alpha-1}\log\!\left(\frac{1}{\epsPA}\right)
  \right)}
  \label{eq:conditiononkeylength:h} \\[0.5em]
  &\leq \epsPA
  \sum_{\{\cobs \mid \lkey(\cobs) > 0\}}
  \Pr(\Ocp)
  2^{\frac{1-\alpha}{\alpha}\left(
    \Halpha(S_1^n | E_n)_{\rho|\Ocp}
    - \fhatfull(\cobs)
  \right)}
  \label{eq:conditiononkeylength:i} \\[0.5em]
  &\leq \epsPA
  \sum_{\cobs}
  \Pr(\Ocp)
  2^{\frac{1-\alpha}{\alpha}\left(
    \Halpha(S_1^n | E_n)_{\rho|\Ocp}
    - \fhatfull(\cobs)
  \right)}
  \label{eq:conditiononkeylength:j} \\[0.5em]
  &= \epsPA\,
  2^{\frac{1-\alpha}{\alpha}\left(
    \Halpha[\fhatfull](S_1^n | \CP_1^n E_n)_\rho
  \right)}
  \label{eq:conditiononkeylength:k} \\[0.5em]
  &\leq \epsPA.
  \label{eq:conditiononkeylength:l}
\end{align}

Here, \cref{eq:conditiononkeylength:a} just restates the quantity we wish to bound. In \cref{eq:conditiononkeylength:b}, we rewrite the sum over all possible values of  $\cobs$, and use the triangle inequality and the fact that the distance between the real and ideal states is zero when the protocol does not produce a key. Hence, we only need to keep events corresponding to a key of non-zero length being generated in the sum over $\cobs$. 
In \cref{eq:conditiononkeylength:c}, we used the Leftover Hashing Lemma (\cref{lemma:LHL}). In \cref{eq:conditiononkeylength:d}, we used the fact that values of $\flagEC,\flagEV,\flagkey$ are uniquely determined by the event $\Ocp$, and therefore can be removed from the conditioning registers without penalty.\footnote{This can be seen formally by observing that when we condition on $\Ocp \wedge \OmegaEV$, one can transform between the state with the $\flagEC,\flagEV,\flagkey$ registers and the state without them using a CPTP map in either direction. In both directions, the CPTP map acts only on the conditioning registers. Therefore, \cref{lemma:DPI} applies both ways, yielding the desired equality.} \cref{eq:conditiononkeylength:e} follows by using $\rho_{| \Ocp \wedge \OmegaEV} = \rho_{( | \Ocp) | \OmegaEV}$, and a direct application of \cref{lemma:conditioning} to get rid of the conditioning on $\OmegaEV$, where we absorb the correction term into the probability before the exponential. \cref{eq:conditiononkeylength:f} follows by using \cref{lemma:EC_cost} to split off the error-correction and error-verification registers. \cref{eq:conditiononkeylength:g} then follows since we can now remove the $\HEV$ register which stored the hash-choice for error verification, since it is independent of the rest of the registers, without penalty, by using data processing in both directions. \cref{eq:conditiononkeylength:h} follows from the fact that the output key length $\lkey(\cobs)$ satisfies \cref{eq:lexpression,eq:fqkdcondition}. \cref{eq:conditiononkeylength:i} is simple algebra. For \cref{eq:conditiononkeylength:j}, observe that we add more (positive) terms to the sum over $\cobs$, and thus the resulting quantity must be larger. \cref{eq:conditiononkeylength:k} follows from the definition of $f$-weighted {\Renyi} entropy from \cref{def:QES}. \cref{eq:conditiononkeylength:l} follows from $\alpha>1$ and the fact that the $\fhatfull$-weighted {\Renyi} entropy is positive, i.e, \cref{eq:conditiononentropy}. This concludes our proof.    
\end{proof}

Thus, the task reduces to finding a tradeoff function $\fhatfull$ such that
\[
\Halpha[\fhatfull](S_1^n \mid \CP_1^n E_n)_{\rho} \geq 0 .
\]
This is precisely the setting in which the MEAT can be applied, as we show in the next subsection. We will first carry out this analysis for a fixed attack by the adversary, and subsequently remove this restriction by performing a worst-case analysis over all possible attacks.

\subsection{Security via application of MEAT} \label{subsec:securityfromMEAT}

\begin{figure}[ht]
\centering
 \MyTikzFigure
  \caption{(Same as \cref{fig:MEAT_Attack}). The evolution of the state through Eve's attack channels $\{ \attack{j} \}$, and Alice and Bob's operations $\{\QKDGmap{j}\}$ for the \nameref{prot:virtualprotocoll}. The evolution of states is also described in \cref{lemma:stateevolution}. Note that the announcements $\CP_j$ are made available to Eve through an explicit copy $\CPhat_j$, which gets merged with $E'_j$ to form $E_j$. The MEAT (\cref{theorem:MEAT}) is applied for the sequence of channels $\{\mathcal{M}_j\}$.}
  \label{fig:MEAT_Attack_repeated}
\end{figure}

\begin{theorem}[Obtaining $\fhatfull$ satisfying \cref{eq:conditiononentropy}] \label{theorem:MEATQKDfirst}
Let $\rho_{S_1^n \CP_1^n E_n}$ be the state obtained in the QKD protocol, i.e, via \cref{eq:stateevolved} from \cref{lemma:stateevolution} or \cref{fig:MEAT_Attack_repeated}. For each $j$, and every value of $\cP_1^{j-1}$, 
let $f_{|\cP_1^{j-1}}$ be a tradeoff function on the register $\CP_j$, and define $\kappafuncgeneric{f_{|\cP_1^{j-1}} }{ \sigma^{(j)}_{A_j} } { \QKDGmap{j} }{ \attack{j} }$ as:
\begin{equation} \label{eq:kappadefined}
   \kappafuncgeneric{f_{|\cP_1^{j-1}} }{ \sigma^{(j)}_{A_j} } { \QKDGmap{j} }{ \attack{j} } \coloneq \inf_{\nu \in \Sigma_j(\attack{j})} \Halpha[f_{|\cP_1^{j-1}}] (S_j | \CP_j E_j \widetilde{E} )_\nu
\end{equation}
where
\begin{equation} \label{eq:Sigmajdefined}
    \Sigma_j (\attack{j})\coloneq \left\{ \QKDGmap{j} \circ \attack{j} \left( \omega_{A_j E_{j-1} \widetilde{E}} \right) \;\middle|\; \omega_{A_j} = \sigmaconstraint{j}  \right\},
\end{equation}
where $\widetilde{E}$ is a purifying register any of the $A_j E_{j-1}$ registers. 
Then the following ``normalized" tradeoff function on $\hat{C}_1^n$:
\begin{equation}
    \fhatfull(\cobs) \coloneq \sum_{j=1}^n \left( f_{| \cP_1^{j-1}} (\cP_j) +    \kappafuncgeneric{f_{|\cP_1^{j-1}} }{ \sigma^{(j)}_{A_j} } { \QKDGmap{j} }{\attack{j} } \right)
\end{equation}
satisfies
\begin{equation}
    \Halpha[\fhatfull](S_1^n | \CP_1^n E_n) \geq 0.
\end{equation}
\end{theorem}
\begin{proof}
   The statement follows from a direct application of \cite[Theorem 4.1a]{arqand_marginal_2025} (reproduced here in \cref{theorem:MEAT}), and we simply describe how that theorem can be applied here. The final state $\rho_{S_1^n \CP_1^n E_c}$  is of the form 
   \begin{equation}
       \rho_{S_1^n \CP_1^n E_c} = \mathcal{M}_n \circ \dots \circ \mathcal{M}_1 \left( \rho_{A_1^n E_0} \right)
   \end{equation}
   where $\rho_{A_1^n E_0}$ is the global source-replaced state defined in \cref{eq:globalsourcereplacedstate}, where we identify $A_j = \Ameas_j \Ashield_j$ as the register that does not leave Alice's lab, and $E_0=(A')_1^n$ as the set of states that leave Alice's lab. Furthermore, $\rho_{A_1^n} = \bigotimes_{j=1}^n \sigma^{(j)}_{A_j}$, and $\mathcal{M}_j   
   = \QKDGmap{j} \circ \attack{j} \in \CPTP(E_{j-1} A_j , S_j \CP_j E_j)
   $. The statement thus follows directly from \cref{theorem:MEAT}, where the symbols $\mathcal{M}_j,E_j, \CP_j,\widetilde{E},\fhatfull,f_{|\cP_1^{j-1}}$ correspond to the same symbols used in that theorem. The only difference is that $A_j$ here should be identified with $A_{j-1}$ in Ref.~\cite{arqand_marginal_2025} (as we have done already in the reproduced \cref{theorem:MEAT}), and we explicitly write $\mathcal{M}_j$ as a concatenation of two maps, $\QKDGmap{j}\circ \attack{j} $.
\end{proof}

Thus, we are now at a stage where, for a given sequence of attack channels $\{ \attack{j} \}$ and operations by Alice and Bob in each round $\{\QKDGmap{j}\}$, an appropriate ``global" tradeoff function $\fhatfull$ satisfying $\Halpha[\fhatfull](S_1^n | \CP_1^n E_n)_{\rho} \geq 0$ can be obtained via \cref{theorem:MEATQKDfirst}. 

\begin{remark} \label{remark:fintuition}
Recall that the statement $\Halpha[\fhatfull](S_1^n | \CP_1^n E_n)_{\rho} \geq 0$ can be understood as the function \(\fhatfull\) lower bounding the entropy \(\Halpha(S_1^n \mid \CP_1^n E_n)_\rho\) in the log-mean-exponential sense (see \cref{eq:lme}). In this light, it is helpful to think of \(f_{|\cP_1^{j-1}}\) as assigning a preliminary “score” to the public announcement in each round. Intuitively, a good choice assigns larger values to test-round announcements that indicate little or no adversarial interference (as opposed to announcements indicating errors), and to generation-round announcements indicating that the round is kept (as opposed to discarded). One is free to start with any such preliminary score that one desires (which may or may not be compatible with the above intuition), and such a score need not have the desired property of lower bounding the relevant {\Renyi} entropy.  This preliminary score must then be reduced by the amount \(\kappafuncgeneric{f_{|\cP_1^{j-1}}}{\sigma^{(j)}_{A_j}}{\QKDGmap{j}}{\attack{j}}\). The resulting per-round contributions can subsequently be combined to yield the global tradeoff function \(\fhatfull\), which indeed satisfies the required property of providing a lower bound on \(\Halpha(S_1^n \mid \CP_1^n E_n)_\rho\) in the log-mean-exponential sense. We discuss the procedure to obtain a good choice of $f_{|\cP_1^{j-1}}$ in \cref{subsec:pickingf} (see also \cref{remark:fcanbelarge}).
\end{remark}

However, the function $\fhatfull$ obtained from \cref{theorem:MEATQKDfirst} depends on the attack performed, and therefore 
this result is still not yet sufficient to yield 
security of the QKD protocol. To circumvent this issue, we require the following definition, that constructs a lower bound on 
$\kappafuncgeneric{f_{|\cP_1^{j-1}} }{ \sigma^{(j)}_{A_j} } { \QKDGmap{j} }{ \attack{j}  }$ by minimizing over all possible attacks. 

\begin{remark} \label{remark:attackset}
Since our goal is to prove security against arbitrary attacks by Eve (subject to $E'j$ having dimension $d_j$), we must in principle optimize
$\kappafuncgeneric{f{|\cP_1^{j-1}}}{\sigma^{(j)}{A_j}}{\QKDGmap{j}}{\attack{j}}$
over all channels $\attack{j} \in \CPTP(E{j-1}, B_j E’_j)$. This is indeed the fundamental task we carry out.

That said, in many scenarios it is possible to restrict attention to a subset of CPTP maps without loss of generality. For example, when squashing maps are used to reduce a protocol with an infinite-dimensional measurement to an equivalent protocol with a finite-dimensional one (see \cref{sec:optics}), Eve’s effective attack channel is given by composing her original channel with the squashing map. Depending on the details of the squashing map, these composed channels may lie in a subset
\[
\attackset{j}(E_{j-1}, B_j E'_j) \subseteq \CPTP(E_{j-1}, B_j E'_j).
\]
This motivates the general definition adopted here.

Note to readers: For a first reading, it is perfectly fine to regard $\attackset{j}(E_{j-1},B_j E'_j)$ as the full set of CPTP maps. The subtleties introduced by squashing-based reductions, which motivate restricting to smaller attack sets, are explained later in \cref{sec:optics}. Readers can return to this point after that discussion to fully appreciate these nuances.
\end{remark}
\begin{definition} \label{def:kappalowergeneric}
    Let $j \in \{1,\dots,n\}$, let  $f_{|\cP_1^{j-1}}$ be a tradeoff function on $\CP_j$, let $\sigma^{(j)}_{A_j} \in \dop{=}(A_j)$, and let $\QKDGmap{j} \in \CPTP(A_j B_j, S_j \CP_j \CPhat{j})$. Let $\attackset{j}(E_{j-1},B_j E_j) \subseteq \CPTP(E_{j-1}, B_j E'_j) $ be a subset of all possible attack channels. Then, we define a value for $  \kappafuncgeneric{f_{|\cP_1^{j-1}} }{ \sigma^{(j)}_{A_j} } { \QKDGmap{j}  }{\attackset{j}}$\footnote{In this work, each $\attackset{j}$ is always associated with specific registers 
$E_{j-1}, B_j, E'_j$, and should formally be written as 
$\attackset{j}(E_{j-1}, B_j E'_j)$.  
While we typically include this explicitly, we occasionally omit the register 
labels to avoid notational clutter.} over the set of attack channels via:
    \begin{equation}\label{eq:kappalowergeneric}
    \begin{aligned}
        \kappafuncgeneric{f_{|\cP_1^{j-1}} }{ \sigma^{(j)}_{A_j} } { \QKDGmap{j}  }{\attackset{j}(E_{j-1},B_j E_j)} &\defvar \inf_{\attack{j} \in \attackset{j}} \kappafuncgeneric{f_{|\cP_1^{j-1}} }{ \sigma^{(j)}_{A_j} } { \QKDGmap{j} }{\attack{j}  } \\
        &\defvar \inf_{\nu \in \Sigma_j (\attackset{j})} \Halpha[f_{|\cP_1^{j-1}}] (S_j | \CP_j E_j \widetilde{E} )_\nu
        \end{aligned}
    \end{equation}
    where
\begin{equation} \label{eq:Sigmajdefined}
    \Sigma_j (\attackset{j}) \coloneq \left\{ \QKDGmap{j} \circ \attack{j} \left( \omega_{A_j E_{j-1} \widetilde{E}} \right) \;\middle|\; \omega_{A_j} = \sigmaconstraint{j}, \; \attack{j} \in \attackset{j}  \right\},
\end{equation}
where $\widetilde{E}$ is a purifying register for any of the $A_j E_{j-1}$ registers. 
\end{definition}

Since
\[
\kappafuncgeneric{f_{|\cP_1^{j-1}}}{\sigma^{(j)}_{A_j}}{\QKDGmap{j}}{\attackset{j}(E_{j-1}, B_j E'_j)}
\;\le\;
\kappafuncgeneric{f_{|\cP_1^{j-1}}}{\sigma^{(j)}_{A_j}}{\QKDGmap{j}}{\attack{j}}, \qquad  \forall \; \attack{j}\in \attackset{j}(E_{j-1}, B_j E'_j)
\]
one might be tempted to use the left-hand side to define $\fhatfull$ and thereby the key length (via \cref{eq:lexpression}), since that value would be the worst case value over all possible attacks. However, this formulation appears to depend on the dimensions assigned to Eve’s side-information registers, and thus we would  require an additional optimization over all possible choices of those dimensions as well. The goal of the next subsection is to present a reformulation of $\kappafuncgeneric{f_{|\cP_1^{j-1}}}{\sigma^{(j)}_{A_j}}{\QKDGmap{j}}{\attackset{j}(E_{j-1}, B_j E'_j)}$ that avoids these issues by getting rid of the dependence on the dimensions of Eve's registers.

\subsection{Reformulating the minimization over attack channels} \label{subsec:reformulatingattackmin}

In order to reformulate $\kappafuncgeneric{f_{|\cP_1^{j-1}}}{\sigma^{(j)}_{A_j}}{\QKDGmap{j}}{\attackset{j}(E_{j-1}, B_j E'_j)}$ in a manner where the dependence on Eve's dimensions disappears, it is convenient to consider two related notions of attack channels, as follows:
\begin{itemize}
	\item	$\attack{j} \in \attackset{j} (E_{j-1}, B_j E'_j) \subseteq \CPTP(E_{j-1}, B_j E'_j)$, which describe Eve’s action from her side-information register $E_{j-1}$ to Bob’s system $B_j$ together with a new side register $E'_j$. In particular, the set $\attackset{j}$ depends on the dimension we allow for Eve’s side-information registers $E'_j$. 
	\item	$\Qattack{j} \in \Qset{j} \subseteq \CPTP(A'_j, B_j)$, which describe maps directly from Alice’s emitted signal $A'_j$ to Bob’s register $B_j$. Intuitively, Eve’s attack in this perspective can be seen as first applying such a channel to the source-replacement state $\sourcesymbol^{(j)}_{A_j A'_j}$, after which we purify the post-attack output state, and give the purifying register to Eve. Notice that $\Qset{j}$ does not depend on the dimensions we allow for Eve's side-information registers $E'_j$.
    \end{itemize}

Instead of considering all attack channels in $\attackset{j} (E_{j-1}, B_j E'_j)$ for the security analysis, we would instead like to consider all maps in $\Qset{j}$. In order to do so, we require $\Qset{j}$ to satisfy a specific property stated in  \cref{def:Qjcondition}. (Informally, we require that applying maps in $\Qset{j}$ and then giving a purification to Eve is at least as powerful as applying maps in $\attackset{j}$).  That is, for every attack in $\attackset{j} (E_{j-1}, B_j E'_j)$, there exists a channel in $\Qset{j}$ such that the outputs on the $A_jB_j$ system is the same. This is formalized in the following definition.

\begin{definition}[Marginal of $\attackset{j}(E'_{j-1} \CPhat_{j},B_j E'_j)$] \label{def:Qjcondition}
Let $\attackset{j}(E'_{j-1} \CPhat_{j},B_j E'_j)  \subseteq \CPTP(E'_{j-1} \CPhat_{j}, B_j E'_j) $ and $\Qset{j} \subseteq \CPTP(A'_j, B_j)$ be attack sets. 
We say that $\Qset{j}$  is a \term{marginal} of $\attackset{j}(E'_{j-1} \CPhat_{j},B_j E'_j)$ if the following \emph{marginalization property} holds: for every $\omega_{A_j E_{j-1} \widetilde{E}}$ such that $\omega_{A_j} = \sourcesymbol^{(j)}_{A_j}$ and $\widetilde{E}$ is a purifying register for the $A_jE_{j-1}$ registers, and every $\attack{j} \in \attackset{j}$, there exists a $\Qattack{j} \in \Qset{j}$ such that 
\begin{equation} \label{eq:Qjcondition}
    \Qattack{j}\left[ \sourcesymbol^{(j)}_{A_j A'_j} \right] 
   = \tr_{E_j \widetilde{E}} \circ \attack{j} \left[ \omega_{A_j E_{j-1} \widetilde{E}} \right].
\end{equation}
where $\sourcesymbol^{(j)}_{A_j A'_j}$ is the source-replacement state (see \cref{eq:globalsourcereplacedstate}). We say that the collection $\left\{\attackset{j}(E'_{j-1} \CPhat_{j},B_j E'_j)\right\}_{E'_{j-1},E'_j}$ satisfies the \term{dimension-independent marginal property} with respect to $\Qset{j}$ if, for all possible choices of the auxiliary dimension of Eve’s registers $E'_j,E'_{j-1}$, the collection $\left\{\attackset{j}(E'_{j-1} \CPhat_{j},B_j E'_j)\right\}_{E'_{j-1},E'_j}$ all possess the same marginal $\Qset{j}$.
\end{definition}

In this way, $\Qset{j}$ is dimension-independent, whereas $\attackset{j}(E'_{j-1} \CPhat_{j}, B_j E'_j) $ depends explicitly on the dimension assigned to $E'_j$. Notice that any state $ \attack{j} \left[ \omega_{A_j E_{j-1} \widetilde{E}} \right]$ on the RHS of \cref{eq:Qjcondition}, can be obtained from some purification of $  \Qattack{j}\left[ \sourcesymbol^{(j)}_{A_j A'_j} \right] $, which is the LHS of \cref{eq:Qjcondition}, by a map acting only on the purifying system. Thus, due to data-processing (\cref{lemma:DPIfweighted}), it suffices to consider purifications of $  \Qattack{j}\left[ \sourcesymbol^{(j)}_{A_j A'_j} \right] $ when performing the infimum over all attacks that appears in $\kappa$ from \cref{eq:kappalowergeneric}. We formalize this in the following lemma.

\begin{lemma}
    \label{lemma:MEATuniform_lowerbound}
    Let $\kappafuncgeneric{f_{|\cP_1^{j-1}} }{ \sigma^{(j)}_{A_j} } { \QKDGmap{j}  }{\attackset{j}(E_{j-1},B_j E'_j)}$ be as defined in \cref{def:kappalowergeneric}, and let $\Qset{j}$ be a marginal of $\attackset{j}(E_{j-1},B_j E'_j)$ in the sense of Definition~\ref{def:Qjcondition}. Then, for any $\alpha\geq 1$ we have:
   \begin{equation}
    \begin{aligned}
        \label{eq:kappalower_oneside}
    \kappafuncgeneric{f_{|\cP_1^{j-1}} }{ \sigma^{(j)}_{A_j} } { \QKDGmap{j}  }{\attackset{j}(E_{j-1},B_j E_j)}&\geq\inf_{\Qattack{j}\in\Qset{j}}  \Halpha[f_{|\cP_1^{j-1}}](S_j| \CP_j \CPhat_j \widehat{E})_{\QKDGmap{j}\left[    \pf\left(\Qattack{j}\left[
\sourcesymbol^{(j)}_{A_j A'_j}\right]\right)\right]}  \\  &=\inf_{\overline{\omega}\in\Sigmab_j\left(\Qset{j}\right)}  \Halpha[f_{|\cP_1^{j-1}}](S_j| \CP_j \CPhat_j \widehat{E})_{\QKDGmap{j}\left[\pf\left(\overline{\omega}_{A_jB_j}\right)\right]}
    \end{aligned}
    \end{equation}
   where $\pf$ is a purifying function of $A_jB_j$ onto $\widehat{E}$, and
    \begin{align}
        &\Sigmab_j\left(\Qset{j}\right) \defvar\left\{\Qattack{j}\left[{\sigma^{(j)}_{A_jA'_j}}\right]
            \;\middle|\;
            \Qattack{j}\in\Qset{j} \right\},
    \end{align}
    where $\sourcesymbol^{(j)}_{A_jA'_j}$ is the source-replaced state from \cref{eq:globalsourcereplacedstate}.
    If $\Qset{j}$ satisfies the dimension-independent marginal property, i.e is the marginal of all $\left\{\attackset{j}(E'_{j-1} \CPhat_{j},B_j E'_j)\right\}_{E'_{j-1},E'_j}$, then \cref{eq:kappalower_oneside} is obtained for any dimensions of Eve's side-information registers.
    Furthermore, the objective function in the RHS of the first line of \cref{eq:kappalower_oneside} is convex in $\Qattack{j}$, and in the second line of \cref{eq:kappalower_oneside} is convex in $\overline{\omega}_{A_jB_j}$.\footnote{If $\Qset{j}$ is convex, then the RHS of \cref{eq:kappalower_oneside} is a convex optimization problem.}
    \end{lemma}
    \begin{proof}
    The first line of \cref{eq:kappalower_oneside} is merely an equivalent reformulation of the second line, obtained by an affine transformation of the input variable $\bar{\omega}$ (which is the post-attack state) to $\Qattack{j}$ (which is the attack channel). Since affine transformations preserve convexity, the convexity of the  objective function in the first line follows. Thus, we focus on proving the inequality in \cref{eq:kappalower_oneside}. In other words, we show that
\begin{equation} \label{eq:kappasameproof}
    \inf_{\attack{j}\in\attackset{j}} \inf_{\omega\in\Sigma_j(\attack{j})}\Halpha[f_{|\cP_1^{j-1}}](S_j| \CP_j E_j \widetilde{E})_{ \QKDGmap{j} \circ \attack{j} \left[ \omega_{A_{j} E_{j-1}\widetilde{E}} \right]} \geq\inf_{\Qattack{j}\in\Qset{j}}  \Halpha[f_{|\cP_1^{j-1}}](S_j| \CP_j \CPhat_j \widehat{E})_{\QKDGmap{j}\left[    \pf\left(\Qattack{j}\left[
    \sourcesymbol^{(j)}_{A_j A'_j}\right]\right)\right]},
\end{equation}
where $\widetilde{E}$ is a register of large enough dimension to accommodate for purification of all $A_jE_{j-1}$ registers.
Let $\left(\omega^*_{A_jE_{j-1}\widetilde{E}},\attack{j}^*\right)$ be a feasible point on the LHS of \cref{eq:kappasameproof}, and let $\omega^*_{A_j B_j E'_j \widetilde{E}} = \attack{j}^* \left[ \omega^*_{A_{j} E_{j-1}\widetilde{E}}\right] $; noting that it satisfies $\omega^*_{A_j} = \sigma^{(j)}_{A_j}$. Since $\Qset{j}$ is a marginal of $\attackset{j}$ in the sense of Definition~\ref{def:Qjcondition}, then \cref{eq:Qjcondition} states that there exists a channel $\Qattack{j}^*\in\Qset{j}$ such that
\begin{align}
    \Qattack{j}^*\left[ \sourcesymbol^{(j)}_{A_j A'_j} \right] 
   = \tr_{E'_j \widetilde{E}} \circ \attack{j} \left[ \omega^*_{A_j E_{j-1} \widetilde{E}} \right]=\omega^*_{A_jB_j}.
\end{align}
 Let $\pf$ be a purifying function of $A_jB_j$ onto $\widehat{E}$. Then, there exists a channel $\mathcal{N}_j \in \CPTP(\hat{E},E'_{j} \widetilde{E})$ such that $\omega^*_{A_j B_j E'_j \widetilde{E}} = \mathcal{N}_j \left[ \pf\left(\Qattack{j}^*\left[\sigma^{(j)}_{A_jA'_j}\right]\right) \right]$ (see \cref{lemma:pur_to_ext}).
Thus, we have:
\begin{align}
           \Halpha[f_{|\cP_1^{j-1}}](S_j| \CP_j E_j \widetilde{E})_{\QKDGmap{j}\circ\attack{j}\left[\omega^*_{A_{j}E_{j-1}\widetilde{E}}\right]}&= \Halpha[f_{|\cP_1^{j-1}}](S_j| \CP_j \CPhat_j E'_j \widetilde{E})_{\QKDGmap{j}\circ\attack{j}\left[\omega^*_{A_{j}E_{j-1}\widetilde{E}}\right]}\nonumber\\
            &=  \Halpha[f_{|\cP_1^{j-1}}](S_j| \CP_j \CPhat_j E'_j \widetilde{E})_{\QKDGmap{j}\circ\mathcal{N}_j\left[ \pf\left(\Qattack{j}^*\left[\sigma^{(j)}_{A_jA'_j}\right]\right) \right]}\nonumber\\
            &=\Halpha[f_{|\cP_1^{j-1}}](S_j| \CP_j \CPhat_j E'_j \widetilde{E})_{\mathcal{N}_j\circ\QKDGmap{j}\left[\pf\left(\Qattack{j}^*\left[\sigma^{(j)}_{A_jA'_j}\right]\right)\right]}\nonumber\\
            &\geq \Halpha[f_{|\cP_1^{j-1}}](S_j| \CP_j \CPhat_j \widehat{E})_{\QKDGmap{j}\left[\pf\left(\Qattack{j}^*\left[\sigma^{(j)}_{A_jA'_j}\right]\right)\right]}.
        \end{align}
        The first line simply rewrites $E_j$ as $\CPhat{j} E'_j$, and the second line replaces $\attack{j} [\omega^*_{A_j E_{j-1} \widetilde{E} }]$ with $ \mathcal{N}_j \left[ \pf\left(\Qattack{j}^*\left[\sigma^{(j)}_{A_jA'_j}\right]\right) \right]$ $\attackset{j} $ since they are equal. The third line holds due to the fact that the maps $\mathcal{N}_j$ and $\QKDGmap{j}$ commute (since they 
        act on different registers), and the last line follows from data processing inequality (\cref{lemma:DPIfweighted}). Thus, for any feasible point on the LHS, we can find a feasible point on the RHS such that the LHS is greater than or equal to the RHS. This proves the first inequality in~\cref{eq:kappalower_oneside}. 
        
     The convexity of the objective function in the second line of \cref{eq:kappalower_oneside} was established in~\cite[Lemma~4.10]{arqand_marginal_2025}. 
    \end{proof}

In the following corollary, we show that regardless of the choice of $\attackset{j}(E_{j-1}, B_j E'_j)$, the set $\Qset{j}$ can always be taken to be $\CPTP(A'_j, B_j)$. 

\begin{corollary}
$\Qset{j}(A'_j, B_j) = \CPTP(A'_j, B_j)$ can be regarded as a marginal of any $\attackset{j}(E_{j-1}, B_j E’j) \subseteq \CPTP(E_{j-1}, B_j E’_j)$ in the sense of Definition~\ref{def:Qjcondition}.
\end{corollary}
    \begin{proof}
        Let $\omega_{A_jE_{j-1}\widetilde{E}}\in\dop{=}(A_jE_{j-1}\widetilde{E})$ be any pure state such that $\omega_{A_j}=\sigma_{A_j}^{(j)}$, and let $\sourcesymbol_{A_jA'_j}^{(j)}$ be the source-replacement state as defined in \cref{eq:globalsourcereplacedstate}. Consider a $\attack{j}\in\attackset{j}$, and let $\omega_{A_jB_jE'_j\widetilde{E}}=\attack{j}\left[\omega_{A_jE_{j-1}\widetilde{E}}\right]$. Then, note that since $\sourcesymbol_{A_jA'_j}^{(j)}$ is pure, the exists a channel $\Qattack{j}\in\CPTP(A'_j,B_j)$ such that $\omega_{A_jB_j}=\Qattack{j}\left[\sourcesymbol_{A_jA'_j}^{(j)}\right]$ (\cref{lemma:pur_to_ext}). Therefore, we have
        \begin{align}
\Qattack{j}\left[\sourcesymbol_{A_jA'_j}^{(j)}\right]=\Tr_{E'_j\widetilde{E}}\circ\attack{j}\left[\omega_{A_jE_{j-1}\widetilde{E}}\right],
        \end{align}
        which is exactly the required  relation (\cref{eq:Qjcondition}). This concludes our proof.
    \end{proof}

\begin{remark}
We note that although \cref{lemma:MEATuniform_lowerbound} only establishes an inequality, this inequality is saturated for all scenarios considered in this paper. Formally,
the inequality in \cref{eq:kappalower_oneside} is saturated whenever the following condition holds:
\begin{align}
&\forall  \Qattack{j} \in \Qset{j}, \;\exists \attack{j} \in \attackset{j}(E_{j-1}, B_j E'_j)\text{ and }\omega_{A_j E_{j-1} \widetilde{E}}\;\text{such that}\;\\
& \Qattack{j}\left[ \sourcesymbol^{(j)}_{A_j A'_j} \right] 
   = \tr_{E_j \widetilde{E}} \circ \attack{j} \left[ \omega_{A_j E_{j-1} \widetilde{E}} \right],
\end{align}
where the registers and sets $\Qset{j}$ and $\attackset{j}$ are defined in that lemma.
It is straightforward to verify that if both $\Qset{j}$ and $\attackset{j}(E_{j-1}, B_j E’_{j-1})$ are taken to be the full set of $\CPTP$ maps, then the above condition is indeed saturated as long as Eve's dimensions are large enough. Indeed, the condition is still satisfied when we  considering some squashed attack sets in
\cref{sec:optics}, although we omit a formal proof of this fact, since it is not required for our analysis.
\end{remark}

Thus, we can instead focus on computing the objective function formulated in terms of $\Qset{j}$, where the dimension of Eve’s side-information registers plays no role. Furthermore, $\Qset{j}$ can \textit{always} be taken as $\CPTP(A’_j, B_j)$ without any loss of generality (although  we will later restrict it in a non-trivial way when introducing the squashing map). We are now ready to bring all the pieces together and present the final security statement for the protocol under study.

\subsection{Final security statement} \label{subsec:finalsecuritystatem}
We now state the final security statement concerning the security of \nameref{prot:abstractqkdprotocol}, in the {\realistic} authentication setting described in \cref{sec:reductionstatement}.
\begin{theorem}[Security statement for \nameref{prot:abstractqkdprotocol} under finite-dimensional Alice–Bob setting]
\label{theorem:abstractsecuritystatement}
Consider the \nameref{prot:abstractqkdprotocol} where Alice’s transmitted systems ($A'_j$) are finite-dimensional and Bob performs finite-dimensional measurements ($B_j$ is finite-dimensional). For this protocol, after applying the source-replacement scheme (\cref{lemma:shieldandsourcereplacement}), let $\sourcesymbol^{(j)}_{A_jA’_j}$ denote the resulting source states in round~$j$, and let $\QKDGmap{j}$ denote the operations Alice and Bob perform in round~$j$.
 (These are the same as those that appear in the corresponding \nameref{prot:virtualprotocoll}, with state evolution described in \cref{lemma:stateevolution}). For each $j$ and every value of $\cP_1^{j-1}$, let $f_{|\cP_1^{j-1}}$ denote a tradeoff function\footnote{See \cref{remark:fintuition} for an intuitive interpretation of this function.} on the register $\CP_j$. Let $\kappaQKDfunc{f_{|\cP_1^{j-1}}}{\sigma^{(j)}_{A_j}}{\QKDGmap{j}}$ be any value satisfying
\begin{equation}
\begin{aligned}
\kappaQKDfunc{f_{|\cP_1^{j-1}}}{\sigma^{(j)}_{A_j}}{\QKDGmap{j}}
&\leq
\inf_{\Qattack{j}\in\Qset{j}}
\Halpha[f_{|\cP_1^{j-1}}](S_j| \CP_j \CPhat_j \widehat{E})_{\QKDGmap{j}\left[\pf\left(\Qattack{j}\left[\sourcesymbol^{(j)}_{A_j A'_j}\right]\right)\right]}, \\
\Qset{j} &= \CPTP(A'_j, B_j),
\end{aligned}
\end{equation}
where $\pf$ is a purifying function of $A_j A'_j$ onto $\widehat{E}$, and define 
\begin{equation} \label{eq:fhatfullQKDdefinition}
\fhatfullQKD(\cobs) \coloneq \sum_{j=1}^n \left( f_{|\cP_1^{j-1}}(\cP_j) + \kappaQKDfunc{f_{|\cP_1^{j-1}}}{\sigma^{(j)}_{A_j}}{\QKDGmap{j}} \right).
\end{equation}
Then, the protocol is $(\epsPA + \epsEV)$-secure (according to \cref{def:qkdsecurityasymmetric}) if the key length $\lkey(\cobs)$ is chosen as
\begin{equation}
 \lkey(\cobs) = \max\left\{ 0 , \floor{ \fhatfullQKD(\cobs)  - \leak(\cobs) - \ceil{ \log(\frac{1}{\epsEV})} -  \frac{\alpha}{\alpha-1} \log(\frac{1}{\epsPA}) + 2 } \right\}.
\end{equation}
\end{theorem}

\begin{proof}
    The required statement follows from an appropriate combination of all the statements we have shown previously, which we recall as follows.

      \paragraph*{Reducing to honest authentication setting.}
From the analysis in \cref{sec:reductionstatement}, summarized in \cref{theorem:reductionstatement}, we find that to prove security of the full protocol (according to \cref{def:qkdsecurityasymmetric}) in the {\realistic} authentication setting, it suffices to prove security of the protocol excluding the final authentication post-processing step, but now considered in the honest authentication setting. In this honest setting, the protocol necessarily has symmetric outcomes, since Alice and Bob always agree on the output key length. Thus we only need to prove security as defined in \cref{def:qkdsecuritysymmetric}.

    \paragraph*{Reducing to  \nameref{prot:virtualprotocoll}.}

    We begin by applying \cref{lemma:modifiedprotocol}, which allows us to construct a modified version of \nameref{prot:abstractqkdprotocol} in which Alice's state preparation is moved earlier in time, such that it occurs before Bob's first measurement. As argued in \cref{lemma:modifiedprotocol}, this can be done without loss of generality, and ensures that the security of this modified protocol implies the security of the original protocol. Next, we invoke \cref{lemma:shieldandsourcereplacement}, which shows that Alice's operations can be viewed as preparing a gobal source-replaced state $ \sourcesymbol_{\Ameas_1^n (\Ashield)_1^n (A')_1^n}$ (that is tensor-product across rounds) and then measuring the $\Ameas_1^n$ systems. Since Alice’s measurements commute with all operations except the public announcements (which, by design, occur only after Bob’s measurement in each round), we can postpone Alice’s measurements to occur simultaneously with Bob’s. 

    After these modifications, the protocol for which we must prove security is given by  \nameref{prot:virtualprotocoll}. In particular, it admits the structure of sequential channels described in \cref{lemma:stateevolution,fig:MEAT_Attack}, where $\{\attack{j}\}$ denotes the sequence of Eve's attack channels, and the maps $\{\QKDGmap{j}\}$ implement Alice and Bob's measurements, postprocessing and public announcements. We note that the arguments so far do not require Alice, Bob or Eve to be finite-dimensional. 
    \paragraph*{Security analysis for state obtained in \nameref{prot:virtualprotocoll}.} At this stage, we fix a particular sequence of attack channels, and analyze the output state corresponding to it. In particular, we also fix the dimensions of all of Eve's side-information registers ($E'_j$) to be some finite value $d_j$. We will later obtain a result that does not depend on any particular sequence of attack channels or dimensions, and holds for any sequence of attack channels that Eve may perform. 

    We first show that the output state is $\epsEV$-correct in \cref{lemma:correct} (regardless of Eve's attack channels). Then, we use \cref{lemma:correctandsecret} to argue that $\epsEV$-correctness and $\epsPA$-secrecy implies $(\epsEV+\epsPA)$-security. Thus, we are now concerned with proving $\epsPA$-secrecy of the state. In \cref{theorem:entropytovarlength}, we show that if 
    \begin{equation} \label{eq:fullprooftempgreater0}
    \Halpha[\fhatfull](S_1^n | \CP_1^n E_n) \geq 0,
    \end{equation} 
    then the function $\lkey(\cobs)$ (see \cref{eq:lexpression}) that determines the length of the output key given by
    \begin{equation}
             l(c_1^n) = \max\left\{ 0 , \floor{ \fhatfullQKD(\cobs)  - \leak(\cobs) - \ceil{ \log(\frac{1}{\epsEV})} -  \frac{\alpha}{\alpha-1} \log(\frac{1}{\epsPA}) + 2  } \right\}
    \end{equation}
    guarantees $\epsPA$-secrecy, provided that  
    \begin{equation} \label{eq:fullprooftempfcondition}
    \fhatfullQKD(\cobs) \leq \fhatfull(\cobs) \quad \forall \cobs.
    \end{equation}  
    Thus, our task reduces to constructing a global tradeoff function $\fhatfullQKD$ satisfying above property.
    
    \paragraph*{Applying MEAT.} To do so, we turn to \cref{theorem:MEATQKDfirst}, where we apply MEAT to the sequence of channels described in \cref{lemma:stateevolution,fig:MEAT_Attack}. This yields a function $\fhatfull$ satisfying $\Halpha[\fhatfull](S_1^n | \CP_1^n E_n) \geq 0$, given by
    \begin{equation}
    \fhatfull(\cobs) \coloneq \sum_{j=1}^n \left( f_{| \cP_1^{j-1}} (\cP_j) +    \kappafuncgeneric{f_{|\cP_1^{j-1}} }{ \sigma^{(j)}_{A_j} } { \QKDGmap{j}}{ \attack{j} } \right), 
\end{equation}
where $f_{| \cP_1^{j-1}}$ are arbitrary tradeoff functions on $\CP_j$ (that we choose), and the constants $\kappafuncgeneric{f_{|\cP_1^{j-1}} }{ \sigma^{(j)}_{A_j} } { \QKDGmap{j}}{ \attack{j} }$ are defined in \cref{eq:kappadefined}. However, both $\kappafuncgeneric{f_{|\cP_1^{j-1}} }{ \sigma^{(j)}_{A_j} } { \QKDGmap{j} }{ \attack{j} }$ and the resulting $\fhatfull$ constructed in this fashion depend on Eve's attack channels, and the dimensions of her registers. We would like to remove this dependence by performing a worst-case analysis over all possible attack channels $\attackset{j}(E_{j-1}, B_j E'_j) = \CPTP(E_{j-1}, B_j E'_j)$ and all possible dimensions of Eve's registers.
\paragraph*{Reformulations to remove dependence on Eve's dimensions and attack channels.} 
To address this, we construct a different kind of attack channels, and denote this set via \(\Qset{j}\). Here, we represent any attack as the application of one such channel to the source-replaced state, and then assume that Eve holds a purification of the resulting output state. We
set $\Qset{j} = \CPTP(A'_j, B_j)$, and apply \cref{lemma:MEATuniform_lowerbound}. This, along with the definition of $\kappaQKDfunc{f_{|\cP_1^{j-1}}}{\sigma^{(j)}_{A_j}}{\QKDGmap{j}}$ in the theorem statement, allows us to obtain
\begin{equation}
\kappafuncgeneric{f_{|\cP_1^{j-1}}}{\sigma^{(j)}_{A_j}}{\QKDGmap{j}}{\attack{j}}
\geq
\inf_{\Qattack{j}\in\Qset{j}}
\Halpha[f_{|\cP_1^{j-1}}](S_j| \CP_j \CPhat_j \widehat{E})_{\QKDGmap{j}\left[\pf\left(\Qattack{j}\left[\sourcesymbol^{(j)}_{A_j A'_j}\right]\right)\right]} \geq \kappaQKDfunc{f_{|\cP_1^{j-1}}}{\sigma^{(j)}_{A_j}}{\QKDGmap{j}},
\end{equation}
 for any attack $\attack{j} \in \CPTP(E_{j-1} , B_j E'_{j-1})$. Thus, if we use $\kappaQKDfunc{f_{|\cP_1^{j-1}}}{\sigma^{(j)}_{A_j}}{\QKDGmap{j}}$  to define 
\begin{equation} 
\fhatfullQKD(\cobs) \coloneq \sum_{j=1}^n \left( f_{|\cP_1^{j-1}}(\cP_j) + \kappaQKDfunc{f_{|\cP_1^{j-1}}}{\sigma^{(j)}_{A_j}}{\QKDGmap{j}} \right),
\end{equation}
then the resultant global tradeoff function $\fhatfullQKD(\cobs)$ satisfies the required property \cref{eq:fullprooftempfcondition}. Moreover,  neither $\kappaQKDfunc{f_{|\cP_1^{j-1}}}{\sigma^{(j)}_{A_j}}{\QKDGmap{j}}$ nor $\fhatfullQKD(\cobs)$ has any dependence on the dimensions of Eve's systems, or her choice of attack. Additionally,  as pointed out in \cref{lemma:MEATuniform_lowerbound}, the resulting optimization problem is finite-dimensional and convex. Thus, $\lkey(\cobs)$, as defined in the theorem statement, is a valid choice for obtaining a $(\epsPA + \epsEV)$-secure protocol when considering any attack in which Eve employs finite-dimensional side registers.

\paragraph*{Relaxing the finite-dimensional Eve assumption.}
We now address the technicality that Eve could, in principle, use infinite-dimensional side registers in her attack. So far, we have proven security for all attacks where Eve’s registers are finite-dimensional. However, since our security statement yields the same security parameter $(\epsPA + \epsEV)$ regardless of the dimensions assigned to Eve’s side registers, we can invoke \cref{thm:finiteToInfiniteFixedA,remark:infiniteEve} to argue that the same security guarantee extends to the infinite-dimensional case. This concludes our proof. 
\end{proof}

Having completed the security analysis for \nameref{prot:abstractqkdprotocol}, we now turn to the next subsection to elaborate on certain aspects, highlight subtleties, and provide additional clarification.

\subsection{Discussion} \label{subsec:discussionabstractproof}

We have now proved \cref{theorem:abstractsecuritystatement}, which guarantees the security of the \nameref{prot:abstractqkdprotocol} against arbitrary attacks by Eve, as long as one can compute a lower bound on a finite-dimensional convex optimization. Our theorem holds for the case where Alice sends finite-dimensional states, and Bob measures finite-dimensional systems. Naturally, this security statement depends on the parameters specified in the protocol, and the secure output key length depends on these parameters. Before proceeding further, let us take stock of what has been accomplished so far and identify the remaining steps needed to complete the analysis and obtain a security proof for a practical decoy-state BB84 protocol. We have two main aspects that need to be addressed.

\begin{itemize}
    \item \textbf{Infinite dimensions of Alice and Bob:} First, recall our earlier caveat: the MEAT is formally applicable only to finite-dimensional systems, even though its statement does not dependent on the dimensions of the underlying systems. However, QKD protocols are implemented using optical systems, which are naturally described by infinite-dimensional Hilbert spaces. In particular, Alice sends infinite-dimensional states, and Bob POVM lives in an infinite-dimensional Hilbert space. 
Ref.~\cite{nahar_postselection_2024} presented an argument to reduce to finite dimensions at the level of the security definition, that is, however it does not apply here due to on-the-fly announcements. Thus we generalize the approach in Ref.~\cite{nahar_postselection_2024} to include on-the-fly announcements. This method is general and potentially compatible with a broad class of squashing maps and source maps. This is undertaken in \cref{sec:optics}. Once this reduction is complete, the resulting finite-dimensional protocol can be analyzed using MEAT, following the procedure outlined \cref{subsec:expressingprotocolassequence,subsec:proofgenericstuff,subsec:securityfromMEAT,subsec:reformulatingattackmin,subsec:finalsecuritystatem}.

\item \textbf{Numerics:} Second, even after reducing to finite dimensions, computing the key rate still requires us to solve a highly non-trivial convex optimization problem. In essence, the difficulty arises from the fact that we aim to obtain a value that is provably \emph{lower} than the infimum --- i.e, a guaranteed lower bound. We outline the numerical requirements and considerations for such computations in \cref{sec:numerics}, and refer the reader to \cite{kamin_renyi_2025,navarro_finite_2025} for some recent work on this topic. 
\end{itemize}

We note that Ref.~\cite{kamin_renyi_2025} is a recent work that applies MEAT to the analysis of the decoy-state BB84 protocol. That analysis computes both fixed-length and variable-length key rates, develops numerical methods for key-rate computation (which we briefly discuss in \cref{sec:numerics}), and additionally incorporates intensity and phase imperfections. However, the analysis in \cite{kamin_renyi_2025} does not include the case of on-the-fly announcements, focusing for simplicity only on the case where announcements are made after all measurements are completed. There is also a technical issue that it performs squashing by applying MEAT directly to the original protocol, reducing to finite dimensions only at the level of the resulting optimization problem. Thus, it applies MEAT to infinite-dimensional systems, which was not sufficiently justified by existing literature.\footnote{We note that this is a common technical gap in many analyses, where statements  proved for arbitrary finite-dimensional systems without explicit dimensional dependence are applied to infinite-dimensional systems. \cref{app:infinitetofinite} provides one approach for making such steps rigorous by showing that results holding for arbitrary finite-dimensional systems may also extend to the infinite-dimensional setting.}

This issue is addressed in the present work. In particular, we provide a formulation in which MEAT is applied only after an explicit reduction to finite-dimensional systems, thereby avoiding the need to invoke the theorem in an infinite-dimensional setting. We note that the key rates obtained in this work coincide with those reported in Ref.~\cite{kamin_renyi_2025}, as we explain later in this manuscript.

Having discussed the above issues—which are addressed in more detail in subsequent sections—we now outline a general recipe for applying the results of the previous subsection to compute QKD key rates. 

\subsubsection{Recipe} \label{subsubsec:recipe}
Since the \nameref{prot:abstractqkdprotocol} is highly general, the security of a wide variety of QKD protocols can be established using the results developed so far. To aid in this task, we  include a practical “recipe’’ for applying our results to compute key rates for a protocol of interest.
We present this recipe in a way that also accommodates squashing maps, source maps, and the optical modeling introduced in \cref{sec:optics}. For a first reading, the reader may safely skip the optical details and return to them after going through \cref{sec:optics}.
\begin{enumerate} 
\item[] \textbf{Recipe}
    \item Specify all details of the QKD protocol, and ensure that it fits the structure described in \nameref{prot:abstractqkdprotocol}.
    \item If required, apply the  source map \cref{lemma:sourcemapsecurityMEAT} to reduce to a protocol with a different state preparation by Alice.
    \item If required, apply the  squashing map \cref{lemma:SquashMapSecurityMEAT}  to reduce to a protocol where Bob performs a different measurement. 
    \item Consider the \nameref{prot:virtualprotocoll}, which is obtained after source-replacement and timing modifications as specified  \cref{subsec:expressingprotocolassequence}.  In particular, this fixes $\sourcesymbol^{(j)}_{A_j}$, $\QKDGmap{j}$ for all $j \in \{1,\dots,n\}$. Note that these quantities are obtained by considering the \nameref{prot:virtualprotocoll} version of the protocol obtained \textit{after} the application of the source-map and squashing transformations.
    \item For all $j\in \{1,\dots,n\}$ and $\cP_1^{j-1}$, specify the tradeoff functions $f_{|\cP_1^{j-1}}$ (or a deterministic procedure for specifying them, see \cref{remark:specifyfatstart}).
    \item  Determine $\attackset{j}(E_{j-1}, B_j E'_j)$ (which depends the details of the squashing map used). From it, determine the corresponding $\Qset{j}$ that satisfies the dimension-independent extension property (\cref{def:Qjcondition}). If we do not use squashing maps, this reduces to $
\Qset{j} = \CPTP(A'_j, B_j)$.
    \item  Compute a lower bound on $\kappafuncgeneric{f_{|\cP_1^{j-1}} }{ \sigma^{(j)}_{A_j} } { \QKDGmap{j}  } {\attackset{j}}$, given by 
    \begin{equation}
        \inf_{\Qattack{j}\in\Qset{j}}
\Halpha[f_{|\cP_1^{j-1}}](S_j| \CP_j \CPhat_j \widehat{E})_{\QKDGmap{j}\left[\pf\left(\Qattack{j}\left[\sourcesymbol^{(j)}_{A_j A'_j}\right]\right)\right]}
    \end{equation}
    where $\pf$ is a purifying function from $A_j B_j$ onto $\widehat{E}$. Denote this lower bound by $\kappaQKDfunc{f_{|\cP_1^{j-1}}}{\sigma^{(j)}_{A_j}}{\QKDGmap{j}}$. 
    \item Define $\fhatfullQKD =  \sum_{j=1}^n \left( f_{| \cP_1^{j-1}} (\cP_j) +    \kappaQKDfunc{f_{|\cP_1^{j-1}} }{ \sigma^{(j)}_{A_j} } { \QKDGmap{j}  } \right)$. 
    \item Set the variable-length decision to be $   \lkey(\cobs) = \max\left\{ 0 , \floor{ \fhatfullQKD(\cobs)  - \leak(\cobs) - \ceil{ \log(\frac{1}{\epsEV})} -  \frac{\alpha}{\alpha-1} \log(\frac{1}{\epsPA}) + 2  } \right\}$. The resulting protocol is $(\epsPA+\epsEV)$-secure.
    \end{enumerate}
We now highlight some subtleties and provide additional clarification in the following remarks. 

\begin{remark} \label{remark:generality} 
Notice that \cref{theorem:abstractsecuritystatement} is fairly general, since the \nameref{prot:abstractqkdprotocol} is itself fairly general. For example, we can straightforwardly accommodate scenarios where Alice's state preparation is different in different rounds, since we allow for the $\sigma^{(j)}_{A_j}$ to vary from round to round, although we do require ``ìndependent"' state preparation across rounds, i.e, $\sigma_{A_1^n} = \sigma^{(1)}_{A_1} \otimes \dots \otimes \sigma^{(n)}_{A_n}$. Interestingly, we can also accommodate fairly contrived protocols where Alice and Bob perform \emph{different} protocol operations in each round - for example, a protocol that alternates between BB84 signal preparation and measurement in odd-numbered rounds and six-state signal preparation and measurement in even-numbered rounds.  This 
manifests in the analysis as $\QKDGmap{j}$ now being different across rounds. This flexibility makes our analysis, which inherits these features from the abstract MEAT statement itself, extremely general. The only additional cost incurred is the need to perform more computations.
\end{remark}

\begin{remark} \label{remark:anyfworks}
Note that when applying the above recipe, one may start with \emph{any} choice of tradeoff functions $f_{| \cP_1^{j-1}}$ that one desires --- there are no constraints whatsoever on the original selection of these functions. This choice determines the values of $\kappaQKDfunc{f_{|\cP_1^{j-1}} }{ \sigma^{(j)}_{A_j} } { \QKDGmap{j}  }$, which in turn influences the normalized tradeoff function $\fhatfullQKD$. While the theorem permits arbitrary choices of the tradeoff $f_{| \cP_1^{j-1}}$, certain choices lead to superior key rates compared to other choices (in fact, a careful choice is typically important to achieve positive key rates). In particular, the \emph{optimal} $f_{| \cP_1^{j-1}}$, in the sense that it generates the best performance for a  fixed honest behaviour of the channel, can be chosen following a procedure described in \cite[Sec.~VI]{kamin_renyi_2025} or~\cite[Sec.~5.3]{arqand_generalized_2024}. Since the initial choice of  $f_{| \cP_1^{j-1}}$ is not relevant for the security claim to hold, and only necessary for ensuring the best performance, we only briefly discuss the procedure to choose   $f_{| \cP_1^{j-1}}$ in \cref{sec:numerics}.
\end{remark}

\begin{remark} \label{remark:adaptivef}
Note that the above recipe, and \cref{theorem:MEATQKDfirst,theorem:abstractsecuritystatement}, allows us to choose \emph{different} tradeoff functions $f_{| \cP_1^{j-1}}$ in different rounds, which can also depend on the observed public announcements of preceding rounds (although one has to determine how this choice is made beforehand). This allows for ``fully adaptive" QKD protocols \cite{zhang_knill_qpe,arqand_generalized_2024}, where the tradeoff functions $f_{| \cP_1^{j-1}}$ can be updated over the course of the protocol, a feature that is important when performing QKD over noisy, or unpredictable channels. This too, is a feature inherited from the abstract MEAT statement itself.  However note that updating  $f_{| \cP_1^{j-1}}$ frequently leads to additional computation cost. The simplest scenario is the one where all $\sigma^{(j)}_{A_j}$s are identical, all $f_{| \cP_1^{j-1}}$s are identical and have no dependence on $\cP_1^{j-1}$, and all $\QKDGmap{j}$s are identical.\footnote{Strictly speaking, these objects act on different registers and are therefore not equal as mathematical objects. Rather, we mean that each $\sigma^{(j)}_{A_j}$ and $\QKDGmap{j}$ is defined on a distinct but isomorphic copy of the same underlying Hilbert space, and coincides with a fixed reference state and map under the corresponding canonical identification. Similarly, we mean that the functions $f_{|\cP_1^{j-1}}$ coincide under the appropriate relabeling of the classical registers. } 
\end{remark}

\begin{remark} \label{remark:specifyfatstart}
Since the function $\fhatfullQKD$ depends on all the $f_{| \cP_1^{j-1}}$ and $\kappaQKDfunc{f_{|\cP_1^{j-1}} }{ \sigma^{(j)}_{A_j} } { \QKDGmap{j}  }$, it might seem as though the computation of $\fhatfullQKD(\cobs)$ requires the computation of all $\kappaQKDfunc{f_{|\cP_1^{j-1}} }{ \sigma^{(j)}_{A_j} } { \QKDGmap{j}  }$ for all the possible inputs, which is prohibitively expensive. However, as argued in Ref.~\cite{zhang_knill_qpe}, when applying the above result in practice, one only needs to compute $\fhatfull(\cobs)$ on the specific sequence $\cobs$ observed in the protocol. This allows for an efficient, iterative procedure: for each round $j$, one examines the past announcements $\cP_1^{j-1}$, makes some choice of $f_{| \cP_1^{j-1}}$ (which depends on those values) and computes $\kappaQKDfunc{f_{|\cP_1^{j-1}} }{ \sigma^{(j)}_{A_j} } { \QKDGmap{j}  }$, then considers a similar computation for the $(j+1)^\text{th}$ round and so on. Note that there is no requirement to finish each of these computations before the next round occurs; in other words, they can be computed at any convenient time between generating the relevant data and the final privacy amplification step --- this is because the computed values are only used to determine the key length for privacy amplification, and not for any other aspect of the protocol.
\end{remark}

\begin{remark} \label{remark:fcanbelarge}
We remark that, when running the numerical routines from  Ref.~\cite{kamin_renyi_2025} for the decoy-state BB84 protocol, one observes that the quantity
$$
f_{|\cP_1^{j-1}}(\cdot)\;+\;
\kappaQKDfunc{f_{|\cP_1^{j-1}}}{\sigma^{(j)}_{A_j}}{\QKDGmap{j}}
$$
can take fairly large positive or negative values. As a consequence, for some realizations of the public announcements \(\cobs\), the globally normalized tradeoff function \(\fhatfullQKD(\cobs)\), and the key length $\lkey(\cobs)$ may exceed the total number of rounds $n$.
This behaviour may appear counterintuitive: the Rényi entropy of the pre-amplification string is always upper bounded by $n$, whereas \(\fhatfullQKD(\cobs)\) has a qualitative interpretation as a lower bound on this quantity ``on average''. Nevertheless, this does not invalidate the security proof. The framework ensures that such extreme values arise only with sufficiently small probability, and our obtained security statement is still valid --- recall for instance that the security definition (\cref{def:qkdsecuritysymmetric}) is averaged over the possible key lengths, and thus key lengths that occur with extremely small probability do not affect it much. (Similar properties hold for the ``log-mean-exponential'' interpretation (\cref{eq:interpretaslogmeanexp}) of the $f$-weighted entropy bounds, as these are also ``averaged'' quantities.) That said, for practical implementations, if one encounters an observation \(\cobs\) for which
\[
\fhatfullQKD(\cobs) > n,
\]
it is natural to propose that its value should be reduced to $n$ (i.e.~the maximal possible {\Renyi} entropy); the protocol still straightforwardly remains secure in this case.\footnote{This can be argued using the monotonicity of $f$-weighted {\Renyi} entropies wrt $f$, see \cref{lemma:Monotonicity_f}.} Importantly, however, even without such corrective adjustments, the formal security guarantees from \cref{theorem:abstractsecuritystatement,theorem:decoystatebb84securitystatement} are fully valid; the adjustment is purely for practical considerations such as implementation difficulties in producing an extremely long output key. 
\end{remark}

We now turn our attention to the remaining steps outlined at the start of this subsection. In the next section, we  consider the quantum optical nature of realistic implementations, which causes Alice’s signal states and Bob’s POVMs to reside in infinite-dimensional Hilbert spaces.

\section{Extending Security to optical protocols} \label{sec:optics}
We will now outline the additional steps required to adapt our analysis to optical QKD implementations. We will do so by reducing the security analysis of the original protocol, where Alice's signal states and Bob's POVMs belong to infinite-dimensional Hilbert spaces, to a protocol where they belong to finite-dimensional Hilbert spaces. We start by focusing on Alice's signal states.

\subsection{Source maps and tagging}

In a decoy-state protocol, Alice prepares fully phase-randomized coherent states (see \cref{eq:AliceSignalStatesDescription}) which are infinite-dimensional. This would lead to an infinite-dimensional shield system (\cref{lemma:shieldandsourcereplacement}), which complicates numerical computations. Furthermore, the MEAT statement requires all systems to be finite-dimensional. To address this, we introduce \emph{source maps}.

Given a particular state preparation, a source map demonstrates that the actual prepared states can be obtained by applying a suitable CPTP map to simpler, ``virtual'' source states. Intuitively, we can then pretend as though the virtual states are being produced, and `give' the source map to Eve. Since one then proves security against arbitrary attacks by Eve, security also holds against an attack where Eve first applies the source map, in which the case the virtual scenario is identical to the original one. Thus, source maps are general tools that allow us to conceptually modify state preparations for theoretical convenience. In particular, they allow us to replace original (often complicated or infinite-dimensional) state preparations with a simple, virtual preparation that facilitates easier theoretical analysis, without affecting security. Note that source maps are an old idea \cite{gottesman_security_2004}, though the nomenclature used often differs.
\begin{restatable}[Source maps]{lemma}{lemmaSourceMaps} \label{lemma:sourcemapsecurityMEAT}
Let $\left\{\{\QKDGmapfullbeforeSR{j}\}_{j=1}^n,\QKDpostprocessingmap, \sigma_{X_1^n (A')_1^n} \right\}$ determine a QKD protocol (see \cref{def:PMQKD}) where Alice prepares the global state
\begin{equation}
    \begin{aligned}\nonumber
        \sigma_{X_1^n (A')_1^n} &=   \sum_{x_1^n} p(x_1^n) \ketbra{x_1^n}_{X_1^n} \otimes \sigma^{(x_1^n)}_{(A')_1^n}.
    \end{aligned}
\end{equation}
Suppose there exists a quantum channel (source map) $\Psi \in  \CPTP((A'')_1^n, (A')_1^n)$ and a set of \emph{virtual states} $\{\xi^{(x_1^n)}_{(A'')_1^n}\}_{x_1^n} \subset \dop{=}( (A'')_1^n )$ such that
\begin{align}
    \sigma_{X_1^n (A')_1^n} &= (\identityMap \otimes \Psi) \left[ \xi_{X_1^n (A'')_1^n}\right] \\
    \text{where } \quad \xi_{X_1^n (A'')_1^n} &= \sum_{x_1^n} p(x_1^n) \ketbra{x_1^n}_{X_1^n} \otimes \xi^{(x_1^n)}_{(A'')_1^n}. 
\end{align}
In other words, $\sigma^{(x_1^n)}_{(A')_1^n} = \Psi\left[\xi^{(x_1^n)}_{(A'')_1^n}\right]$ for all  $x_1^n$. Then if the protocol $\left\{\{\QKDGmapfullbeforeSR{j}\}_{j=1}^n,\QKDpostprocessingmap, \xi_{X_1^n (A'')_1^n} \right\}$
using the virtual states is $\epssecure$‑secure, the protocol $\left\{\{\QKDGmapfullbeforeSR{j}\}_{j=1}^n,\QKDpostprocessingmap, \sigma_{X_1^n(A')_1^n} \right\}$ using the real states is also $\epssecure$‑secure.
\end{restatable}
\begin{proof}
If the virtual protocol $\left\{\{\QKDGmapfullbeforeSR{j}\}_{j=1}^n,\QKDpostprocessingmap, \xi_{X_1^n (A'')_1^n} \right\}$ is $\epssecure$-secure, then for all attack channels $\{\attack{j}\}_{j=2}^n$ and $\attack{1}' \in \CPTP((A'')_1^n, B_1 E_1)$, we have (as in \cref{def:qkdsecurityChannelVersion})
\begin{align} \label{eq:EpsilonSecurityTau}
    \left\|
        \left(
            \QKDpostprocessingmap \circ \QKDGmapfullbeforeSR{n} \circ \attack{n} \circ \cdots \circ \QKDGmapfullbeforeSR{1}
            - \idealmap \circ \QKDpostprocessingmap \circ \QKDGmapfullbeforeSR{n} \circ \attack{n} \circ \cdots \circ \QKDGmapfullbeforeSR{1}
        \right)
        \circ \left(\identityMap_{X_1^n} \otimes \attack{1}'\right)\left[\xi_{X_1^n (A'')_1^n}\right]
    \right\|_1 \leq \epssecure,
\end{align}
where we used \cref{lemma:stateevolutionmodifiedprotocol} to represent the output state obtained in the protocol. We treat the first attack channel $\attack{1}'$ separately from the rest of the attack channels in our proof, since this attack channel is the one that can implement the source map. Consider the restricted subset of channels $  \mathcal{C}_\Psi $, where we first apply the source map $\Psi$ before applying an attack channel $\attack{1} \in \CPTP((A')_1^n, B_1 E_1)$
\begin{align} \label{eq:SourceMapChannelSubset}
    \mathcal{C}_\Psi \coloneqq \left\{\attack{1} \circ \Psi \,\vert\, \attack{1} \in \CPTP((A')_1^n, B_1 E_1) \right\}.
\end{align}
Since \cref{eq:EpsilonSecurityTau} holds for all $\attack{1}'$, it also holds for any $\attack{1}' \in \mathcal{C}_\Psi$. Thus,  for any $\attack{1} \in \CPTP({A'}_1^n, B_1 E_1)$, we set $\attack{1}' = \attack{1} \circ \Psi$. With this identification, we have
\begin{align} \label{eq:tempsourceproof}
    \left(\identityMap_{X_1^n} \otimes \attack{1}'\right)[\xi_{X_1^n (A'')_1^n}]
    = \left(\identityMap_{X_1^n} \otimes \attack{1} \circ \Psi\right)[\xi_{X_1^n (A'')_1^n}]
    = \left(\identityMap_{X_1^n} \otimes \attack{1}\right)[\sigma_{X_1^n (A')_1^n}].
\end{align}
Now, using \cref{eq:tempsourceproof,eq:EpsilonSecurityTau} we obtain
\begin{align}
    \left\|
        \left(
            \QKDpostprocessingmap \circ \QKDGmapfullbeforeSR{n} \circ \attack{n} \circ \cdots \circ \QKDGmapfullbeforeSR{1}
            - \idealmap \circ \QKDpostprocessingmap \circ \QKDGmapfullbeforeSR{n} \circ \attack{n} \circ \cdots \circ \QKDGmapfullbeforeSR{1}
        \right)
        \circ \left(\identityMap_{X_1^n} \otimes \attack{1}\right)\left[\sigma_{X_1^n (A')_1^n}\right]
    \right\|_1 \leq \epssecure,
\end{align}
for all attack channels $\attack{1} \in \CPTP((A')_1^n, E_1 B_1)$, and for all $\{\attack{j}\}_{j=2}^n$.
This proves that the real protocol is $\epssecure$-secure.
\end{proof}

We note that the registers $A'$ and $A''$ are treated interchangeably throughout this manuscript. After applying the above lemma on source maps, the signal preparation is described as occurring in the $A''$ registers rather than the original $A'$ registers, and all analysis in \cref{sec:proof} now should be undertaken with the $A''$ register instead of the $A'$ register.    Note that \cref{lemma:sourcemapsecurityMEAT} is stated in considerable generality and places no restrictions on the form of the states prepared in either the real or virtual protocols. In particular, the real and virtual states may be correlated across the rounds. However, security proofs typically require that the state prepared in the virtual protocol be expressed as a tensor product: $\xi_{X_1^n (A'')_1^n} = \bigotimes_{j=1}^n \xi_{X_j(A'')_j}^{(j)}$.

\subsubsection{Tagging source map}
 Having explained the use of source maps, we will now construct a concrete source map and finite-dimensional virtual states for the real states produced in the decoy-state BB84 instantiation of the  \nameref{prot:abstractqkdprotocol}. The protocol described in \cref{subsec:detailprotdesc} is one where the real state prepared by Alice is IID, so that the state she prepares in each round is given by (\cref{eq:AliceSignalStatesDescription})
\begin{equation}
    \sigma_{(a,\mu,\testgenflag)} = \sum_{N=0}^\infty e^{-\mu}\frac{\mu^N}{N!}\ketbra{N}_a,
\end{equation}
where Alice's setting choice is given by  her choice of polarization $a$ and intensity $\mu$ and whether it is a test or generation round. We can construct a finite-dimensional virtual source, which emits tagged states \cite{gottesman_security_2004} where the space with photon number greater than $\ndecoy$ emits classical tags via the following lemma. 
\begin{lemma}[Tagged laser source]\label{lemma:tagging}
    Let $\sigma_{(a,\mu,\testgenflag)} = \sum_{N=0}^\infty e^{-\mu}\frac{\mu^N}{N!}\ketbra{N}_a$ be the real state prepared by Alice corresponding to setting choice $a,\mu$. Let the virtual state prepared by Alice be
    \begin{align} \label{eq:taggedStates}
        \xi_{(a,\mu,\testgenflag)} = \sum_{N=0}^\ndecoy e^{-\mu}\frac{\mu^N}{N!} \ketbra{N}_a + \left(1-\sum_{N=0}^\ndecoy e^{-\mu}\frac{\mu^N}{N!}\right) \ketbra{a,\mu},
    \end{align}
    where $\{\ket{a,\mu}\}_{a,\mu}$ form an orthonormal basis for a space orthogonal to the span of $\{\ket{N}\}_{N=0}^\ndecoy$.
    Then there exists a channel $\Psi_{\mathrm{tag}}$ such that $\Psi_{\mathrm{tag}}[\xi_{(a,\mu,\testgenflag})] = \sigma_{(a,\mu,\testgenflag)}$ for all setting choices $a,\mu,\testgenflag$.
\end{lemma}
\begin{proof}
    Define $\Psi_{\mathrm{tag}}$ to be the channel that projects onto $\ketbra{a,\mu}$ and prepares $\frac{1}{\left(1-\sum_{N=0}^\ndecoy e^{-\mu}\frac{\mu^N}{N!}\right)}\sum_{N=\ndecoy+1}^\infty e^{-\mu}\frac{\mu^N}{N!} \ketbra{N}_a$ for all $a$, $\mu$; and that acts as the identity on the space spanned by $\{\ket{N}\}_{N=0}^\ndecoy$. It is then straightforward to verify that $\Psi_{\mathrm{tag}}[\xi_{a,\mu}] = \sigma_{a,\mu}$.
\end{proof}

Thus, as shown in Lemma \ref{lemma:sourcemapsecurityMEAT} it is sufficient to show that the virtual protocol where Alice prepares the finite-dimensional states $\{\xi_{a,\mu}\}_{a,\mu}$ in each round is an $\epssecure$-secure PMQKD protocol. In this case the $n$-round source map $\Psi$ is constructed from the tagging source map $\Psi_{\mathrm{tag}}$ as $\Psi = \Psi_{\mathrm{tag}}^{\otimes n}$.

\subsection{Squashing maps}

We will now turn our attention to Bob, who still has infinite-dimensional POVM elements.
We address this through the use of squashing maps \cite{beaudry_squashing_2008,tsurumaru_squash_2010,fung_universal_2011,gittsovich_squashing_2014,zhang_security_2021,nahar_postselection_2024,nahar2025imperfect} where the measurements can be replaced with finite-dimensional protocols for the purposes of the security proof.
\begin{restatable}[Squashing maps]{lemma}{lemma:squashingmapsecurity} \label{lemma:SquashMapSecurityMEAT}
Let $\left\{\{\QKDGmapfullbeforeSR{j}\}_{j=1}^n,\QKDpostprocessingmap, \xi_{X_1^n (A'')_1^n} \right\}$ be a QKD protocol (see \cref{def:PMQKD}) where Bob measures his received state in round $j$ with the POVM $\left\{M_{i}^{(B_j)}\right\}$. That is, we have, $\QKDGmapfullbeforeSR{j} = \QKDmapfullwithoutBobMeas{j}\circ\measChannel{\left\{M_{i}^{(B_j)}\right\}}$, where $\measChannel{\left\{M_{i}^{(B_j)}\right\}} \in \CPTP(B_j, Y_j)$ and $\QKDmapfullwithoutBobMeas{j} \in \CPTP(X_j Y_j, S_j X_j Y_j \CP_j \CPhat{j})$. Suppose there exists quantum channels (squashing maps) $\Lambda_j \in \CPTP(B_j, Q_j)$ and measurement channels $\measChannel{\left\{F_{i}^{Q_j}\right\}} \in \CPTP(Q_j,Y_j)$ such that
\begin{align} \label{eq:squashingCondition}
    \measChannel{\left\{F_{i}^{(Q_j)}\right\}} \circ \Lambda_j = \measChannel{\left\{M_{i}^{(B_j)}\right\}} \quad \text{for all } j.
\end{align}
Moreover, suppose that the squashed protocol $\left\{\left\{\QKDmapfullwithoutBobMeas{j}\circ\measChannel{\left\{F_{i}^{(Q_j)}\right\}}\right\}_{j=1}^n,\QKDpostprocessingmap, \xi_{X_1^n (A'')_1^n} \right\}$ using the measurement channel $\measChannel{\left\{F_{i}^{(Q_j)}\right\}}$ is $\epssecure$-secure against all attacks $\attackSquash{j} \in \attackset{j}$, where 
\begin{align} \label{eq:attackSetConstraint}
    \attackset{j} \supset \Lambda_j \circ \CPTP(E_{j-1},B_jE_j) \quad \text{for all } j.
\end{align}
Then the real protocol $\left\{\left\{\QKDmapfullwithoutBobMeas{j}\circ\measChannel{\left\{M_{i}^{(B_j)}\right\}}\right\}_{j=1}^n,\QKDpostprocessingmap, \xi_{X_1^n (A'')_1^n} \right\}$ using the measurement channel $\measChannel{\left\{M_{i}^{(B_j)}\right\}}$ is $\epssecure$-secure against all attacks $\attack{j} \in \CPTP(E_{j-1},B_jE_j)$ in round $j$.
\end{restatable}

\begin{proof}
    We start with Eve's attack  $\attack{j}\in\CPTP(E_{j-1},B_jE_j)$ for the real protocol\newline $\left\{\left\{\QKDmapfullwithoutBobMeas{j}\circ\measChannel{\left\{M_{i}^{(B_j)}\right\}}\right\}_{j=1}^n,\QKDpostprocessingmap, \xi_{X_1^n (A'')_1^n} \right\}$ and reduce its security to that of the squashed protocol as follows
    \begin{align}
        \nonumber&\Big\|
        \Big(
            \QKDpostprocessingmap \circ \QKDmapfullwithoutBobMeas{n}\circ\measChannel{\left\{M_{i}^{B_n}\right\}} \circ \attack{n} \circ \cdots \circ \QKDmapfullwithoutBobMeas{1}\circ\measChannel{\left\{M_{i}^{B_1}\right\}}\circ\attack{1}\\
            \nonumber & 
            \phantom{\Big\|
            \Big(\QKDpostprocessingmap \circ \QKDmapfullwithoutBobMeas{n}\circ\measChannel{\left\{M_{i}^{B_n}\right\}}}
            - \idealmap \circ \QKDpostprocessingmap \circ \QKDmapfullwithoutBobMeas{n}\circ\measChannel{\left\{M_{i}^{B_n}\right\}} \circ \attack{n} \circ \cdots \circ \QKDmapfullwithoutBobMeas{1}\circ\measChannel{\left\{M_{i}^{B_1}\right\}}\circ\attack{1}
            \Big)
            \left[\xi_{X_1^n (A'')_1^n}\right]
        \Big\|_1\\
        =& \nonumber\Big\|
        \Big(
            \QKDpostprocessingmap \circ \QKDmapfullwithoutBobMeas{n}\circ\measChannel{\left\{F_{i}^{Q_n}\right\}} \circ \Lambda_n \circ \attack{n} \circ \cdots \circ \QKDmapfullwithoutBobMeas{1}\circ\measChannel{\left\{F_{i}^{Q_1}\right\}} \circ \Lambda_1\circ\attack{1}\\
            \nonumber & 
            \phantom{\Big\|
            \Big(\QKDpostprocessingmap \circ \QKDmapfullwithoutBobMeas{n}\circ\measChannel{\left\{M_{i}^{B_n}\right\}}}
            - \idealmap \circ \QKDpostprocessingmap \circ \QKDmapfullwithoutBobMeas{n}\circ\measChannel{\left\{F_{i}^{Q_n}\right\}} \circ \Lambda_n \circ \attack{n} \circ \cdots \circ \QKDmapfullwithoutBobMeas{1}\circ\measChannel{\left\{F_{i}^{Q_1}\right\}} \circ \Lambda_1\circ\attack{1}
            \Big)
            \left[\xi_{X_1^n (A'')_1^n}\right]
        \Big\|_1\\
        =& \nonumber\Big\|
        \Big(
            \QKDpostprocessingmap \circ \QKDmapfullwithoutBobMeas{n}\circ\measChannel{\left\{F_{i}^{Q_n}\right\}} \circ \attackSquash{n} \circ \cdots \circ \QKDmapfullwithoutBobMeas{1}\circ\measChannel{\left\{F_{i}^{Q_1}\right\}} \circ\attackSquash{1}\\
            \label{eq:squashedProtSecurityReduction} & 
            \phantom{\Big\|
            \Big(\QKDpostprocessingmap \circ \QKDmapfullwithoutBobMeas{n}\circ\measChannel{\left\{M_{i}^{B_n}\right\}}}
            - \idealmap \circ \QKDpostprocessingmap \circ \QKDmapfullwithoutBobMeas{n}\circ\measChannel{\left\{F_{i}^{Q_n}\right\}} \circ \attackSquash{n} \circ \cdots \circ \QKDmapfullwithoutBobMeas{1}\circ\measChannel{\left\{F_{i}^{(Q_j)}\right\}} \circ \circ\attackSquash{1}
            \Big)
            \left[\xi_{X_1^n (A'')_1^n}\right]
        \Big\|_1,
    \end{align}
    where the first equation follows from \cref{eq:squashingCondition}, and the second equation follows by relabelling $\attackSquash{j}\coloneqq \Lambda_j \circ \attack{j}$. Note that $\attackSquash{j}\in \attackset{j}$ follows from \cref{eq:attackSetConstraint}. Since this holds for every attack $\attack{j} \in \CPTP(E_{j-1}, B_j E_j)$ for the real protocol, minimizing the final expression in \cref{eq:squashedProtSecurityReduction} over all
$\attackSquash{j} \in \attackset{j} \subseteq \CPTP(E_{j-1}, B_j E_j)$
is guaranteed to yield a value that is less than or equal to the result of minimizing the first expression in \cref{eq:squashedProtSecurityReduction} over all
$\attack{j} \in \CPTP(E_{j-1}, B_j E_j)$.
This completes the proof, by identifying these minimizations with the corresponding definitions of security given in \cref{def:qkdsecurityChannelVersion}. 
\end{proof}

\begin{remark}
    Eve's set of squashed attacks can always be unrestricted, i.e, $\attackset{j} = \CPTP(E_{j-1},Q_jE_j)$ is always valid as it satisfies \cref{eq:attackSetConstraint}. This unrestricted special case is essentially the case studied in \cite[Lemma 6]{nahar_postselection_2024}.
    Using this special case, we could already apply the weight-preserving flag-state squasher to make Bob's systems finite-dimensional as described in \cite[Section IV. B. 1.]{nahar_postselection_2024}. However, it is easier to address detector imperfections via the flag-state squasher \cite[Theorem 1]{zhang_security_2021}, as described in \cite{nahar2025imperfect}, hence we choose to work with that squasher instead. 
\end{remark}

\cref{lemma:SquashMapSecurityMEAT} allows us to do two things. First, it allows us to analyse a squashed QKD protocol that uses some (presumably finite-dimensional) squashed POVM $\left\{F_{i}^{(Q_j)}\right\}$ instead of the actual QKD protocol that uses POVM $\left\{M_{i}^{(B_j)}\right\}$, where the POVMs are related via \cref{eq:squashingCondition}. Second, for the analysis of this squashed protocol, it allows us to restrict Eve's attacks to be in some set $\attackset{j}$ as long as \cref{eq:attackSetConstraint} holds.

\begin{remark} \label{remark:sourcemaplemma}
Recall that \cref{lemma:sourcemapsecurityMEAT}, which deals with source maps, is formulated assuming that attack channel sets $\attackset{j}$ comprise all CPTP maps acting on relevant Hilbert spaces, while \cref{lemma:SquashMapSecurityMEAT}, which deals with squashing maps, necessitates restricting the attack channel sets $\attackset{j}$ to subsets smaller than the full set of CPTP maps (see \cref{remark:attackset}). This apparent discrepancy, however, does not pose a fundamental issue. 

    This issue can be resolved, as is done in this work, by applying \cref{lemma:sourcemapsecurityMEAT} \emph{prior} to \cref{lemma:SquashMapSecurityMEAT}, ensuring that any restrictions on $\attackset{j}$ emerge only after the use of \cref{lemma:sourcemapsecurityMEAT}. Alternatively, one may revise the proof of \cref{lemma:sourcemapsecurityMEAT} explicitly to accommodate restricted attack channel sets $\attackset{j}$. In this case, one only requires that attack sets for the virtual and real protocols satisfy $\attackset{1,\mathrm{real}} \circ  \Psi \subseteq \attackset{1,\mathrm{virtual}} $.
 \end{remark}

We now give a brief description of a specific class of squashing maps called the flag-state squasher \cite[Theorem 1]{zhang_security_2021} before describing its application via \cref{lemma:SquashMapSecurityMEAT}.

\subsubsection{Flag-state squasher}

The flag-state squasher construction from Ref.~\cite{zhang_security_2021} applies to POVMs that are block-diagonal. Importantly, it applies to detection setups using threshold detectors, which are block-diagonal in the photon-number as described in \cref{subsec:detProtMeas}. Note that we drop the subscript $j$ (which denotes the protocol round) in the following, since we perform the same measurement in each round. 

Consider a photon-number cutoff $\nFSS$. Then, Bob's POVM element $M_i^{B}$ corresponding to any outcome $i$ can be written as a direct sum of two blocks: 
$$M^{(B)}_i = M_{i,m\leq \nFSS} \oplus M_{i,m > \nFSS},$$
where the blocks correspond to photon numbers of less than or equal to $\nFSS$ photons and greater than $ \nFSS$ photons. We refer to the subspace of less than oe equal to $\nFSS$ photons as the ``preserved subspace''. We let $\preservedSubspace,\nonpreservedSubspace \in \Pos(B) $ denote the projectors onto the preserved subspace and the space orthogonal to it.

The flag-state squasher $\Lambda_j\in\CPTP(B,Q)$ then connects (in the sense of \cref{eq:squashingCondition}) this infinite-dimensional POVM to a finite-dimensional one:
$$F^{(Q)}_i = M_{i,m\leq \nFSS} \oplus \ketbra{i},$$
where $\{\ket{i}\}$ is an orthonormal set of ``flag'' states spanning the subspace orthogonal to the preserved subspace.

Intuitively, this map does the following:
\begin{enumerate}
    \item Makes a QND measurement to check if the state is in the preserved subspace $\preservedSubspace$ or outside the preserved subspace $\nonpreservedSubspace$. In other words, check if the incoming state has less than or equal to $\nFSS$ or more than $\nFSS$ photons.
    \item If the state is in the preserved subspace, i.e, has $\leq \nFSS$ photons, the map acts trivially.
    \item If the state is outside the preserved subspace, i.e, has more than $\nFSS$ photons, the map completes the measurement, observes outcome $i$, and prepares a flag $\ketbra{i}$. In effect, this allows Eve to know the measurement outcome. 
\end{enumerate}
Importantly, the construction in the proof of \cite[Theorem~1]{zhang_security_2021} guarantees that the flag-state squashing map $\Lambda$ satisfies
\begin{align}
    \Lambda^\dag(F^{(Q)}_i) &= M^{(B)}_i, \label{eq:FSSConstraint}\\
    \Lambda^\dag(\pibarFlag) &= \nonpreservedSubspace, \label{eq:FSSPreservedSubspaceConstraint}\
 \end{align}
where $\pibarFlag\in\Pos(Q)$ is the projector onto the flag subspace.
\cref{eq:FSSConstraint} justifies referring to this map as a flag-state \textit{squasher}, as it guarantees that \cref{eq:squashingCondition} holds. \cref{eq:FSSPreservedSubspaceConstraint} is useful in the construction of Eve's set of restricted attacks on the squashed protocol $\attackset{}$ as described below.

First note that if Eve's squashed attack had no restriction, the protocol with squashed POVM elements would not be able to guarantee security. This is because in this case, Eve could always measure the state before sending Bob a flag of the measurement outcome she obtained. Thus, Eve would have full knowledge of Bob's measurement results, and no security is possible. This is typically addressed \cite[Section 3.4.2]{li_application_2020} by bounding the weight of the state outside the preserved subspace prior to squashing, and then restricting Eve's attack to satisfy this bound on the weight in the flag-space after squashing. Such a bound is obtained for passive detection setups by identifying a POVM element $\Gammacc$ which has non-zero minimum eigenvalue outside the preserved subspace $\lambdamin\left(\nonpreservedSubspace\Gammacc^{(B)}\nonpreservedSubspace\right)\eqqcolon \lambdamin$, as described in \cite[Eq. (27)]{nahar_postselection_2024}. This can be stated equivalently as the operator inequality $\Gammacc^{(B)} \geq \lambdamin\nonpreservedSubspace$. 

For linear optical setups, Ref.~\cite[Section VII. F.]{kamin2024improved} determines a generic method to find a non-zero bound on $\lambdamin$ when the outcome corresponding to $\Gammacc$ is chosen to contain all multi-clicks. For instance, for the polarization-encoded BB84 setup described in \cref{subsec:detProtMeas}, we can obtain $\lambdamin\geq 1-t^{\nFSS+1}-(1-t)^{\nFSS+1}$. (In this context, the operator inequality $\Gammacc^{(B)} \geq \lambdamin\nonpreservedSubspace$ represents the fact that if we do not observe multiple detectors clicking, we likely have a small number of photons.)
We can now use this to construct the set of Eve's restricted attacks that does not make explicit reference to her attacks on the infinite-dimensional optical protocol.

\begin{remark} \label{rem:noActiveMEAT}
    The method in Ref.~\cite[Section VII. F.]{kamin2024improved} to lower bound $\lambdamin$ crucially relies on the setup being a passive linear optical setup. In particular, lower bounding $\lambdamin$ for setups with active-basis choice remains an open problem. In fact, there is no known way to bound the weight outside the preserved subspace for active basis-choice BB84.\footnote{There is some work \cite{trushechkinSecurity2022} to bound this weight. However, this work, which applies to the asymptotic regime, does not bound $\lambdamin$ as is done for passive setups. Thus, the application of this work to MEAT is not straightforward.} Note that although \cite[Sec.~VII.F]{kamin2024improved} applies to lossy detection setups, it computes the $\lambdamin$ after pulling out loss\footnote{That is, the optical detection setup is first viewed a unitary transformation  implementing loss, followed by some other unitary transformation. The $\lambdamin$ computed in \cite[Sec.~VII.F]{kamin2024improved} corresponds to the POVM corresponding to the second unitary transformation, which does not include the loss.}, and therefore changes the POVM element in doing so. A method for computing $\lambdamin$ without altering the POVM elements was presented in \cite[Sec.~III]{wang2025phase}, which extends the method from \cite[Sec.~VII.F]{kamin2024improved}. 
\end{remark}

\begin{lemma} \label{lemma:FSSEveAttackSet}
    Let $\measChannel{\left\{M_{i}^{(B_j)}\right\}}$ be a block-diagonal measurement so that $\measChannel{\left\{F_{i}^{(Q_j)}\right\}}$ is the flag-state squashed measurement \cite[Theorem 1]{zhang_security_2021} with $\measChannel{\left\{F_{i}^{(Q_j)}\right\}}\circ \Lambda = \measChannel{\left\{M_{i}^{(B_j)}\right\}}$, where $\Lambda$ is the flag-state squasher \cite[Theorem 1]{zhang_security_2021}. Further, let $\Gammacc^{(B)} \geq \lambdamin \nonpreservedSubspace$ for some outcome $\mathrm{o}$.    
    Then the restricted set of attacks can be constructed as 
    \begin{equation}
        \attackset{j} \coloneqq \left\{\attackSquash{j}\in \CPTP(E_{j-1},Q_jE_j)\; \Big\vert \;\attackSquash{j}^\dag\left(\finGammacc^{(Q)}\otimes\id_{E_j}\right) \geq \lambdamin \attackSquash{j}^\dag\left(\pibarFlag\otimes\id_{E_j}\right) \right\}
    \end{equation}
    such that $\attackset{j} \supset \Lambda \circ \CPTP(E_{j-1},B_jE_j)$.
    Further, a marginal of this set, as defined in \cref{def:Qjcondition}, is given by
    
\begin{equation}
    \Qset{j} \coloneqq \left\{\Qattack{j}\in \CPTP(A',Q_j)\; \Big\vert \;\Qattack{j}^\dag\left(\finGammacc\right) \geq \lambdamin \Qattack{j}^\dag\left(\pibarFlag\right) \right\}.
    \end{equation}
\end{lemma}
\begin{proof}
    To prove the set inclusion $\attackset{j} \supset \Lambda \circ \CPTP(E_{j-1},B_jE_j)$, consider any $\attackSquash{j} \in \Lambda \circ \CPTP(E_{j-1},B_jE_j)$, such that $\attackSquash{j} = \Lambda \circ \attack{j}$ for some $\attack{j} \in \CPTP(E_{j-1},B_jE_j)$. Then we show that $\attackSquash{j}$ must be an element of the constructed $\attackset{j}$ as follows:
    \begin{align}
        \nonumber \attackSquash{j}^\dag\left(\finGammacc^{(Q)}\otimes\id_{E_j}\right) &= \attack{j}^\dag \circ \Lambda^\dag\left(\finGammacc^{(Q)}\otimes\id_{E_j}\right)\\
        \label{eq:2} &= \attack{j}^\dag\left(\Gammacc^{(B)}\otimes\id_{E_j}\right)\\
        \nonumber &\geq \lambdamin\ \attack{j}^\dag\left(\nonpreservedSubspace\otimes\id_{E_j}\right)\\
        \label{eq:4} &= \lambdamin\ \attack{j}^\dag \circ \Lambda^\dag\left(\pibarFlag\otimes\id_{E_j}\right)\\
        \nonumber &= \lambdamin\ \attackSquash{j}^\dagger\left(\pibarFlag\otimes\id_{E_j}\right),
    \end{align}
    where \cref{eq:2} follows from \cref{eq:FSSConstraint}, and \cref{eq:4} follows from \cref{eq:FSSPreservedSubspaceConstraint}. All other (in)equalities follow from definitions or conditions in the theorem statement.

    We now prove that $\Qset{j}$ is a marginal of $\attackset{j}$ (in the sense of \cref{def:Qjcondition}). Consider any $\attack{j}\in\attackset{j}$ and $\omega_{A_j E_{j-1} \widetilde{E}}$ such that $\omega_{A_j} = \sourcesymbol^{(j)}_{A_j}$. Since $\sourcesymbol^{(j)}_{A_j A'_j}$ is pure, there exists a channel $\mathcal{N}\in\CPTP(A_j',E_{j-1}\widetilde{E})$ that maps $\sourcesymbol^{(j)}_{A_j A'_j}$ to $\omega_{A_j E_{j-1} \widetilde{E}}$. We can use this to construct the channel $\Qattack{j}\in\Qset{j}$ as $\Qattack{j} \coloneqq \Tr_{E_j\widetilde{E}}\circ\attack{j}\circ\mathcal{N}$. This is clearly a channel, as it is constructed by composing channels. Thus, we show that this channel is in the constructed set $\Qset{j}$ by showing that it satisfies the required constraint $\Qattack{j}^\dag\left(\finGammacc\right) \geq \lambdamin \Qattack{j}^\dag\left(\pibarFlag\right)$ as follows
    \begin{align}
        \nonumber \Qattack{j}^\dag\left(\finGammacc^{(Q)}\right) &= \mathcal{N}^\dag\circ \attack{j}^\dag \circ \Tr_{E_j\widetilde{E}}^\dag \left[\finGammacc^{(Q)}\right]\\
        \label{eq:traceAdjoint} &= \mathcal{N}^\dag\circ \attack{j}^\dag\left[\finGammacc^{(Q)} \otimes \mathbb{I}_{E_j\widetilde{E}}\right]\\
        \label{eq:attackSetConstraintUsed} &  \geq \lambdamin\mathcal{N}^\dag\circ \attack{j}^\dag\left[\pibarFlag \otimes \mathbb{I}_{E_j\widetilde{E}}\right]\\
        \label{eq:traceAdjoint2} &= \lambdamin\mathcal{N}^\dag\circ \attack{j}^\dag \circ \Tr_{E_j\widetilde{E}}^\dag \left[\pibarFlag \right] = \Qattack{j}^\dag\left(\pibarFlag\right),
    \end{align}
    where \cref{eq:traceAdjoint,eq:traceAdjoint2} follow from the definition of the partial trace channel and \cref{eq:attackSetConstraintUsed} follows from the definition of $\attackset{j}$. This completes the proof.
\end{proof}

Note that our approach uses similar intuition to the approach in \cite[Appendix B]{kamin_renyi_2025}, though it differs from the analysis there in an important way. \cite[Appendix B]{kamin_renyi_2025} applies the flag-state squasher after converting the problem to a single-round optimization, while we opt to apply it before using the MEAT statement. Thus, our description of squashing \cref{lemma:SquashMapSecurityMEAT} is independent of the proof technique that will be used to prove security.
\begin{remark}
    Although \cref{lemma:SquashMapSecurityMEAT} is independent of proof technique, not all restrictions on Eve's attacks can be included into all proof techniques. For instance, the restriction on Eve's attacks described in \cref{lemma:FSSEveAttackSet} cannot be straightforwardly used with the postselection technique \cite{christandl_postselection_2009,nahar_postselection_2024}.
\end{remark}
\subsection{Final Security Statement for the decoy-state BB84 protocol}

\begin{theorem} \label{theorem:decoystatebb84securitystatement}[Security statement for the decoy-state BB84 protocol: \nameref{prot:abstractqkdprotocol} instantiated as described in \cref{subsec:detailprotdesc}]
Consider the QKD protocol described in \nameref{prot:abstractqkdprotocol} as instantiated in \cref{subsec:detailprotdesc}, denoted by $\left\{\{\QKDGmapfullbeforeSR{j}\}_{j=1}^n,\QKDpostprocessingmap, \sigma_{X_1^n (A')_1^n} \right\}$ (see \cref{def:PMQKD}). Consider the QKD protocol obtained after the application of the tagging source map and the flag-state squasher to this protocol, i.e, the protocol where Alice sends the tagged states  $(\xi_{(a,\mu,\testgenflag)})_{A''}$ instead of $(\sigma_{(a,\mu,\testgenflag)})_{A'}$, and Bob measures using squashed POVMs $\{F^{(Q)}_i\}_i $ instead of $\{M^{(B)}_i\}_i$. In other words, consider the QKD protocol  described by $\left\{\left\{\QKDmapfullwithoutBobMeas{j}\circ\measChannel{\left\{F_{i}^{(Q_j)}\right\}}\right\}_{j=1}^n,\QKDpostprocessingmap, \xi_{X_1^n (A'')_1^n} \right\}$, which is obtained after using source maps (\cref{lemma:sourcemapsecurityMEAT}) and squashing maps (\cref{lemma:SquashMapSecurityMEAT}). For this protocol, after applying the source-replacement scheme, let $\sourcesymbol^{(j)}_{A_jA''_j}$ denote the resulting source states in round~$j$, and let $\QKDGmap{j}$ denote the operations Alice and Bob perform in round~$j$.
 (These are the same as those that appear in the corresponding \nameref{prot:virtualprotocoll}, with state evolution described in \cref{lemma:stateevolution}). For each $j$ and every value of $\cP_1^{j-1}$, let $f_{| \cP_1^{j-1}}$ denote a tradeoff function on the register $\CP_j$. 
Let $\kappaQKDfunc{f_{|\cP_1^{j-1}}}{\sigma^{(j)}_{A_j}}{\QKDGmap{j}}$ be any value satisfying
\begin{equation} \label{eq:decoykappacompute}
\begin{aligned}
\kappaQKDfunc{f_{|\cP_1^{j-1}}}{\sigma^{(j)}_{A_j}}{\QKDGmap{j}}
&\leq
\inf_{\Qattack{j}\in\Qset{j}}
\Halpha[f_{|\cP_1^{j-1}}](S_j| \CP_j \CPhat_j \widehat{E})_{\QKDGmap{j}\left[\pf\left(\Qattack{j}\left[\sourcesymbol^{(j)}_{A_j A''_j}\right]\right)\right]} \\
\Qset{j} &= \left\{\Qattack{j}\in \CPTP(A''_j,Q_j)\; \Big\vert \;\Qattack{j}^\dag\left(\finGammacc\right) \geq \lambdamin \Qattack{j}^\dag\left(\pibarFlag\right) \right\}
\end{aligned}
\end{equation}
where $\Gammacc^{(B)},\finGammacc^{(Q)},\lambdamin$ are as defined in \cref{lemma:FSSEveAttackSet}, and define 
\begin{equation} 
\fhatfullQKD(\cobs) \coloneq \sum_{j=1}^n \left( f_{|\cP_1^{j-1}}(\cP_j) + \kappaQKDfunc{f_{|\cP_1^{j-1}}}{\sigma^{(j)}_{A_j}}{\QKDGmap{j}} \right).
\end{equation}

Then, the protocol is $(\epsPA + \epsEV)$-secure (according to \cref{def:qkdsecurityasymmetric}) if the key length $\lkey(\cobs)$ is chosen as
\begin{equation}
 \lkey(\cobs) = \max\left\{ 0 , \floor{\fhatfullQKD(\cobs)  - \leak(\cobs) - \ceil{ \log(\frac{1}{\epsEV})} -  \frac{\alpha}{\alpha-1} \log(\frac{1}{\epsPA}) + 2 } \right\}.
\end{equation}
\end{theorem}

%

\begin{proof}
   The required statement again follows from an appropriate combination of all the statements we have shown previously, and is similar to the proof of \cref{theorem:abstractsecuritystatement}.
\paragraph*{Reducing to honest authentication setting.} This part of the proof remains the same as in the proof of \cref{theorem:abstractsecuritystatement}.

\paragraph*{Applying source-maps and squashing maps.} We first apply \cref{lemma:sourcemapsecurityMEAT}, with the source map given by the tagging map in \cref{lemma:tagging}. We then apply \cref{lemma:SquashMapSecurityMEAT}, with the squashing map given by the flag state squasher from \cref{lemma:FSSEveAttackSet}.  This allows to reduce the security analysis to that of $\left\{\left\{\QKDmapfullwithoutBobMeas{j}\circ\measChannel{\left\{F_{i}^{(Q_j)}\right\}}\right\}_{j=1}^n,\QKDpostprocessingmap, \xi_{X_1^n (A'')_1^n} \right\}$. Notice that we are now in the regime where Alice sends finite-dimensional states, and Bob performs finite-dimensional measurements. Thus we simply repeat the proof of \cref{theorem:abstractsecuritystatement} from this point forth. 

\paragraph*{Security analysis for state obtained in \nameref{prot:virtualprotocoll}.}  This part of the proof remains the same as in the proof of \cref{theorem:abstractsecuritystatement}.

\paragraph*{Applying MEAT.}  This part of the proof remains the same as in the proof of \cref{theorem:abstractsecuritystatement}.

\paragraph*{Reformulations to remove dependence on Eve's dimensions and attack channels.}  This part of the proof remains the same as in the proof of \cref{theorem:abstractsecuritystatement}, except that we choose $\Qset{j}$ to be as stated in the theorem statement, and not $\CPTP(A'_j, Q_j)$.

\paragraph*{Relaxing the finite-dimensional Eve assumption.}  This part of the proof remains the same as in the proof of \cref{theorem:abstractsecuritystatement}.

This concludes our proof. 
   
\end{proof}

\section{Numerics} \label{sec:numerics}
\newcommand{\Fobs}{\mbf{F}^{\operatorname{obs}}}
To generate key rates from \cref{theorem:abstractsecuritystatement,theorem:decoystatebb84securitystatement}, we must compute the quantity $\kappaQKDfunc{f_{|\cP_1^{j-1}}}{\sigma^{(j)}_{A_j}}{\QKDGmap{j}}$ defined in \cref{eq:decoykappacompute}. We have shown in \cref{lemma:MEATuniform_lowerbound} that this quantity can be expressed as a finite-dimensional convex optimization problem, making it suitable for numerical computation. If the protocol of interest is such that Alice and Bob perform the same operation in each round, then all the maps $\QKDGmap{j}$ are identical and states $\sigma^{(j)}_{A_j}$ are identical. Furthermore, one can choose the same tradeoff function $f_{|\cP_1^{j-1}}$ for all $j, \cP_1^{j-1}$. Thus the dependence on $j$ disappears and in what follows
we drop the subscript $j$ and write it more compactly as $\kappaQKDfunc{f}{\sigma_A}{\QKDGmap{}}$:
\begin{equation} \label{eq:decoykappacomputenumerics}
    \begin{aligned}
        \kappaQKDfunc{f}{\sigma_A}{\QKDGmap{}}
        &\leq
        \inf_{\Qattack{}\in\Qset{}}
        \Halpha[f](S| \CP \CPhat \widehat{E})_{\QKDGmap{}\left[\pf\left(\Qattack{}\left[\sourcesymbol_{A A''}\right]\right)\right]}, \\
        \Qset{} &= \left\{\Qattack{}\in \CPTP(A'',Q)\; \Big\vert \;\Qattack{}^\dag\left(\finGammacc\right) \geq \lambdamin \Qattack{}^\dag\left(\pibarFlag\right) \right\}.
    \end{aligned}
\end{equation}

The main challenge in performing these computations is that we require a value guaranteed to be a \emph{lower bound} on the true infimum. Consequently, some optimization approaches such as gradient descent are not immediately suitable, as they do not provide guarantees of global optimality. Fortunately, several works have developed  numerical methods that attempt to yield values below the true QKD infimum \cite{kamin_renyi_2025,winick_reliable_2018,hu2022robust,navarro_finite_2025,he2024qics,lorente2025quantum,chung2025generalized,kossmann2024optimising,wang2019characterising} for a variety of relevant optimization problems. Among these,  Ref.~\cite{kamin_renyi_2025} applies directly to the specific problem we consider. We note that Ref.~\cite{navarro_finite_2025} presents an alternative numerical approach; however, it applies only to fixed-length protocols using MEAT and therefore does not address the optimization problem that arises here. Moreover, we note that \emph{any} method that reliably computes lower bounds on \cref{eq:decoykappacomputenumerics} suffices to compute key rates. Furthermore, while the reliability of the numerical methods is essential to ensure that the computed key rates are correct, it is important to note that the security analysis itself remains fully valid and independent of any numerical procedure. The proofs in \cref{theorem:abstractsecuritystatement,theorem:decoystatebb84securitystatement} concern only the abstract formulation of the key length expression and do not rely on, or make assumptions about, how the corresponding optimization problem is solved in practice.

We first briefly outline the approach used in Ref.~\cite{kamin_renyi_2025,winick_reliable_2018}. This discussion serves only to illustrate an approach to find reliable\footnote{These lower bounds are, in principle, reliable up to the numerical precision of the underlying floating-point implementation (e.g., MATLAB or similar numerical software).} lower bounds on the key rate using convex optimization techniques. In \cref{subsec:pickingf} we also briefly describe a method to pick a good function $f$.

\subsection{Computing $\kappaQKDfunc{f}{\sigma_A}{\QKDGmap{}}$ from a given tradeoff function $f$ using Frank-Wolfe algorithm}\label{subsec:computingkappa}

Let us fix a tradeoff function $f$, which we recall from \cref{remark:anyfworks} can be taken to be \textit{any} function. In \cref{subsec:pickingf} we will outline a method to pick a good function $f$.
We now outline the procedure from Ref.~\cite{kamin_renyi_2025}.
This is an extension of the framework in \cite{winick_reliable_2018} to exactly the objective function of interest to us.
The framework is essentially an application of the Frank-Wolfe algorithm \cite{frank1956algorithm}. Importantly, this uses the following property of any differentiable\footnote{In our case the function is not differentiable everywhere. This is remedied in \cite{kamin_renyi_2025} by applying a perturbation to construct a new function that is differentiable --- importantly, it is possible to rigorously account for the impact of this perturbation to ensure that  one still obtains a reliable lower bound (see \cite[Lemma 28]{kamin_renyi_2025} for details). This perturbation can be avoided \cite{navarro_finite_2025} by using facial reduction methods \cite{hu2022robust} instead.} convex function $g$:
\begin{equation}
    g(X^*) \geq g(X) - \Tr\left[X^\dag\nabla g(X)\right] + \min_{Y\in \mathcal{D}}\Tr\left[ Y^\dag \nabla g(X)\right],
\end{equation}
where $g(X^*)$ is the minimum value of the function, $\mathcal{D}$ is the domain over which we wish to optimise the function $g$, and $X$ is any point in the domain.

Thus, given any point $X$ in the domain, the task is to lower bound $\min_{Y\in \mathcal{D}}\Tr\left[ Y^\dag \nabla g(X)\right]$. When the domain $\mathcal{D}$ consists of linear or semidefinite constraints, this optimisation problem is a positive semidefinite problem (SDP). This can be lower bounded, 
for example, 
by looking at the dual problem and using any dual feasible point in the dual objective function\footnote{Note that Ref.~\cite{kamin_renyi_2025} does not explicitly evaluate the dual SDP, instead supposing that the SDP solver used found a sufficiently small gap between the original SDP value and its dual value.} as described in \cite{winick_reliable_2018}. More concretely, in our case the domain $\mathcal{D}$ is the set of attack channels $\Qset{}\subset \CPTP(A'',Q)$, and the function $g:\CPTP(A'',Q) \xrightarrow[]{} \mathbb{R}$ is defined by $g(\Qattack{}) = \Halpha[f](S| \CPhat  \widehat{E})_{\Tr_{\CP} \circ \QKDGmap{}\left[\pf\left(\Qattack{}\left[\sourcesymbol_{A A''}\right]\right)\right]},$ as defined in \cref{eq:decoykappacomputenumerics}.
    
Thus, the procedure from  Ref.~\cite{kamin_renyi_2025} reduces the optimization problem to a form where the gradient can be easily computed. We now list the key elements in this procedure, referring the interested reader directly to Ref.~\cite{kamin_renyi_2025} for further details. Note that while we try to keep our notation consistent with Ref.~\cite{kamin_renyi_2025}, some discrepancies are inevitable. Furthermore, we use the notation after squashing and source maps: so Bob measures the $Q$ systems, and Alice sends out states in $A''$ register, which are finite-dimensional. 

\begin{enumerate}
    \item \textbf{Choi–Jamiołkowski isomorphism}: The original optimisation is a function of the attack channel $\Qattack{}$, constrained to be in the set $\Qset{}$. This is rephrased in terms of the Choi–Jamiołkowski isomorphism that bijectively maps the channels to PSD matrices satisfying linear constraints. This rephrasing allows one to more directly view the final optimisation as an SDP.
    \item \textbf{Up-arrow to down-arrow {\Renyi} entropy} \cite[Corollary 9]{kamin_renyi_2025}: The \(\Halpha\) entropies involve an additional supremum compared to their down-arrow counterparts \(\HalphaDown\) (see \cref{def:sandwiched entropy}). Consequently, numerical computations are significantly easier to perform using \(\HalphaDown\) and \(\HalphaDown[f]\), rather than \(\Halpha\) and \(\Halpha[f]\). Since \(\HalphaDown \leq \Halpha\) and \(\HalphaDown[f] \leq \Halpha[f]\), and we are interested in a reliable lower bound on \cref{eq:decoykappacomputenumerics}, a lower bound on the down-arrow entropy gives us a lower bound on the up-arrow entropy.
    \item \textbf{Duality relations}: In order to compute gradients of $\HalphaDown[f]$, we would need to compute gradients of $\HalphaDown$. One way to compute this is via entropic duality relations \cite[Theorem 8]{kamin_renyi_2025}.
    \item \textbf{Block-diagonal simplifications for decoy-state QKD}: The block-diagonal structure of the prepared states can be utilised to speed up the numerics.
    In particular, as described in \cite[Appendix D]{li_application_2020}, the attack channel $\Qattack{}$ (and hence the corresponding Choi state) can be viewed as block-diagonal $\Qattack{} = \bigoplus_m \Qattack{m}$. This block-diagonal structure reduces the size of the optimisation problem as described in \cite[Sections VIII. and IX.]{kamin_renyi_2025}.
\end{enumerate}

In addition to these steps,
Ref.~\cite{kamin_renyi_2025} imposes some 
conditions in order to make the optimization problem more tractable.
In terms of the notation in this work, these conditions can be expressed as follows. First, it requires that the possible announcements in the $\CPhat$ register can be partitioned into two disjoint sets $\testset$ and $\genset$, where $\testset$ contains all announcements corresponding to rounds that Alice designates as \emph{test} rounds, while $\genset$ contains all announcements where she designates the round for \emph{key generation}. Furthermore, for each round the probability of it being a test round is a fixed value $\gamma$, i.e.~we have $
\sum_{\cP \in \testset} \Pr(\cP) = \gamma$ and $\sum_{\cP \in \genset} \Pr(\cP) = 1-\gamma$. Our protocol as described in \cref{subsec:detailprotdesc} indeed satisfies these requirements, as the value of the test/generation flag $\testgenflag$ is included in the public announcement register $\CPhat$ (see \cref{eq:fannounce}), and the same testing probability is indeed used in all rounds.
Ref.~\cite{kamin_renyi_2025} furthermore restricts the analysis to choices of tradeoff function where $f(\cP) $ is a fixed value  for all $\cP\in\genset$. This restriction simplifies the gradient computations later. 

Finally, we note that our formulation in this work involves two copies of the public announcement register $\CP,\CPhat$, one of which is given to Eve in each round. It is straightforward to argue that one of the copies can be removed from \cref{eq:decoykappacomputenumerics}, for instance by noting that $\CPhat$ can be computed from $\CP$ so we can apply data-processing  (\cref{lemma:DPIfweighted}) to remove it without changing the value of $\Halpha[f]$.

\subsection{Choosing a  good tradeoff function} \label{subsec:pickingf}

 
Ref.~\cite{kamin_renyi_2025} presents the following approach to obtaining the optimal choice for the tradeoff function, under some partially heuristic\footnote{We emphasize that while some of these assumptions are heuristic, these heuristics merely affect the ``performance'' of the protocol (in terms of expected key length), not any aspect of its security. This is again because \emph{every} choice of $f$ yields a secure protocol, via the security proof we have presented --- the use of some heuristics in the following steps merely means that we may not find the choice that truly yields the highest expected key rate.} assumptions, as follows. For notational convenience, let us write $\Fobs_{\cP_1^n}$ to denote the frequency distribution corresponding to the string $\cP_1^n$ (i.e.~exactly the frequency distribution $\freq_{\cP_1^n}$ in \cref{def:freq}, in slightly more compact notation).
Then, $\fhatfullQKD$ in \cref{eq:fhatfullQKDdefinition} can be written as follows, writing $\mbf{f}$ to denote the tradeoff function $f$ viewed as a vector $\mbf{f}\in\mathbb{R}^{|\mathcal{\CP}|}$:
\begin{align}
    \label{eq:fhatfull_obs}
    \fhatfullQKD(\cobs)=n\left(\mbf{f}\cdot\Fobs_{\cP_1^n}+\kappaQKDfunc{f}{\sourcesymbol_{A}}{\QKDGmap{}}  \right).
\end{align}
Recall that, by \cref{theorem:abstractsecuritystatement}, the key length $\lkey(\cobs)$ is given by
\begin{align}\label{eq:n_key_length}
 \lkey(\cobs) = \max\left\{ 0 , \fhatfullQKD(\cobs)  - \leak(\cobs) - \ceil{ \log(\frac{1}{\epsEV})} -  \frac{\alpha}{\alpha-1} \log(\frac{1}{\epsPA}) + 2  \right\}.
\end{align}
Observe that the expression of the key length in the above depends on the choice of the tradeoff function. Therefore, a natural question would be to find the optimal choice for the tradeoff function.

\newcommand{\leakfreq}{g_{\mathrm{EC}}}
\newcommand{\coarseCP}{\overline{C}}
\newcommand{\coarsecP}{\overline{c}}
\newcommand{\coarsef}{\overline{f}}
\newcommand{\coarsefvec}{\overline{\mbf{f}}}
\newcommand{\coarseqhon}{\overline{\mbf{q}}^{\operatorname{hon}}}

Let us consider the expected value of the key length with respect to announcement registers $\CP_1^n$, with respect to an honest behaviour of the channel connecting Alice and Bob, which is assumed to be IID. Given that the honest behaviour is IID, the expected value of $\Fobs_{\CP_1^n}$ is simply given by 
\begin{align}\label{eq:expectedFobs}
\mathop{\mathbb{E}}\limits_{\CP_1^n}\left(\Fobs_{\CP_1^n}\right) = \mbf{q}^{\operatorname{hon}},
\end{align}
where $\mbf{q}^{\operatorname{hon}}$ is the probability distribution produced by the honest behaviour on a single $\CP$ register.
Next, note that assuming the function $\leak$ has the specific structure $\leak(\cP_1^n) = \leakfreq(\Fobs_{\cP_1^n})$ for a concave function $\leakfreq: \mathbb{R}^{|\alphCP|} \to \mathbb{N}$, from \cref{eq:n_key_length,eq:expectedFobs}
one can obtain the following bound on the expected key length:
\begin{align}\label{eq:expected_key}
    \mathop{\mathbb{E}}\limits_{\CP_1^n}\left(\lkey(\cobs)\right)\geq\mbf{f}\cdot\mbf{q}^{\operatorname{hon}}+\kappa^{\operatorname{QKD}}- \leakfreq(\mbf{q}^{\operatorname{hon}}) - \ceil{ \log(\frac{1}{\epsEV})} -  \frac{\alpha}{\alpha-1} \log(\frac{1}{\epsPA}) + 2 .
\end{align} 

Then in principle, the optimal choice of tradeoff function in the above formula corresponds to the solution of the following optimization problem
\begin{align}\label{eq:opt_f}
    \sup_f\left(\mbf{f}\cdot\mbf{q}^{\operatorname{hon}}+\kappa^{\operatorname{QKD}}\right).
\end{align}
It was shown in~\cite[Lemma~4.12]{arqand_marginal_2025} 
that the optimal solution for $\mbf{f}$ in the above optimization problem is given by the optimal dual solution to the following constrained optimization:
\begin{equation}\label{eq:opt_f_dual}
\begin{aligned}
    \inf_{\bsym{\lambda}\in\mathbb{P}(\CP), \Qattack{}\in \Qset{}} &\qquad \frac{\alpha}{\alpha-1}D\left(\bsym{\lambda}\Vert\bsym{\nu}^{\Qattack{}}_{\CP} \right)+\sum_{\cP\in\supp\left(\bsym{\nu}^{\Qattack{}}_{\CP}\right)}\lambda(\cP)\Halpha(S|E\widetilde{E})_{\nu^{\Qattack{}}_{|\CP}},\\
    \text{ s.t.} &\qquad \mbf{q}^{\operatorname{hon}}-\bsym{\lambda} = \mbf{0},
    \end{aligned}
\end{equation}
where $\nu^{\Qattack{}}\defvar\QKDGmap{}\left[\pf\left(\Qattack{}\left[\sourcesymbol_{A A'}\right]\right)\right]$; informally, one can think of the state $\nu^{\Qattack{}}$ as a single-round state produced by the attack $\Qattack{}$, apart from an extra purification before Alice and Bob's measurements. (This is conceptually the same as the states described in in \cref{lemma:MEATuniform_lowerbound}; we merely introduce the notation $\nu^{\Qattack{}}$ to denote the objective function here more compactly.)
This result was obtained by performing a duality analysis on the optimization in~\cref{eq:opt_f} and showing that the resulting dual problem could be written as the optimization  in \cref{eq:opt_f_dual}. 

In summary, the optimal choice of the tradeoff function $\mbf{f}$ in~\cref{eq:opt_f} is given by the optimal value of the dual variable to the constraint $\mbf{q}^{\operatorname{hon}}-\bsym{\lambda} = \mbf{0}$ in \cref{eq:opt_f_dual}. 
We highlight that in  \cref{eq:opt_f_dual}, if the constraint is substituted into the objective to eliminate the $\bsym{\lambda}$ variable, we obtain an extremely similar optimization problem to the one that is relevant for computing secure key lengths for fixed-length protocols (\cref{eq:halpha}) --- essentially the only difference is that the latter requires specifying and optimizing over an acceptance set $\Sacc$, rather than just a single honest probability distribution $\mbf{q}^{\operatorname{hon}}$. Hence many numerical techniques that are relevant to the fixed-length case are also relevant to the variable-length case, and vice versa.

However, Ref.~\cite{kamin_renyi_2025} does not tackle exactly the optimization problem in~\cref{eq:opt_f} over all possible $f$, because as noted above, that work effectively restricts the allowed choices of $f$ to those where $f(\cP)$ takes the same value for all $\cP\in\genset$. With this restriction in mind, 
for any such choice of $f$, let us introduce a new function $\coarsef: \testset \cup \{\gen\} \to \mathbb{R}$ defined by $\coarsef(\cP)=f(\cP)$ for all $\cP\in\testset$ and $\coarsef(\gen)=f(\cP^\star)$ where $\cP^\star$ is some arbitrarily chosen value in $\genset$ (recall the value of $f(\cP^\star)$ is the same for any such choice). We also again write $\coarsefvec$ to denote that function viewed as a real-valued tuple. 
Optimizing over choices of $f$ under the stated restriction is clearly equivalent to optimizing over choices of $\coarsef$.
Furthermore,
let us similarly introduce a new register $\coarseCP$ that is a ``coarse-graining'' of $\CP$, by setting $\coarseCP=\CP$ whenever $\CP\in\testset$ but $\coarseCP=\gen$ whenever $\CP\in\genset$. Let $\coarseqhon$ be the distribution on $\coarseCP$ induced by $\mbf{q}^{\operatorname{hon}}$.\footnote{The formulation we present here is in fact a somewhat different perspective from Ref.~\cite{kamin_renyi_2025} --- in that work, a distinct register from $\CP$ was used for the public announcements, with $\CP$ being instead used to denote the ``coarse-grained'' version of the public announcements. Put another way, $\coarseCP$ in this work corresponds to $\CP$ in that work.}

Now observe that in \cref{eq:opt_f} we can rewrite $\mbf{f}\cdot\mbf{q}^{\operatorname{hon}}$ as $\coarsefvec\cdot\coarseqhon$; an analogous rewriting can be performed for $\kappa^{\operatorname{QKD}}$ to express it in terms of $\coarsef$ rather than $f$, but we do not spell out the details here.
The goal is now to optimize over choices of $\coarsefvec$. By again applying~\cite[Lemma~4.12]{arqand_marginal_2025} (up to some differences in notation), the optimal choice of $\coarsefvec$ is given by the optimal dual solution to 
\begin{align}
\begin{aligned}
    \inf_{\bsym{\lambda}\in\mathbb{P}(\coarseCP), \Qattack{}\in \Qset{}} &\qquad \frac{\alpha}{\alpha-1}D\left(\bsym{\lambda}\Vert\bsym{\nu}^{\Qattack{}}_{\coarseCP} \right)+\sum_{\coarsecP\in\supp\left(\bsym{\nu}^{\Qattack{}}_{\coarseCP}\right)}\lambda(\coarsecP)\Halpha(S|
    \CP \widehat{E}
    )_{\nu^{\Qattack{}}_{|\coarsecP}},\\
    \text{ s.t.}&\qquad  \coarseqhon-\bsym{\lambda} = \mbf{0},
    \end{aligned}
\end{align}
where we always implicitly extend the state with the classical coarse-grained register $\coarseCP$ described above. This was the method used in Ref.~\cite{kamin_renyi_2025} (together with a further relaxation from $\Halpha$ to $\HalphaDown$ as discussed in \cref{subsec:computingkappa}) to find optimal choices of $\coarsef$, and thus $f$ (under the stated restriction); this computation was somewhat easier compared to handling \cref{eq:opt_f_dual} directly, as it allowed for writing somewhat simpler formulas for the objective function and its gradient.
(Again, for fixed-length protocols  one can also find valid bounds on the key length using a similar coarse-graining to simplify the computations; we do not discuss the details here.)

\subsection{Key Rate Plots}
First, we note that the key length expressions obtained in this work for the decoy-state BB84 protocol are exactly identical to those in \cite[Eq.(97)]{kamin_renyi_2025}. Furthermore, the resulting optimization defining $\kappa^\mathrm{QKD}$ in \cref{eq:decoykappacomputenumerics} can be shown to coincide with the quantity $\kappa$ appearing in \cite[Eq.(122)]{kamin_renyi_2025}, under the restrictions and simplifications of that work described above. For decoy-state protocols, the relevant comparison is \cite[Eq.(147)]{kamin_renyi_2025}, where our approach yields a very slightly tighter bound\footnote{The differences in the constraints arise from a different treatment of terms corresponding to pulses containing more than $\ndecoy$ photons. Since such events occur with very low probability they do not lead to any appreciable difference in the key rate.} due to the fact that we use source maps to reduce Alice to finite-dimensions, while Ref.~\cite{kamin_renyi_2025} relaxes the constraints corresponding to higher photon numbers in the optimization problem instead. 

The protocol analyzed in Ref.~\cite{kamin_renyi_2025} uses exactly the same signal states, measurements, and public announcements as the protocol specified in this work (see \cref{subsec:detailprotdesc}), although it assumes that all public announcements occur after all states have been sent and measured. Consequently, the key length plots reported in \cite[Figs. 5,6]{kamin_renyi_2025} apply directly to our analysis as well. We refer the reader to that reference for the corresponding plots, and more details on the numerical methods.

\section{Handling imperfections} \label{sec:imperfections}

We now turn to imperfections in the quantum devices, which are inevitable in practical implementations. The proof technique developed in this work can, in principle, be extended to accommodate a wide variety of such imperfections. We only present a outline some approaches here, leaving a detailed analysis to future work. See Ref.~\cite{kamin_renyi_2025} for an analysis of source imperfections using MEAT.

Imperfections can be broadly classified into source and detector imperfections. We first discuss how the framework can be adapted when these imperfections are independent across the rounds of the QKD protocol, and then comment on the additional complications that arise in the presence of correlated imperfections.

\subsection{Independent imperfections}

By independent imperfections, we refer to scenarios in which the prepared quantum states or the implemented POVMs are not known exactly, but only approximately. The precise notion of ``approximate'' depends on the specific type of imperfection and the modeling assumptions adopted. For instance, one might assume that the prepared states are close in trace distance to some set of ideal reference states, or that the POVM elements correspond to an optical detection setup whose detector efficiencies and dark count rates are only known to be in some characterized range.

\subsubsection{Source imperfections}

We first assume that the states prepared $\{\rho_x\}$ belong to some subset $S_x$ for each setting choice $x$.
There are then two primary ways to address such source imperfections:
\begin{enumerate}
    \item \underline{Source maps:} We can use \cref{lemma:sourcemapsecurityMEAT} to prove security, much like the analysis to address the infinite-dimensional signal states via tagging. In particular, the virtual states need to be chosen such that there exists a channel $\Psi$ which acts on the virtual state to produce the real, imperfect state. Then, as stated in \cref{lemma:sourcemapsecurityMEAT}, it is sufficient to prove security for the protocol in which the virtual states are prepared.
    Note that this method of addressing imperfections is \emph{independent of proof-technique}, and not specific to MEAT.
    In this work, we already saw the tagging \cite{gottesman_security_2004} source map. Other source map constructions to address phase imperfections in QKD \cite{nahar_imperfect_2023}, and to design specialised protocols \cite{wang2019practical} robust to source imperfections already exist, though more constructions for other imperfections remains an interesting open question.  
    \item \underline{Adjusting fixed marginal constraint:} Note that the description of the states prepared enters the security proof only in the description of the fixed marginal constraint $\sigma_A$. Thus, imperfect state preparation such that the prepared states belong to some subset translates to this marginal not being fixed, but belonging to a subset that corresponds to the allowed state preparations. This can be incorporated into the numerical analysis (\cref{eq:decoykappacompute}) by optimising over all marginals in this set. Note that some relaxation of the resulting optimization may be necessary to ensure that it can be formulated as or bounded by a convex optimization. 
    Such an analysis was performed for intensity imperfections in decoy-state QKD \cite{kamin_renyi_2025}, and more generally\footnote{In this context, the generality often comes at the cost of performance. Thus, it is often advantageous to address the special case separately, as was done for intensity imperfections.} assuming only that the prepared states are close in fidelity to some reference states \cite{pereira2025optimal}. The latter case was shown to be applicable to side-channel attacks such as the Trojan Horse Attack in \cite[Methods]{curras-lorenzo_security_2024}.
\end{enumerate}

\subsubsection{Detector imperfections}

Analogously to the case of source imperfections, we first assume that the POVM representing the measurement in each round belongs to some well-defined subset of POVMs. Then, there are two primary ways to address detector imperfections:
\begin{enumerate}
    \item \underline{Noise channels:} Analogously to source maps, \cref{lemma:SquashMapSecurityMEAT} can be used to address detector imperfections. Once again, this involves proving the existence of a channel, termed noise channel, connecting the ideal and real POVMs. See \cite{nahar2025imperfect} for examples of such constructions, which build on the flag-state squasher. Applying these constructions require the the restricted set of attacks $\Qset{}$ in \cref{eq:decoykappacompute} to be adjusted depending on the extent of the imperfection.

    Moreover, noise channels can be used with the simple squasher \cite{gittsovich_squashing_2014} to prove security for active basis-choice BB84. This sidesteps the issue described in \cref{rem:noActiveMEAT} by avoiding the use of the flag-state squasher.
    \item \underline{Continuity bounds:} The description of the POVM elements appears in the objective function of the optimization problem in \cref{eq:decoykappacompute}, in the description of the protocol channel $\QKDGmap{}$. Hence, if it is known that the relevant subset of POVMs consists of elements that are close to a fixed reference POVM, this information can be used to infer that the state generated as the output of the protocol map $\QKDGmap{}$ is close to the reference state obtained from the protocol map defined from the fixed POVM (for any attack). Such deviations can then be incorporated as penalties to the key rate through continuity bounds for the ($f$-weighted) {\Renyi} entropies (see for e.g. Ref.~\cite{kamin_renyi_2025}).
\end{enumerate}

\subsection{Correlated imperfections}

The problem of establishing security in the presence of correlated source or measurement imperfections is substantially more challenging.
This is not an artifact of the MEAT framework, but is a challenging problem across proof techniques, with partial progress for sources \cite{nagamatsu2016security,pereira2020quantum,zapatero2021security,curras2023security,curras-lorenzo_security_2024,pereira2024quantum,marwah2024proving,li2025secure,mizutani_quantum_2019} and detectors \cite{burenkov2010security,tupkary_phase_2024,nahar2025imperfect,wang2025phase}.

There are, however, some promising directions in the literature that attempt to tackle correlated effects. For instance, \cite[Chater 7]{nahar_phd_2025} constructs an effective correlated noise channel for certain types of correlated detector scenarios, after which the MEAT framework can still be applied. On the source side, certain works \cite{curras2023security,curras-lorenzo_security_2024} have proposed that one can ``interleave’’ multiple QKD sub-protocols, effectively restoring independence between the rounds. This interleaving technique reduces the analysis to that of independent imperfections. Recent work \cite{li2025secure} refined this idea so that the interleaving is applied only for the entropy bounds, avoiding a complete decomposition into subprotocols. Further, Ref.~\cite{marwah2024proving} proposed the notion of a ``source test'' which effectively characterizes the source during the QKD protocol. While these represent promising directions, more work is required for a tight treatment of correlated effects without any hardware modifications and qubit assumptions.

\section{Conclusion} \label{sec:conclusion}
In this work, we provide a rigorous and complete security proof  for a wide class of QKD protocols. In particular, our proof can be applied to a fully specified decoy-state BB84 protocol with polarization encoding, which can in turn be obtained by specifying additional details of the protocol described in \cref{subsec:abstractprotdesc,subsec:detailprotdesc}. In obtaining this general result, we also resolve several issues that, while previously noticed, had not yet been fully
incorporated into a single complete QKD security proof. This includes the subtleties associated with classical authentication, timing of messages, and asymmetric aborts, as well as technical challenges arising from Eve’s potential use of infinite-dimensional systems. 

The analysis presented in this work constitutes a very general and flexible application of the MEAT technique to QKD security analysis. It begins by considering an abstract formulation of a generic QKD protocol (\nameref{prot:abstractqkdprotocol}) and establishes security under the assumption of finite-dimensional state preparation and measurements. This framework is then extended to practical optical implementations (which inherently involve infinite-dimensional systems) through the use of standard techniques such as squashing (\cref{lemma:SquashMapSecurityMEAT}) and source maps (\cref{lemma:sourcemapsecurityMEAT}). The framework is general and can, in principle, accommodate new classes of source and squashing maps that have not yet been formalized in the literature - for instance, those designed to model correlated imperfections. The entire analysis supports on-the-fly announcements, a feature that is important to efficiently utilize classical memory in implementations, and fully adaptive key rates (\cref{remark:adaptivef}), an aspect especially relevant for satellite QKD. Moreover, much of our analysis—such as the use of squashing, source maps, and the treatment of an
infinite-dimensional Eve—is not tied to the specific formulation of MEAT used here. These
components are likely to remain valid for future variations in QKD proof techniques, such as the
new EAT variants, for example.

As an example of the modularity and flexibility of our approach, consider the security proof of
the six-state protocol \cite{bruss_optimal_1998} implemented using, for instance, a single-photon
source. This can be obtained straightforwardly by (1) expressing the protocol as an instance of
our \nameref{prot:abstractqkdprotocol}, (2) applying the appropriate squashing and source map\footnote{Source maps are not required if the single-photon source is assumed to have ideal properties.},
and (3) evaluating the resulting optimization problem using Ref.~\cite{kamin_renyi_2025}. This procedure is outlined as a recipe in
\cref{subsubsec:recipe}. In fact, the present framework can be applied to essentially the same protocol considered in Ref.~\cite{mizutani2025protocolleveldescriptionselfcontainedsecurity}, which likewise aims to provide a self-contained and rigorous security analysis of decoy-state BB84, by following the steps outlined above.
The only minor distinction is that Ref.~\cite{mizutani2025protocolleveldescriptionselfcontainedsecurity} employs dual-universal$_2$
  hash families for privacy amplification, whereas we require  universal$_2$
  hash families.
  
More generally, as discussed in \cref{remark:adaptivef}, our results also extend to fully adaptive protocols - a feature that is not currently available in any other proof technique. Moreover, it may prove particularly valuable for scenarios such as satellite QKD, where the channel behaviour changes with time. While this work provides the complete theoretical framework required for such analyses, a detailed study of various ways to update the optimal trade-off function and its impact on achievable key rates in time-varying channels is left for future work.

Another important direction for future work is the detailed treatment of device imperfections. Although we have shown that a wide class of imperfections can, in principle, be incorporated within our framework, we have not performed explicit computations here. Moreover, one would ultimately need to construct mathematical models for various side-channels (such as those identified in Ref.~\cite{bsiImplementation}), and integrate them along with device imperfections into a unified, coherent analysis. Achieving this is an open and formidable task. Nonetheless,  the framework and results presented in this work provide the essential tools and foundations for such an endeavor.

In summary, this work establishes a rigorous, extensible, and implementation-oriented foundation for the security analysis of practical QKD protocols. We hope that the methods developed here will serve as a basis for future efforts to further close the gap between theoretical security proofs and real-world QKD systems.

\section*{Acknowledgments}
This article was written as part of the Qu-Gov project, which was commissioned by the German Federal Ministry of Finance. We thank the Bundesdruckerei --- Innovation Leadership and Team for their support and encouragement. In particular, we thank Holger Eble of the  Bundesdruckerei and Tobias Hemmert of the Federal Office for Information Security (BSI) for providing valuable feedback and comments on multiple drafts of this manuscript.   This work was conducted at the Institute for Quantum Computing, University of Waterloo, which
is funded by the Government of Canada through ISED. DT was partially funded by the Mike and Ophelia Lazaridis Fellowship.

We would like to thank Lars Kamin for a large number of valuable discussions on decoy-state methods, security proofs using MEAT, and associated numerics, specifically regarding Ref.~\cite{kamin_renyi_2025}. We thank Zhiyao Wang, Jerome Wiesemann, Martin Sandfuchs, and Aodh\'an Corrigan for valuable feedback on early drafts of this work. We are also grateful to John Burniston for helpful discussions on the numerical methods used to compute key rates. We thank Takaya Matsuura for pointing out the subtle difference between ideal universal$_2$ hash families and universal$_2$ hash families in the context of their use in the Leftover Hashing Lemma.
\bibliography{bibliography}

\appendix

\newcommand{\N}{\mathbb{N}} 
\newcommand{\ran}[1]{\mathrm{range}\left(#1\right)} 
\newcommand{\projChannel}[1]{\mathrm{P}_{#1}} 
\newcommand{\flagFinite}{\mathrm{flag}} 
\newcommand{\idChannel}{\mathrm{id}}
\newcommand{\linearOperators}[1]{\mathcal{L}\left(#1\right)} 
\newcommand{\traceClass}[1]{\mathcal{T}\left(#1\right)} 

\section{Infinite to finite argument} \label{app:infinitetofinite}

The goal of this appendix is to take a QKD security proof that holds for all finite dimensions of underlying systems, and lift it to a proof that holds for infinite dimensions.
First we define some notation that will be used through this section.

We denote the set of linear operators acting on a Hilbert space $\mathcal{H}$ by $\linearOperators{\mathcal{H}}$. The set of ``trace-class'' operators $\traceClass{\Hinf}$ is the subset of linear operators acting on $\Hinf$ which have finite trace. In particular, note that density operators are a subset of trace-class operators, i.e., $\dop{=}\left(\Hinf\right)\subset\traceClass{\Hinf}$.
Let $\Phi: \traceClass{\Hinf} \xrightarrow[]{} \traceClass{\Hinf}$ be a linear map on the set of trace-class operators acting on $\Hinf$. We say $\Phi$ is bounded if $\norm{\Phi}_{1\xrightarrow[]{}1}\coloneqq \sup_{X\in\traceClass{\Hinf}} \frac{\norm{\Phi(X)}_1}{\norm{X}_1} < \infty$. In particular, note that channels, and differences of channels are examples of such bounded linear maps.

Let $\{\Pi_d\}_{d\in\N}\subset \Pos\left(\Hinf\right)$ be a set of nested, finite-dimensional projectors such that $\overline{\bigcup_{d\in \N} \ran{\Pi_d}} = \Hinf$. Here, $\overline{X}$ denotes the closure of a set $X$, $\ran{\Pi_d}$ denotes the range of $\Pi_d$, and by ``nested'' we mean that $\ran{\Pi_d} \subset \ran{\Pi_{d'}}$ for all $d\leq d'$.
\begin{definition}[Projection channels] \label{def:projChannels}
    Let $\{\Pi_d\}_{d\in\N}\subset \Pos\left(\Hinf\right)$ be a set of projectors. Define the projection channels $\projChannel{d}: \traceClass{\Hinf} \xrightarrow{} \traceClass{\Hinf \oplus \mathbb{C}}$ corresponding to the set of projectors $\{\Pi_d\}_{d\in\N}$ as
\begin{equation}
    \projChannel{d}[X] = \Pi_d X \Pi_d + \left(1-\Tr\left[\Pi_d X\right]\right)\ \ketbra{\flagFinite},
\end{equation}
where $\ket{\flagFinite}$ is orthogonal to $\Hinf$.
\end{definition}
It is easy to verify that these projection channels are indeed completely positive and trace-preserving.
\begin{remark} \label{remark:abuseOfNotation}
    There are some comments to be made at this point regarding a minor abuse of notation we shall use. At various points in what follows, we shall take channels (such as the protocol channels $\QKDGmap{j}$ and attack channels $\attack{j}$) that are defined on $\traceClass{\Hinf}$, and apply them on some output state of $\projChannel{d}$, which would be finite-dimensional. In each such case, we implicitly extend the channels by pinching across the decomposition $\Hinf \oplus \mathrm{span}(\ket{\flagFinite})$ and acting as the identity on the flag block as follows:
    \begin{align*}
        \ketbra{\flagFinite}{v} &\mapsto 0\\
        \ketbra{v}{\flagFinite} &\mapsto 0\\
        \ketbra{\flagFinite} &\mapsto \ketbra{\flagFinite},
    \end{align*}
    for any $\ket{v}\in\Hinf$. Additionally, we might write $\projChannel{d}$ as acting on some finite-dimensional space $\mathbb{C}^{d'}$. In such cases, we implicitly interpret $\mathbb{C}^{d'}$ as embedded in $\Hinf$.
\end{remark}
Note that any trace-class operator $\rho\in\traceClass{\Hinf}$ can be approximated by its finite projections \cite[Theorem 4.3]{gohberg2013classes}, i.e.,
\begin{equation} \label{eq:approxProp}
    \lim_{d\xrightarrow[]{}\infty} \Pi_d \rho \Pi_d \xrightarrow[]{\norm{\cdot}_1} \rho,
\end{equation}
where the projections $\{\Pi_d\}_{d\in\mathbb{N}}$ can be taken to be any set of nested, finite-dimensional projectors such that $\overline{\bigcup_{d\in \N} \ran{\Pi_d}} = \Hinf$.
Abusing notation similarly as in \cref{remark:abuseOfNotation}, we observe that \cref{eq:approxProp} implies that
\begin{equation} \label{eq:approxPropChannel}
    \lim_{d\xrightarrow[]{}\infty} \projChannel{d}[ \rho] \xrightarrow[]{\norm{\cdot}_1} \rho,
\end{equation}
where $\rho$ is understood to be embedded in $\traceClass{\Hinf\oplus\mathbb{C}}$.
This gives us the following lemma.
\begin{lemma} \label{lemma:pointwiseLimitOfProjChannels}
    Let $\{\Pi_d\}_{d\in\N}\subset \Pos\left(\Hinf\right)$ be any set of nested, finite-dimensional projectors such that $\overline{\bigcup_{d\in \N} \ran{\Pi_d}} = \Hinf$, and $\{\projChannel{d}\}_{d\in \N}$ be the set of projection channels corresponding to these projectors (as defined in \cref{def:projChannels}). Then,
    \begin{equation} \label{eq:pointwiseLimitOfProjChannels}
        \lim_{d \xrightarrow[]{}\infty} \norm{\Phi\circ\projChannel{d}\circ\Psi\left[\sigma\right]}_1 = \norm{\Phi\circ\Psi\left[\sigma\right]}_1,
    \end{equation}
    for any trace-class operator $\sigma\in\traceClass{\Hinf}$, and any bounded linear maps $\Phi$,$\Psi: \traceClass{\Hinf}\xrightarrow[]{}\traceClass{\Hinf}$.
\end{lemma}
\begin{proof}
    First we remark that
    \begin{equation} \label{eq:strongConvergenceOfProjChannels}
        \lim_{d \xrightarrow[]{}\infty} \norm{\Phi\circ\left(\idChannel-\projChannel{d}\right)\circ\Psi\left[\sigma\right]}_1 \leq \lim_{d \xrightarrow[]{}\infty} \norm{\Phi}_{1\xrightarrow{}1}\norm{\left(\idChannel-\projChannel{d}\right)\circ\Psi\left[\sigma\right]}_1
        = 0,
    \end{equation}
    which follows from the boundedness of $\Phi$, and \cref{eq:approxPropChannel}.
    The rest of the proof follows by using \cref{eq:strongConvergenceOfProjChannels} with the triangle inequality of $\norm{\cdot}_1$ to upper and lower bound $\underset{d \xrightarrow[]{}\infty}{\lim} \norm{\Phi\circ\projChannel{d}\circ\Psi\left[\sigma\right]}_1$ by $\norm{\Phi\circ\Psi\left[\sigma\right]}_1$, along with the sandwich theorem.
    First, we lower bound as follows
    \begin{align}
        \lim_{d \xrightarrow[]{}\infty} \norm{\Phi\circ\projChannel{d}\circ\Psi\left[\sigma\right]}_1 &= \lim_{d \xrightarrow[]{}\infty} \norm{\Phi\circ\left(\idChannel-\projChannel{d}\right)\circ\Psi\left[\sigma\right]-\Phi\circ\Psi\left[\sigma\right]}_1\\
        \label{eq:UsingTriangleIneqLowerBound} &\geq \lim_{d \xrightarrow[]{}\infty} \abs{\norm{\Phi\circ\left(\idChannel-\projChannel{d}\right)\circ\Psi\left[\sigma\right]}_1-\norm{\Phi\circ\Psi\left[\sigma\right]}_1}\\
        &= \abs{\lim_{d \xrightarrow[]{}\infty}\norm{\Phi\circ\left(\idChannel-\projChannel{d}\right)\circ\Psi\left[\sigma\right]}_1-\norm{\Phi\circ\Psi\left[\sigma\right]}_1}\\
        \label{eq:UsingStrongConvergenceOfProjChannelsForLowerBound} &= \norm{\Phi\circ\Psi\left[\sigma\right]}_1,
    \end{align}
    where \cref{eq:UsingTriangleIneqLowerBound} is the (reverse) triangle inequality of the 1-norm, and \cref{eq:UsingStrongConvergenceOfProjChannelsForLowerBound} follows from \cref{eq:strongConvergenceOfProjChannels}.
    Similarly, we upper bound as follows
    \begin{align}
        \lim_{d \xrightarrow[]{}\infty} \norm{\Phi\circ\projChannel{d}\circ\Psi\left[\sigma\right]}_1 &= \lim_{d \xrightarrow[]{}\infty} \norm{\Phi\circ\left(\idChannel-\projChannel{d}\right)\circ\Psi\left[\sigma\right]-\Phi\circ\Psi\left[\sigma\right]}_1\\
        \label{eq:UsingTriangleIneqUpperBound} &\leq \lim_{d \xrightarrow[]{}\infty} \norm{\Phi\circ\left(\idChannel-\projChannel{d}\right)\circ\Psi\left[\sigma\right]}_1+\norm{\Phi\circ\Psi\left[\sigma\right]}_1\\
        \label{eq:UsingStrongConvergenceOfProjChannelsForUpperBound} &= \norm{\Phi\circ\Psi\left[\sigma\right]}_1,
    \end{align}
    where \cref{eq:UsingTriangleIneqUpperBound} is the triangle inequality of the 1-norm \cref{eq:UsingStrongConvergenceOfProjChannelsForUpperBound} follows from \cref{eq:strongConvergenceOfProjChannels}. Thus, using the sandwich theorem\footnote{The sandwich theorem states that if a function is bounded above and below by two functions with the same limit, then it shares that limit.}, we obtain \cref{eq:pointwiseLimitOfProjChannels}.
\end{proof}

We will now use \cref{lemma:pointwiseLimitOfProjChannels} in QKD security proofs. Recall that from the security definition in \cref{def:qkdsecurityChannelVersion}, we wish to find an upper bound $\epssecure$ on
\begin{equation}
    \left\|
        \left(
            \QKDpostprocessingmap \circ \QKDGmapfullbeforeSR{n} \circ \attack{n} \circ \cdots \circ \QKDGmapfullbeforeSR{1}\circ\attack{1}
            - \idealmap \circ \QKDpostprocessingmap \circ \QKDGmapfullbeforeSR{n} \circ \attack{n} \circ \cdots \circ \QKDGmapfullbeforeSR{1}\circ\attack{1}
        \right)
        \left[\sigma_{X_1^n (A'')_1^n}\right]
    \right\|_1 \leq \epssecure,
\end{equation}
for a QKD protocol defined by $\left\{\{\QKDGmapfullbeforeSR{j}\}_{j=1}^n,\QKDpostprocessingmap, \sigma_{X_1^n (A')_1^n} \right\}$, and for any sequence of attacks $\attack{j}\in \CPTP(E_{j-1},B_jE'_j)$. For full generality, we set $E'_j,E_j = \mathcal{H}_\infty$.   

%
\newcommand{\epsSecureds}{\epssecure_{d_0,\dots,d_n}(\attack{1},\dots, \attack{n})}
\newcommand{\epsSecureFixedA}{\epssecure(\attack{1},\dots, \attack{n})}
We now prove the main result of this appendix.
\begin{theorem} \label{thm:finiteToInfiniteFixedA}
    Let $\{\Pi_d\}_{d\in\N}\subset \Pos\left(\Hinf\right)$ be any set of nested, finite-dimensional projectors such that $\overline{\bigcup_{d\in \N} \ran{\Pi_d}} = \Hinf$, and $\{\projChannel{d}\}_{d\in \N}$ be the set of projection channels corresponding to these projectors (as defined in \cref{def:projChannels}).
    Given a QKD protocol defined by $\left\{\{\QKDGmapfullbeforeSR{j}\}_{j=1}^n,\QKDpostprocessingmap, \sigma_{X_1^n (A')_1^n} \right\}$, 
    let us define
    \begin{equation}
        \epsSecureFixedA\coloneqq \norm{\left(\idChannel-\idealmap\right)\circ \QKDpostprocessingmap \circ \QKDGmapfullbeforeSR{n} \circ \attack{n} \circ \cdots \circ \QKDGmapfullbeforeSR{1}\circ\attack{1}
        \left[\sigma_{X_1^n (A'')_1^n}\right]}_1,
    \end{equation}
    and suppose 
    $\underset{d_n\xrightarrow[]{}\infty}{\lim}\dots\underset{d_0\xrightarrow[]{}\infty}{\lim}\sup_{\attack{1},\dots,\attack{n}} \epsSecureds$ exists, where
    \begin{equation}
        \epsSecureds\coloneqq \norm{\left(\idChannel-\idealmap\right)\circ \QKDpostprocessingmap \circ \QKDGmapfullbeforeSR{n} \circ \projChannel{d_n} \circ \attack{n} \circ \cdots \circ \QKDGmapfullbeforeSR{1}\circ\projChannel{d_1}\circ\attack{1}\circ\projChannel{d_0} \left[\sigma_{X_1^n (A'')_1^n}\right]}_1.
    \end{equation}
    Then,
    $$\sup_{\attack{1},\dots,\attack{n}}
    \epsSecureFixedA
    \leq \lim_{d_n\xrightarrow[]{}\infty}\dots\lim_{d_0\xrightarrow[]{}\infty}\sup_{\attack{1},\dots,\attack{n}} \epsSecureds,$$
    where the suprema can be taken over any subset $\attackset{j}$ of channels.
    
\end{theorem}
\begin{proof}
First note that given a fixed sequence of attack channels $\{\attack{j}\}_{j=1}^n$, we have
\begin{equation} \label{eq:securityFixedAFiniteToInfinite}
    \lim_{d_n\xrightarrow[]{}\infty}\dots\lim_{d_0\xrightarrow[]{}\infty}
    \epsSecureds =
    \epsSecureFixedA
    ,
\end{equation}
as can easily be seen by repeated applications of \cref{lemma:pointwiseLimitOfProjChannels} (noting that differences of CPTP linear maps are bounded linear maps). 
The claimed result immediately follows by noting
\begin{align}
        \label{eq:usingThm}\sup_{\attack{1},\dots,\attack{n}} \epsSecureFixedA &= \sup_{\attack{1},\dots,\attack{n}} \lim_{d_n\xrightarrow[] {}\infty} \dots \lim_{d_0\xrightarrow[] {}\infty} \epsSecureds\\
        \label{eq:swappingLimAndSup}&\leq  \lim_{d_n\xrightarrow[] {}\infty} \dots \lim_{d_0\xrightarrow[] {}\infty} \sup_{\attack{1},\dots,\attack{n}} \epsSecureds,
    \end{align}
    where \cref{eq:usingThm} follows from \cref{eq:securityFixedAFiniteToInfinite}, and \cref{eq:swappingLimAndSup} holds because moving a supremum into any limit always yields a valid upper bound (as long as the relevant limits all exist):
    \begin{align}
    \text{For any set $S$ and function $f:\mathbb{N} \times S \to \mathbb{R}$, \quad}
        \sup_{x \in S} \lim_{m\to\infty} f(m,x) &\leq \sup_{x \in S} \lim_{m\to\infty} \sup_{x' \in S} f(m,x') \label{eq:boundwithsupremum}\\
        &= \lim_{m\to\infty} \sup_{x \in S} f(m,x) \label{eq:relabelsupremum},
    \end{align}
    where \cref{eq:boundwithsupremum} follows from the trivial upper bound $f(m,x) \leq \sup_{x' \in S} f(m,x')$, and in \cref{eq:relabelsupremum} we simply remove the superfluous outer supremum and relabel the inner supremum.
\end{proof}
We observe that for the above theorem, it can be seen from the proof that in fact the order in which we take the limits does not matter, as long the limit of $\sup_{\attack{1}\dots\attack{n}}\epsSecureds$ exists for that ordering.
Essentially we have proven that if we have security proof techniques that hold for all finite dimensions, 
then we can obtain security for the infinite-dimensional case by taking a suitable limit of the finite-dimensional results.

\begin{remark} \label{remark:infiniteEve}
    In the rest of the text, we make use of the above result only when Alice and Bob act on finite-dimensional spaces. In this case, we can restrict the projections in \cref{thm:finiteToInfiniteFixedA} to act only on Eve's space. For this special case, \cref{thm:finiteToInfiniteFixedA} states that a security proof assuming Eve's system has dimensions $\{d_0+1, \dots , d_n+1\}$\footnote{The `$+1$' is for the extra flag space.} with security parameter $\epssecure_{d_0,\dots,d_n}$ extends to a security proof with no assumptions on Eve's space with security parameter $\lim_{d_n\xrightarrow[] {}\infty} \dots \lim_{d_0\xrightarrow[] {}\infty}\epssecure_{d_0,\dots,d_n}$.
\end{remark}

\end{document}